\documentclass[12pt]{article}

\usepackage[utf8]{inputenc}
\usepackage[T1]{fontenc}
\usepackage[english]{babel}
\usepackage{eurosym}
\usepackage{amssymb, amsmath, amssymb, amsmath}
\usepackage[mathscr]{eucal}
\usepackage{float}
\usepackage{makeidx}
\usepackage{fancybox}
\usepackage[notref,notcite,final]{showkeys}
\usepackage[none]{hyphenat}
\usepackage[hypcap=false]{caption}
\usepackage{lmodern} 
\usepackage{graphicx}
\graphicspath{ {images/} }
\usepackage{hyperref} 
\usepackage{tablefootnote}
\usepackage{siunitx}
\usepackage{subcaption}
\usepackage{setspace}
\usepackage{enumitem}
\usepackage{isotope}
\usepackage{svg}
\usepackage{bm}
\usepackage{cancel}
\usepackage{braket}
\usepackage{dsfont}
\usepackage{bbold}
\usepackage{cite}

\usepackage{scalerel}

\usepackage[a4paper, margin=15mm]
{geometry}

\newcommand{\ketbra}[2]{\ensuremath{\ket{#1}\!\bra{#2}}}

\begin{document}
\setlength{\abovedisplayskip}{15pt}
\setlength{\belowdisplayskip}{15pt}

\title{Master in Physics of Complex Systems \\ \vfill
Design of Novel Coupling Mechanisms \\ between\\ Superconducting Flux Qubits \vfill}
\author{MSc Thesis\\ Author: Gabriel Jaumà Gómez\\ Advisor: Juan José García Ripoll}


\newpage

\begin{figure}
    \centering
    \includegraphics[width=3cm]{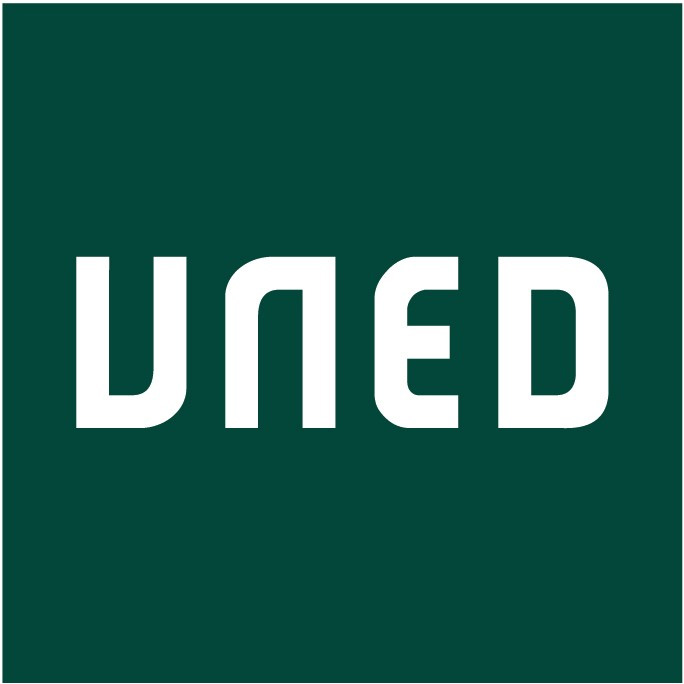}
\end{figure}
\maketitle\thispagestyle{empty}

\pagebreak

\begin{abstract}
    We have analyzed and proposed coupling mechanisms between Three Josephson Junction Flux Qubits (3JJQ). For this, we have developed a numerical method to extract the effective Hamiltonian of a system of coupled qubits via the Schrieffer-Wolff transformation (SWT). This method is a more efficient version of the one proposed by Ref.~\cite{consani2020effective}. We then give a comprehensive introduction to the 3JJQ, and study it analytically by approximating its potential with a Harmonic well. With a clear understanding of the 3JJQs, we use the SWT to gain intuition about their effective dipolar interaction with the electromagnetic field, and use that intuition to propose and study analytically and numerically the capacitive coupling of two 3JJQs via a non-tunable capacitor, and the inductive coupling of two 3JJQs via a tunable Josephson Junction (dc-SQUID), showing that we are able to reproduce non-stoquastic Hamiltonians in the strong-coupling regime.
\end{abstract}

\pagebreak
\tableofcontents
\clearpage\pagenumbering{arabic}\pagebreak

\section{Introduction}

The field of Quantum Information Science (QIS) was born at the end of the XX century and has been flourishing theoretically ever since. In the last decades the technological obstacles that hindered its experimental realization have started to crumble, giving way to an explosion in variety, complexity and applicability of the field. Propelled by said explosion this thesis will theoretically analyze and design the experimental implementation (with a certain level of abstraction) of \textbf{interacting pairs of qubits}—the smallest useful amount of quantum information—made of superconducting circuits, with the specific goal of trying to expand the available interactions of these systems. We will focus on a specific type of superconducting qubit: the three Josephson junction flux qubit (3JJQ) \cite{orlando,orlando2}, also called the persistent current qubit.  

This project is relevant to three sub-fields of QIS: quantum simulation, adiabatic quantum computation, and gate-based quantum computation. First, quantum simulation \cite{simulator1,simulator2,simulator3} studies the development and analysis of experimental setups that reproduce diverse quantum models. This opens endless possibilities to probe the quantum world since those models might be practically impossible to come by in Nature or at least difficult to control. The relation between quantum simulation and this project is as follows: many quantum models, for example the transverse field Ising model,  are based on pairwise interacting spins, which are two-level systems that can be simulated with qubits, thus, increasing the range of realizable qubit-qubit interactions will increase the range of simulable models, for example from the Ising model to the Heisenberg model. Second, adiabatic quantum computation \cite{adiabatic1,adiabatic2,adiabatic3,adiabatic4} is based on the idea that one can design a Hamiltonian whose ground state describes the solution of a given problem. This ground state can be found via the quantum annealing procedure, thus, increasing the range of available interactions increases the range of Hamiltonians that can be implemented and solved by quantum annealing. In a ideal scenario we know that the problem of finding the ground state of an arbitrary Hamiltonian belongs to the complexity class QMA complete \cite{biamonte2008realizable}, and hence any problem or algorithm governed by the rules of quantum mechanics can be translated to this. We can see that quantum simulation and adiabatic quantum computation are very closely related: the ability to simulate arbitrary quantum Hamiltonians would open the door for universal quantum computation. Third and finally, quantum gates can be constructed with interacting pairs of flux qubits\cite{gate1,gate2}, thus, creating and controlling new interactions between superconducting qubits is relevant for this field.

Nowadays the literature thrives with a substantial number of implementations of qubits, each one based on a different physical support: photons, electrons, nucleus, atoms, quantum dots, superconducting circuits, etc. Why have we chosen to study qubits based on superconducting circuits, and not on another platform? Within the scope of this thesis there are four requirements\footnote{Ref. \cite{divincenzo2000physical} gives a more general and profound discussion of this topic.} that we will demand from a qubit: 
\begin{enumerate}
\item The qubit must have two well-defined, distinct and measurable states which we will label as $|0\rangle$ and $|1\rangle$. 
\item We must be able to create arbitrary superpositions of the qubit states that are protected from decoherence and perdure in time.
\item The qubit must have an anharmonic spectrum, \textit{i.e.} the energy gap $\hbar\omega_{01}$ between the qubit states must not be an integer fraction of the gap to other neighboring states.
\item It must be possible to create and tune interactions between qubits.
\end{enumerate}
One of the features that sets apart superconducting circuits from the other physical platforms for qubits is that they are macroscopic. If we were only concerned about requirements one, two and three any microscopic entity of the list would have been an excellent candidate. This is because we can constrain their infinite range of available quantum states to a two level system that is inherently anharmonic, and additionally their microscopic nature makes them intrinsically easier to isolate from the macroscopic world to preserve their quantum state. Nevertheless, this positive trait becomes negative when considering the fourth requisite, because the interaction degrees of freedom of  microscopic entities are strictly limited by the physical laws that govern them, \textit{e.g.} one cannot add charge to some photons and make them interact electrostatically. This of course does not mean that microscopic entities are invalid as qubits, but it does mean that the complexity of the macroscopic world is a great advantage for this endeavor: with superconducting circuits one can in principle use any imaginable circuit to create qubits, and then connect them with any other imaginable circuit to obtain and control a certain type of interaction.

To build an intuition for the rest of the document lets consider the superconducting flux qubit that will be thoroughly discussed in the following pages, the three Josephson junction flux qubit or persistent current qubit. Roughly speaking, a persistent current qubit is superconducting loop of Aluminum where a current of electrons flows without dissipation. This system can be in an infinite number of clockwise or counterclockwise current states, with the only condition that the total magnetic flux threading the loop at any time must be an integer multiple of the magnetic flux quantum $\Phi_0=h/2e$\footnote{In section \ref{s. 3JJQ} and appendix. \ref{a. Josephson junction} we will thoroughly discuss this idea.}. Since this system has an infinite number of equally spaced states available it would not be a good candidate for a quibt, hence, the solution is to interrupt the superconducting loop with Josephson junctions. The role of these junctions is to introduce anharmonicity in the spectrum of the qubit, $\omega_{01}\ne\omega_{12}$, ensuring that in practice the system will only have two states available. This states are actually two degenerate ground states: clockwise and counterclockwise currents of equal intensity that ensure that the total flux quanta threading the superconducting loop is an integer number.

With these qubits in mind lets qualitatively imagine how one could produce and control interactions among them. Since these qubits are in a sense magnets, one way to make them interact is simply by placing them together and allowing the magnetic flux of each to thread the loop of the other in a mutual inductance interaction. If we could control the mutual inductance coefficient of the qubits we could then control this interaction. Another way to make these qubits interact would be to connect them with a wire. This way the charges on one qubit would affect the other. If we then interrupted the wire with a capacitor, controlling its capacitance would allow us to control the intensity of this interaction. The kind of questions that we will try to answer in this thesis are: Are these interactions fundamentally different? Can we achieve interactions in all the degrees of freedom of the qubit's Hamiltonian? What is their physical origin? How do they scale with the parameters of the qubit?

To give a more specific motivation for this work we have to start by asking the question: Why have we chosen flux qubits among the different flavours of superconducting qubits? The answer to this question comes in two parts. First, we have chosen the flux qubit because it is the second most used superconducting qubit, and the most used qubit—the transmon qubit\footnote{A transmon qubit is a superconducting circuit whose two available states are states of different charge in a superconducting island.}—has already been widely studied regarding qubit-qubit interactions and is currently being used by the largest players in the field (Google, IBM) at the core of their gate-based quantum computers. Second, we have chosen the flux qubit because of the recent publication of several works regarding the qubit-qubit interactions between flux qubits.

The scientific team of D-Wave published an article\cite{ozfidan2020demonstration} stating that they had demonstrated a non-stoquastic Hamiltonian with capacitively coupled rf-SQUIDS, a type of flux qubits. A Hamiltonian is said to be non-stoquastic \textit{in a certain base} if all its off-diagonal elements are real and positive. What is the importance of demonstrating a non-stoquastic Hamiltonian? This question is truly about whether a Hamiltonian can capture or not all of the richness of the quantum realm. If a Hamiltonian $H$ is stoquastic in a certain base one can shown that its partition function $Z(\beta) =
\text{Tr} \exp(-\beta H)$ can be written as a sum of products of non-negative weights, and hence the estimation of equilibrium properties of such Hamiltonian thorugh stochastic Monte Carlo methods is exempt of the well known sign problem. Note, however, that stoquasticity and the absence of a sign problem does not necessarily imply polynomial-time convergence of standard Monte-Carlo methods \cite{obshastings2013obstructions,obsjarret2016adiabatic,obsbringewatt2018diffusion}.

As we have highlighted, an important fact regarding the stoquasticity of a  Hamiltonian is that it is a base-dependent feature. Any Hamiltonian has at least one base in which it is stoquastic, its base of eigenstates. This does not mean that a Hamiltonian which is non-stoquastic in a specific base can be easily translated to a base where it is stoquastic and hence simulable without the sign problem. 

Quantum Monte Carlo algorithms are usually defined in a local base, \textit{i.e.} a base in which all base vectors are product states of individual qubit states. Local basis are used because otherwise the representation of the base vectors requires exponential resources, thus, the useful way to determine whether a Hamiltonian would be hard to simulate is to determine if there exist a transformation to a local basis in which the Hamiltonian is stoquastic. Several studies\cite{stoqhalverson2020efficient,stoqklassen2020hardness,stoqioannou2020sign,halverson2020efficient} have elucidated under which conditions such transformation is computationally easy and hence the system can be efficiently simulated with a quantum Monte Carlo algorithm. Precisely in this line of thought and with great importance for this work, A. Ciani and B.M. Terhal recently showed \cite{stoqciani2021stoquasticity} that if the capacitive coupling between rf-SQUIDS demonstrated by D-Wave\cite{ozfidan2020demonstration} is sufficiently small, the non-stoquasticity of the effective qubit Hamiltonian can be avoided by performing a canonical transformation prior to projecting onto the effective qubit Hamiltonian. Thus, the main questions concerning the current state of the art that we will try to answer is: Does this apply to 3JJQs? How big must the capacitive coupling be to render the transformation \cite{stoqciani2021stoquasticity} invalid? How can we achieve tunable non-stoquastic Hamiltonians with 3JJQs?

A sufficient condition \cite{biamonte2008realizable} to ensure that there exists no efficient transformation to make stoquastic a non-stoquastic two-qubit Hamiltonian is: i) with the presence of the single-qubit fields $\sigma^x$ and $\sigma^z$;  and ii) with the presence of two-qubit local interactions of the type $\sigma^x\sigma^x$ and $\sigma^z\sigma^z$. 
$$
\hat{H} = h_{1} \sigma_{1}^{x}+h_{2} \sigma_{2}^{x}+ \Delta_{1} \sigma_{1}^{z}+\Delta_{2} \sigma_{2}^{z}+  J_{xx} \sigma_{1}^{x} \sigma_{2}^{x}+J_{zz} \sigma_{1}^{z}\sigma_{2}^{x}
$$
If these interactions were tunable and if we could scale this system to an arbitrary number of qubits, then we would find ourselves in front of a programmable Hamiltonian that could be used for universal adiabatic quantum computation via quantum annealing.

As mentioned, a relevant part of the stoquastic/non-stoquastic argument is that it it regards Hamiltonians expressed in a local basis, thus, the first step in order to determine the stoquasticity of a Hamiltonian is to obtain an effective Hamiltonian express in a basis of products between singe qubit operators. This is not an obvious task since the Hamiltonian of a quantum circuit is usually extracted from its Lagrangian and written in a infinite base of charge or flux variables of the nodes of the circuit. Throughout the literature the effective Hamiltonian is usually obtained via perturbation theory. This method is problematic because the perturbation parameter is related to the intensity of the coupling and hence one can only consider weakly-coupled systems, which is precisely the regime that we want to exit to avoid the transformation proposed by \cite{stoqciani2021stoquasticity}. However, there exists a method called the Schrieffer-Wolff transformation \cite{schrieffer1966relation,bravyi2011schrieffer,consani2020effective} that is equivalent to the summation of all the orders of perturbation theory, and hence is an exact transformation which allows to study coupling schemes of arbitrary intensity. Nevertheless, this transformation involves infinite-sized matrices and the literature lacks an efficient numeric recipe.

The structure of this document is as follows. In section \ref{s. Heff} we will present two different ways to obtain an effective Hamiltonian , first thorugh perturbation theory and then with the Schrieffer-Wolff transformation, giving an explicit efficient numeric recipe. The rest of the document will be devoted to study flux qubits and couplings between them. In sections \ref{s. 3JJQ} we will introduce general flux qubits and the 3JJQ, and then we will use the harmonic approximation to explain the physical origin and scaling of the qubit's properties. In section \ref{s. Electromagnetic interaction with flux qubits} we will analyze the interaction of these qubits with the electric and magnetic fields, and in section \ref{s. Couplings} we will use this knowledge to  propose schemes that create the desired interactions. Finally we will validate these designs, studying them in the strong interaction regime to determine their stoquasticity by calculating their exact properties and interaction terms using the numerical scheme proposed in section \ref{s. Heff}.
\pagebreak

\section{Low-energy effective Hamiltonian}
\label{s. Heff}
A qubit is a quantum entity whose ground and first-excited energies—the low-energy subspace or qubit subspace—are clearly separated from the rest of the spectrum. There are two ways in which this can happen. If the qubit has a positive anharmonicity\footnote{Anharmonicity is usually defined as $(\omega_{12}-\omega_{01})/\omega_{01} $, thus $\omega_{12}>\omega_{01}$ means $(\omega_{12}-\omega_{01})/\omega_{01} >0$.}  ,  $\omega_{12}>\omega_{01}$, then the qubit states are the slow degrees of freedom of the system and hence can be measured and interacted with via low frequency electromagnetic radiation without shifting the system towards the high space of the spectrum. This is the case of the persistent current qubits that we will consider in this document. If the qubit has a negative anharmonicity, $\omega_{12}<\omega_{01}$, then the qubit states are the fast degrees of freedom and hence can also be isolated from the rest of the spectrum. This is the case of the transmon qubits and capacitively shunted flux qubits that will not be discussed in this document.

\begin{figure}[H]
    \centering
    \includegraphics[width=0.95\linewidth]{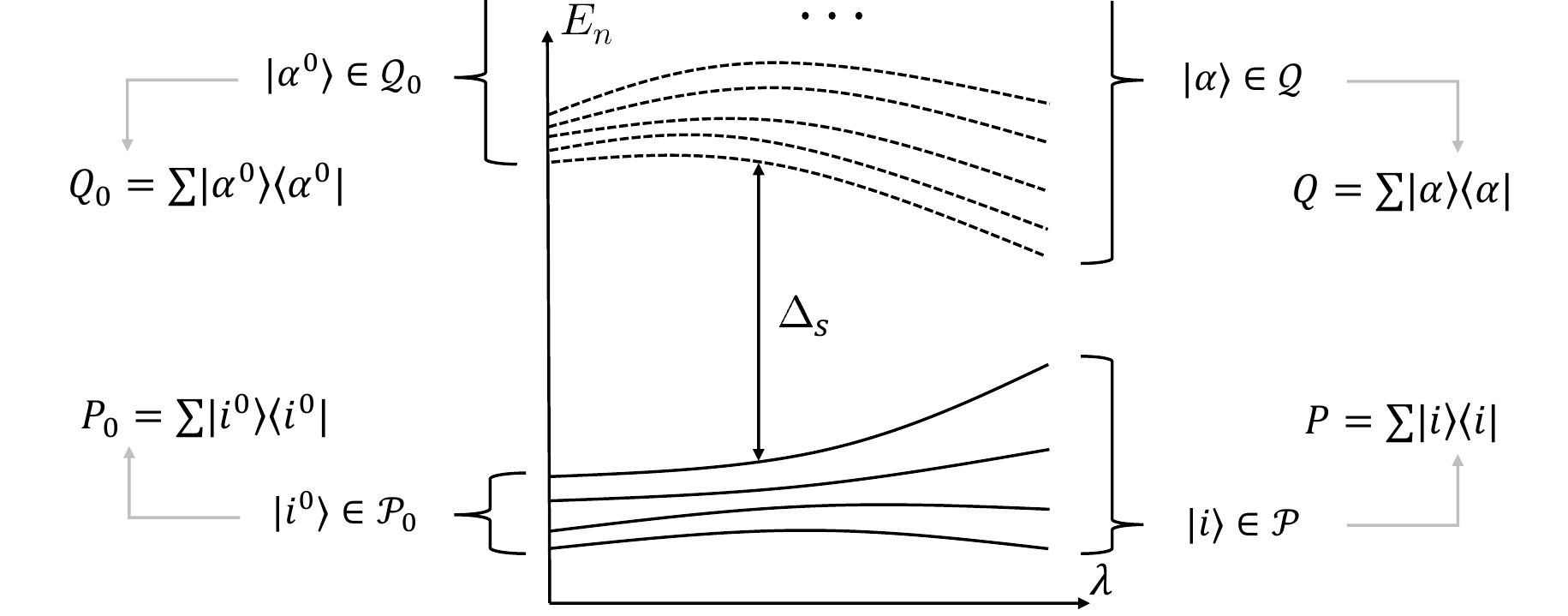}
    \caption{Spectrum, subspaces and projectors of the perturbed Hamiltonian \eqref{e. perturbed hamiltonian} as a function of $\lambda$.  }
    \label{f. subespacios y proyectores}
\end{figure}

Calling $\mathcal{P}_0$ and $\mathcal{Q}_0$ to the low and high energy subspaces of a quantum system whose Hamiltonian is $H_0$, and calling $P_0$ and $Q_0$ to the projectors of the respective subspaces, allows us obtain the low-energy Hamiltonian or qubit Hamiltonian, $H_0^q$, simply by projecting $H_0$ in $\mathcal{P}_0$, \textit{i.e.} simply by writing $H_0$ in its base of eigenstates and keeping only the first two terms:
$$
H_0^q = P_0H_0P_0 = \sum_{i=0,1} E_i^0 \ketbra{i^0}{i^0}, \quad \ket{i^0} \in \mathcal{P}_0\,.
$$
Since a qubit is a two-level system we can always express any operator in the qubit space as a linear combination of the Pauli matrices:
$$
\quad \sigma^{I}=\left[\begin{array}{cc}
1 & 0 \\
0 & 1
\end{array}\right]\,,
\sigma^{x}=\left[\begin{array}{ll}
0 & 1 \\
1 & 0
\end{array}\right], \quad \sigma^{y}=\left[\begin{array}{cc}
0 & -i \\
i & 0
\end{array}\right], 
\quad \sigma^{z}=\left[\begin{array}{cc}
1 & 0 \\
0 & -1
\end{array}\right]\,.
$$
For instance, the qubit Hamiltonian can always be expressed in the eigenstate base as function of the qubit's gap, $\Delta^0= E_1^0- E_0^0$:
\begin{equation}
\begin{gathered}
H_\text{0}^q
=
\left[\begin{array}{cc}
E_0^0 & 0 \\
0 & E_1^0\\
\end{array}\right]
=
\frac{1}{2}\left[\begin{array}{cc}
 (E_1^0+E_0^0)-\Delta^0 & 0 \\
0 &  (E_1^0+E_0^0)+\Delta^0\\
\end{array}\right]
= 
\frac{\Delta^0}{2}\sigma^z + \frac{1}{2}(E_1^h+E_0^h)\sigma_I\, ,\\
H_\text{0}^q= \frac{\Delta^0}{2}\sigma^z\,.
\end{gathered}
\label{e. qubit hamiltonian}
\end{equation}
Note that in the last part of this eq. we have chosen to ignore the term proportional to the identity because an energy offset only adds an undetectable global shift to the phase of the wavefunctions. This is what we mentioned in the introduction as a local qubit base. How can we write the Hamiltonian of two qubits in a local qubit base? In this case we have to define a new base as the tensor product of the basis of the qubits,
\begin{equation*}
\{\ket{0},\ket{1}\} \otimes \{\ket{0},\ket{1}\} =\{\ket{00}, \ket{01}, \ket{10}, \ket{11}\}
\end{equation*}
and then express the Hamiltonian of the complete system in this base. If we want to use a local qubit base then we will have to leave the resulting Hamiltonian as a sum of Kronecker products of two Pauli matrices, the first defined in the subspace of the first qubit and the second in the subspace of the second qubit.  For example, if the two qubits are non-interacting, the simplest case, then we can write the system's Hamiltonian as:
\begin{equation*}
H_\text{0}^{1+2} = H_\text{0}^{1}\otimes\mathbb{1}^{2} +\mathbb{1}^{1}\otimes H_\text{0}^{2} = \frac{\Delta^0_{1}}{2}\sigma^z_{1}\otimes\sigma^I_{2} +\frac{\Delta^0_{2}}{2}\sigma^I_{1} \otimes\sigma^z_{2}\,.
\end{equation*}
Throughout this thesis we will often discuss equations just as the one shown above, thus, we will simplify the notation omitting the Kronecker product symbol and also omitting the identity operator. For example, the equation above would be written as:
\begin{equation*}
H_\text{0}^{1+2} = \frac{\Delta^0_{1}}{2}\sigma^z_1 +\frac{\Delta^0_{2}}{2}\sigma^z_2\,.
\end{equation*}
What would happen if the two qubits were interacting, for example, through a perturbative operator acting on both of the qubit's subspaces? In this case the Hamiltonian of the system would be written as the sum of the non-interacting qubits plus the interaction:
\begin{equation}
    H = H_\text{0}^{1+2} + \lambda H_\text{int} = H_0 + \lambda V\,.
    \label{e. perturbed hamiltonian}
\end{equation}
If the perturbation $\lambda V$ does not modify the spectrum so much as to prevent the distinction of a low energy subspace—\textit{i.e.} as long as $\Delta_s$ shown in fig. \ref{f. subespacios y proyectores} is positive— then we can always repeat the same process as before: diagonalize the Hamiltonian and keep the eigenstates corresponding to the low energy subspace. This, however, is not useful, since our goal is to study the Hamiltonian of the coupled system in a local qubit basis, defining the local qubit basis as the tensor product of the qubit's basis \textit{before} the interaction. Thus, our problem can be stated as follows: how can we obtain a low-energy \textbf{effective} Hamiltonian, $H_\text{eff}$, whose spectrum matches that of the low-energy subspace of $H$, but which is expressed in the low-energy subspace of $H_0$, $\mathcal{P}_0$. The difficulty to define $H_\text{eff}$ in $\mathcal{P}_0$ arises from the fact that the perturbation $V$ couples the low and high eigenstates of the original Hamiltonian $H_0$, thus, $\mathcal{P}\, \bcancel{=}\, \mathcal{P}_0$ but rather $\mathcal{P}\in \mathcal{P}_0\cup\mathcal{Q}_0$.

\noindent The analysis in the following pages will be general and applicable to any quantum system with a distinguishable low-energy subspace, however, we can state that the need to obtain a low-energy effective Hamiltonian is particularly notorious in the study of quantum computing hardware, where one has to bridge the gap between experimental quantum computing and quantum information science, translating measurements of real devices into ensambles of interacting two-level quantum entities.

\noindent To summarize, our goal is to obtain a low-energy effective Hamiltonian $H_\text{eff}$ defined by three requirements:

\begin{enumerate}[label=(\alph*)]
    \item The effective Hamiltonian $H_\text{eff}$ must be entirely expressed in $\mathcal{P}_0$.
    \item The energy spectrum of the effective Hamiltonian $H_\text{eff}$ must match the low-energy spectrum of the perturbed Hamiltonian $H$.
    \item  The effective Hamiltonian $H_\text{eff}$ must be Hermitian.
\end{enumerate}

\noindent This problem will be tackled from two perspectives. First we will develop a perturbative Schrieffer Wolf transformation which allows to obtain a $H_\text{eff}$ with a precision up to the desired order, and then we will introduce a numerical scheme to calculate an exact Schrieffer-Wolff transformation which is equivalent to the summation of all the orders of the previous perturbation theory.

The so-called Schrieffer-Wolff transform (SWT) was originally introduced in \cite{schrieffer1966relation}, generalized for quantum many-body systems by \cite{bravyi2011schrieffer}, and applied to superconducting circuits as intended in this work by \cite{consani2020effective}. This transformation satisfies conditions (a), (b) and (c) and can be qualitatively understood as follows \cite{bravyi2011schrieffer}: 
\begin{enumerate}
    \item Define $R_\mathcal{P}$ as the reflection operator that flips the sign of all the vectors of the subspace $\mathcal{P}$ and leaves the vectors of the orthogonal subspace $\mathcal{Q}$ intact,
     $R_\mathcal{P} = P - Q = 2P - I\,. $
    \item In the one-dimensional case if one has two non-orthogonal vectors $\psi$ and $\phi$, the double reflection $R_\phi R_\psi$ is a rotation that rotates the two-dimensional subspace generated by $\psi$ and $\phi$ an angle $\theta$, being $\theta$ the double of the angle between $\psi$ and $\phi$. Therefore, the operator $\sqrt{R_\phi R_\psi}$ is a rotation leading from $\psi$ to $\phi$.
    \item For higher dimensions the rotation described above has the same effect but is harder to visualize. Explicitly, one can show that two non-orthogonal subspaces $\mathcal{P}$ and $\mathcal{P}_0$ have a direct rotation from one to the other if their projectors $P$ and $P_0$ satisfy $||P-P_0||<1$, which turns out to be equivalent to both subspaces having the same dimension\cite{bravyi2011schrieffer}. This condition will be the only limit on the magnitude of the perturbation parameter, which is natural since otherwise we would not have a low-energy subspace to extract. In this case, the rotation between the subspaces can be defined as:
    \begin{equation}
        U = \sqrt{R_{\mathcal{P}_0} R_\mathcal{P}} = \sqrt{(P_0 - Q_0)(P - Q) }
        \label{e. def SWT}
    \end{equation}
    \item The rotation $U$ is the SWT transformation and by definition it guarantees
        \begin{equation}
            \begin{aligned}
            UPU^\dagger &= P_0 \quad \rightarrow \quad UP =P_0U\,, \\
            UQU^\dagger & = Q_0\quad \rightarrow \quad UQ = Q_0U\,.
            \end{aligned}
            \label{e. SWT on P and Q}
        \end{equation}
    \end{enumerate}
As a summary, the SWT is a rotation that maps operators from $\mathcal{P}$ to $\mathcal{P}_0$ as long as both subspaces have the same dimension, thus, applying it to the low energy perturbed Hamiltonian, $H_{low}=PHP$, yields a new Hermitian operator $H_{\text{eff}}$ defined in $\mathcal{P}_0$ and whose energy spectrum is that of $H_{low}$, exactly what we desired.
\begin{equation}
    H_{\text{eff}} =U H_{low} U^\dagger = UPHPU^\dagger = P_0UHU^\dagger P_0
    \label{e. H_eff}
\end{equation}
This method, of course, is numerically incomplete since $U$ is defined in terms of the infinite matrices $Q$ and $Q_0$. One way to cope with this is to truncate this matrices to a certain size\cite{consani2020effective}, however, this matrices will become very large for systems with multiple qubits, and a convergence study is necessary each time it is applied to determine the truncation of the matrix $Q$. In section \ref{s. Heff SWT} we will introduce a numerical scheme more efficient than this, but before we will derive a perturbative SWT that will allow us to perform analytical estimates throughout the thessis.

\subsection{Perturbative Schrieffer-Wolff transformation}
\label{s. Heff perturbation} 

To clarify the following derivation we are going to introduce a shorthand notation for the decomposition of a matrix in the subspaces $\mathcal{P}_0$ and $\mathcal{Q}_0$. With this notation the decomposition of any matrix $M$ can be written as:
\begin{equation*}
    M = P_0MP_0+P_0MQ_0+Q_0MP_0+Q_0MQ_0 = M^{P_0P_0} + M^{P_0Q_0} + M^{Q_0P_0} + M^{Q_0Q_0}\,.
\end{equation*}
Let's show an example where this notation is practical. If a matrix only had off-diagonal elements, that is $M = P_0MQ_0+Q_0MP_0 = M^{P_0Q_0}+M^{Q_0P_0}$, then it would be easy to see, for example, that $P_0M = P_0(P_0MQ_0+Q_0MP_0) = P_0MQ_0 = M^{P_0Q_0}$. The rule is that if you are projecting on the left with $P_0$ you only keep the matrix elements whose left super-index is $P_0$, and so on. 

With this in mind we can start the derivation. Condition (a) can be stated in terms of the projectors $P_0$ and $Q_0$:
\begin{equation}
    H_\text{eff}^{P_0P_0}= H_\text{eff}\,, \quad H_\text{eff}^{P_0Q_0}= H_\text{eff}^{Q_0P_0} = 0\,.
\label{e. Pert c)}
\end{equation}
Conditions b) and c) can be guaranteed if we assume that there exists a SWT transformation $U$ that maps $H$ to $H_\text{eff}$,
\begin{equation}
    H_\text{eff} = U H U^\dagger\,,
    \label{e. Pert ab)1}
\end{equation}
and then impose that $U$ must be a unitary transformation which, by definition, preserves the inner product, ensuring that $H_\text{eff}$ has the same eigenvalues as $H$. A way to impose the unitarity of $U$ is to define it as the exponential of a anti-Hermitian generating matrix $S$:
\begin{equation}
    U = e^{S}\,, \quad S^{\dagger} = -S\,.
    \label{e. Pert ab)2}
\end{equation}
While equations \eqref{e. Pert c)},\eqref{e. Pert ab)1} and \eqref{e. Pert ab)2} ensure conditions (a), (b) and (c), they do not entirely determine the transformation $U$, since it is always possible to construct an alternative solution $U'=TU$, where $T$ can be any unitary transformation acting only on $\mathcal{P}_0$. To overcome this under-constrained situation we can impose the additional condition, which is not obligatory for the definition of $H_\text{eff}$, that the matrix $U$ must not have matrix elements inside $\mathcal{P}_0$ nor inside $\mathcal{Q}_0$, but only matrix elements between $\mathcal{P}_0$ and $\mathcal{Q}_0$, that is:
\begin{equation}
    S^{P_0P_0} = S^{Q_0Q_0} = 0\,,\quad \rightarrow\quad S = S^{P_0Q_0}+S^{Q_0P_0}\,, \quad S^{P_0Q_0} =- (S^{Q_0P_0})^\dagger.
\end{equation}
Under this definition we guarantee that $S^\dagger=-S$. Having reached this point we have all the necessary constrains to obtain the transformation $U$ from the perturbation theory perspective, which starts by expanding the matrix $S$ in powers of $\lambda$:
\begin{equation}
    S = \lambda S_1 + \lambda^2 S_2 + \cdots + \lambda^n S_n + \cdots \,.
    \label{e. expansion S}
\end{equation}
The following step is to realize that equation \eqref{e. Pert ab)1} can be written in terms of commutators using an expansion similar to the Baker–Campbell–Hausdorff formula:
\begin{equation}
    H_\text{eff} = UHU^\dagger = e^{S}He^{-S} = H + [S,H] + \frac{1}{2!}[S,[S,H]] + \frac{1}{3!}[S,[S,[S,H]]] + \cdots \,. 
    \label{e. expansion commutators}
\end{equation}
Introducing here the expansion of $S$ one obtains an expansion of the effective Hamiltonian in powers of $\lambda$. For instance, the expansion up to first order is:
\begin{equation}
    H_{\text{eff},1} = H_0 + \lambda V +[\lambda S_1, H_0] 
        \label{e. first order expansion}
\end{equation}
To solve for $H_\text{eff,1}$ we simply have to project this equation in the subspace $\mathcal{P}_0$:
\begin{equation}
H_{\text{eff},1}^{P_0P_0}= H_{\text{eff},1} = H_0^{P_0P_0} + \lambda V^{P_0P_0} + [\lambda S_1, H_0]^{P_0P_0} \,.\
\end{equation}
Here we can see the usefulness of our notation: it allows us to easily check that $[\lambda S_1, H_0]^{P_0P_0}=S_1^{P_0Q_0}H_0^{P_0P_0} -H_0^{P_0P_0}S_1^{Q_0P_0} =0$, because the subspaces $\mathcal{P}_0$ and $\mathcal{Q}_0$ are orthogonal and hence $H_0^{P_0P_0}S_1^{Q_0P_0}=S_1^{P_0Q_0}H_0^{P_0P_0}=0$. With this we can write the first order effective Hamiltonian as:
\begin{equation}
    H_{\text{eff},1} = H_0^{P_0P_0} + \lambda V^{P_0P_0} = P_0H_0P_0 + \lambda P_0VP_0\,.
    \label{e. first order correction}
\end{equation}
For the second order corrections we have to expand equation \eqref{e. expansion commutators}, keep terms up to order $\lambda^2$ and project in the subspace $\mathcal{P}_0$ :
\begin{equation}
     H_{\text{eff},2} = H_{\text{eff},1} + \left[ \lambda^{2} S_{2}, H_{0}\right]^{P_0P_0}+\left[ \lambda S_{1}, \lambda V\right]^{P_0P_0}+\frac{1}{2}\left[ \lambda S_{1},\left[ \lambda S_{1}, H_{0}\right]\right]^{P_0P_0}
\end{equation}
Since we already know $H_\text{eff,1}$ we just have to find the value of the three new commutators. The first commutator will disappear just as $[\lambda S_1, H_0]^{P_0P_0}$ did. To find the value of the second commutator we have to calculate $S_1^{P_0Q_0}$ and $S_1^{Q_0P_0}=-(S_1^{P_0Q_0})^\dagger$. This can be done projecting eq. \eqref{e. first order expansion} and using $H_{\text{eff},1}^{P_0Q_0}=0$:
\begin{equation}
    \lambda V^{P_0Q_0} + [\lambda S_1, H_0]^{P_0Q_0} = \lambda V^{P_0Q_0} + \lambda\left( S_1^{P_0Q_0}H_0^{Q_0Q_0}-H_0^{P_0P_0}S_1^{P_0Q_0}\right) = 0
    \label{e. VPQ}
\end{equation}
Writing the matrices $H_0^{P_0P_0}$ and $H_0^{Q_0Q_0}$ in their eigenvector expansion we can find the matrix elements of $S_1^{P_0Q_0}$,
\begin{equation}
    \braket{i^0|S_1|\alpha^0} = \frac{\braket{i^0|V|\alpha^0}}{E_{i^0}-E_{\alpha^0}}\,,
\end{equation}
Which allows us to finally write the second commutator as:
\begin{equation}
    \left[ \lambda S_{1}, \lambda V\right]^{P_0P_0} = \lambda^2\sum_{i,j,\alpha} \left(\frac{1}{E_{i^0}-E_{\alpha^0}}+\frac{1}{E_{j^0}-E_{\alpha^0}} \right)\braket{i^0|V|\alpha^0}\braket{\alpha^0|V|j^0} \ket{i^0}\bra{j^0}
    \label{e. second commutator}
\end{equation}
The third commutator can be expanded as:
\begin{equation}
\frac{1}{2}\left[ \lambda S_{1},\left[ \lambda S_{1}, H_{0}\right]\right]^{P_0P_0} = \frac{1}{2}\left( \lambda S_1^{P_0Q_0}\left[ \lambda S_{1}, H_{0}\right]^{Q_0P_0} -\left[\lambda S_{1}, H_{0}\right]^{P_0Q_0}\lambda S_1^{Q_0P_0} \right) \,.
\end{equation}
Substituting the value of $\left[\lambda S_{1}, H_{0}\right]^{Q_0P_0}$ from eq. \eqref{e. VPQ} we can finally write the third commutator,
\begin{equation}
\frac{1}{2}\left[ \lambda S_{1},\left[ \lambda S_{1}, H_{0}\right]\right]^{P_0P_0} = \frac{1}{2}\left(-\lambda S_1^{P_0Q_0}\lambda V^{Q_0P_0} +\lambda V^{P_0Q_0}\lambda S_1^{Q_0P_0}   \right) = -\frac{1}{2}\left[\lambda S_{1}, \lambda V\right]^{P_0P_0} \,,
\end{equation}
which is exactly minus half of the second commutator \eqref{e. second commutator}, thus, we can finally write the second order effective Hamiltonian as:
\begin{equation}
H_{\text{eff},2} = H_{\text{eff},1} + \frac{\lambda^2}{2} \sum_{i,j,\alpha} \left(\frac{1}{E_{i^0}-E_{\alpha^0}}+\frac{1}{E_{j^0}-E_{\alpha^0}} \right)\braket{i^0|V|\alpha^0}\braket{\alpha^0|V|j^0} \ket{i^0}\bra{j^0}
\end{equation}

\subsection{Non-perturbative Schrieffer-Wolff transformation}
\label{s. Heff SWT} 
In this section we propose a method to obtain a numerically exact SWT in a more efficient manner than \cite{consani2020effective}, making use of the fact that in the equation \eqref{e. H_eff} $U$ does not appear isolated but as $UP$ or $PU^\dagger$. If we take into account that $U$ leads from $P$ to $P_0$ through the equation \eqref{e. SWT on P and Q}, we can see that the only terms that are going to play a role in this transformation are those whose form is:
\begin{equation}
    UP = \sum_{i,j} A_{ij} \ketbra{i^0}{j}  =P_0AP
    \label{e. Anm}
\end{equation}
Under this approach we don't have to find the full transformation $U$ but rather the matrix A. This goal can be achieved considering that:
\begin{equation*}
    \begin{aligned}
    P_0 U^2 P &= (P_0 U)(UP) = (P_0 U)^2 = (UP)^2\,, \\
    P_0 U^2 P &= P_0 R_{\mathcal{P}_0} R_\mathcal{P} P = P_0 (P_0 - Q_0)(P - Q) P = P_0P\,. 
    \end{aligned}
\end{equation*}
From this we obtain that $(UP)^2 =P_0P$. Introducing here the definition of $A$ and the matrix of scalar products of the subspaces $B=P_0P$
\begin{equation}
\begin{gathered}
    (UP)^2 = (P_0AP)(P_0AP)= P_0AB^\dagger AP=\\
    P_0P =P_0P_0PP=P_0BP \quad \rightarrow \\
    A B^\dagger A = B
\end{gathered}
\end{equation}
The value of $A$ can be extracted if we introduce the singular value decomposition of $B = W \Sigma V^\dagger$, 
\begin{equation}
A (V\Sigma W^\dagger) A =W \Sigma V^\dagger\quad\rightarrow\quad A  = W V^\dagger\,,
\end{equation}
and substituted into equation \eqref{e. H_eff} to finally obtain the effective Hamiltonian:
\begin{equation}
    H_{\text{eff}} = UPHPU^\dagger = P_0APHPA^\dagger P_0\,.
\end{equation}
Note that this scheme only requires the SWT decomposition of $B=P_0P$, without the need to truncate the matrices $Q$ and $Q_0$ and perform operations that will always be more costly.

Consani \textit{et al.}~\cite{consani2020effective}  computed $U$ directly from its definition \eqref{e. def SWT}. Since these would require to obtain the full set of eigenstates of the Hamiltonian of the system, to make the method more affordable they express $H$ in the basis of eigenstates of the uncoupled qubits with up to $N_T$ states---a number $N_T$ determined by convergence---. In this basis, $H$ is approximately diagonalized to recover the interacting eigenstates and $P$, and $U$ is computed using equation \eqref{e. def SWT}. This step dominates the complexity of the algorithm, due to working with matrices of size $N_T\times N_T$.

Instead, we have proposed to compute the rank-$d$ matrices $P_0U P$ and $PU^\dagger P_0$, using only the $d$ lowest energy eigenstates of $H_0$ and $H$ that compound the low energy subspace. Thus, instead of computing $U$ in the full basis, we only need to estimate $d \times d$ matrices. The cost of the algorithm is now dominated by the calculation of the $d$ eigenstates, which will always be smaller than $N_T$ and does not require a convergence study.

\pagebreak

\section{Three Josephson Junctions Flux Qubit}
\label{s. 3JJQ}

As mentioned in the introduction, the main goal of this work is to study flux qubit-qubit couplings. The specific qubit that we have chosen is the three Josephson junctions flux qubit \cite{orlando, orlando2} (3JJQ), thus, before immersing ourselves in the complexities of the qubit-qubit interactions we will devote this section to give a general view of flux qubits and a specific view of the 3JJQ and its properties.


The quantum observables that describe a superconducting circuit below its critical temperature are the electric fluxes and electric charges of its nodes, $\hat{\phi}_i$ and $\hat{q}_i= \dot{\phi}_i$. These observables define a classification of superconducting qubits: at one extreme we have charge qubits, superconducting circuits with well-defined charge states (\textit{i.e.} an excess/lack of Cooper pairs on a superconducting island); at the other extreme we have flux qubits, superconducting circuit with well-defined electric flux states.

Flux qubits make use of two physical phenomenal to address the need for two well-defined low-energy states and an anharmonic spectrum. First, flux qubits have a superconducting loop threaded by an external magnetic field. Thanks to the quantization of the magnetic field this external flux causes the appearance of two degenerate ground states with distinguishable current distributions. Second, the superconducting loop of flux qubits is interrupted by Josephson junctions which have a non-linear inductance (see appendix \ref{a. Josephson junction}) that modifies the inductive energy of the circuit and produces an anharmonic spectrum.



\begin{figure}[H]
  \centering
  \hfill%
  \subcaptionbox{}{\includegraphics[width=.33\textwidth]{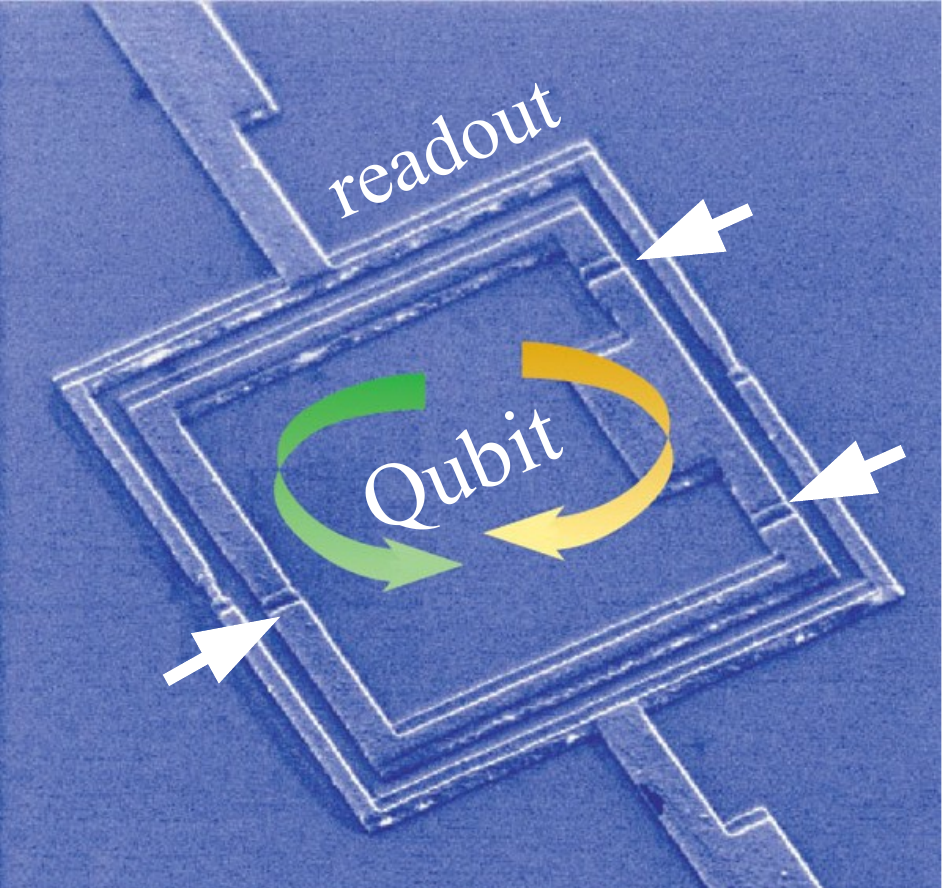}\vspace{0.5cm}}\hfill%
  \subcaptionbox{}{\includegraphics[width = 0.5\linewidth]{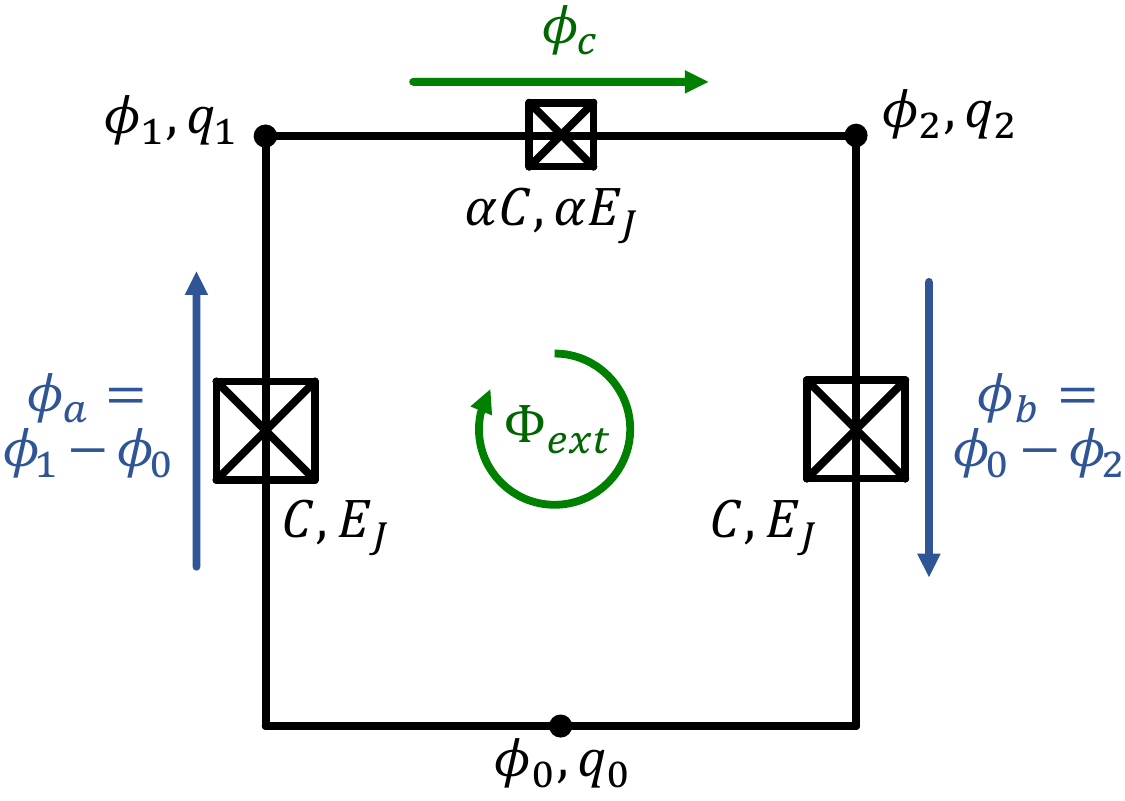}}\hfill%
\medskip
  \caption{\textbf{(a)} Picture of a 3JJQ immersed in a dc-SQUID for readout \cite{clarke2008superconducting}, the arrows point to the Josephson junctions. \textbf{(b)} Lumped-element model of a 3JJQ. The crossed squares are the representation of Josephson junctions, and should be regarded as a capacitor and a Josephson inductance in parallel (see appendix \ref{a. Josephson junction} for more details). }
  \label{f. circuito 3JJ}
\end{figure}

Fig. \ref{f. circuito 3JJ} depicts the circuit of a 3JJQ. It consists of a superconducting loop interrupted by three Josephson junctions, two identical and one $\alpha$ times smaller, typically $\alpha\sim0.7$. 
\noindent Following appendices \ref{a. Josephson junction} and \ref{a. circuit hamiltonian} the Hamiltonian of a 3JJQ can be expressed in terms of the electric fluxes and electric charges of the nodes of the circuit, $\hat{\phi}_i$ and $\hat{q}_i= \dot{\phi}_i$, quantum observables that obey the canonical commutation relations, $[\hat{\phi}_i,\hat{q}_j]=i\delta_{ij}$. 

To understand the potential that governs this Hamiltonian it is more convenient to switch from the node operators to the difference between the node operators across the junctions, the branch operators shown in blue and green in fig.\ref{f. circuito 3JJ}. This would mean that the Hamiltonian has three independent variables, $\phi_a$, $\phi_b$ and $\phi_c$ however, we can invoke the flux quantization condition and remove one of the electric flux variables.

\begin{figure}[H]
\includegraphics[width=\linewidth]{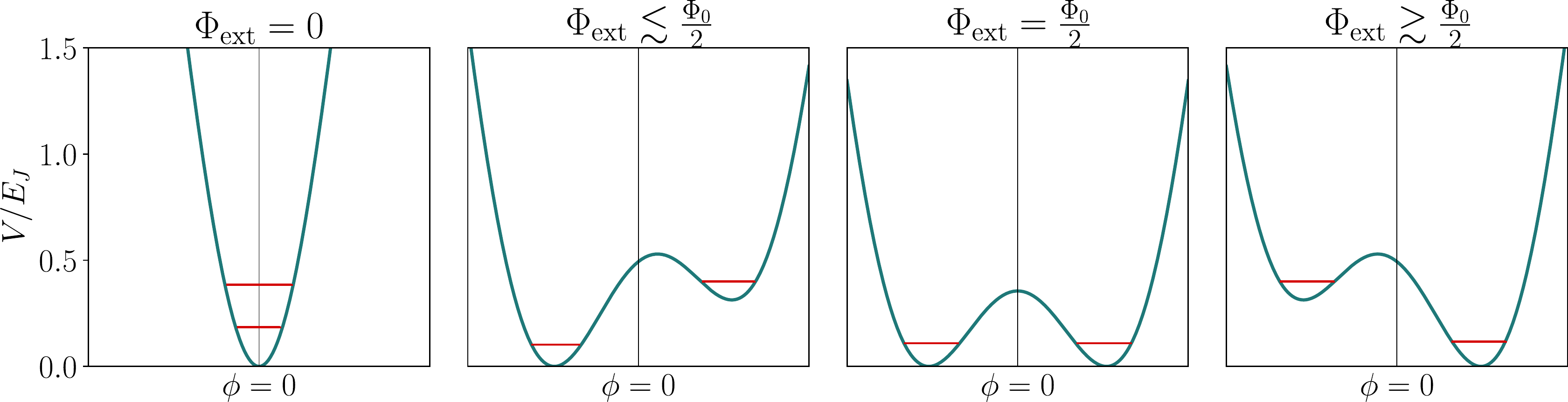}
\hspace*{2.6cm} (a) \hfill (b) \hfill (c) \hfill (d)  \hspace*{1.5cm}
\caption{Schematic potential and energies of a flux qubit for different values of the external flux.}
\label{f. potential vs phi}
\end{figure}

The flux quantization condition states that the total magnetic flux threading the loop must be an integer number of the magnetic flux quantum $\Phi_0=h/2e$. This total magnetic flux is the sum of the external magnetic flux, $\Phi_\text{ext}$, and the magnetic flux due to the induced supercurrents in the loop which appear to expel the external magnetic field piercing the bulk of the superconductor, as explained in the Meissner effect. The induced magnetic flux can be written in terms of the electric fluxes across the components of the loop, and result in the following flux quantization condition:
\begin{equation}
    \oint_C\nabla\phi\cdot dl +\Phi_\text{ext} = \phi_a+\phi_b+\phi_c + \Phi_\text{ext} = n\Phi_0 
    \label{e. flux quantization condition}
\end{equation}
This allows us to write the flux across the small junction as $\phi_c = (n\Phi_0-\Phi_\text{ext}) - (\phi_a+\phi_b) = \Phi - (\phi_a+\phi_b) $, where we have defined $\Phi = n\Phi_0-\Phi_\text{ext} = \phi_a+\phi_b+\phi_c = $ as the flux due to the externally induced current in the loop, which we will just sometimes call external flux. Finally we can write the Hamiltonian as: 
\begin{equation}
\begin{aligned}
&\hat{H}=\frac{1+\alpha}{2(1+2\alpha)C}(\hat{q}_a+\hat{q}_b)^2-\frac{\alpha}{(1+2\alpha)C}\hat{q}_a\hat{q}_b+V(\phi_a,\phi_b)\,,\\
&V(\phi_a,\phi_b) = -E_J\left[\cos{\left(\frac{\phi_a}{\varphi_0}\right)}+\cos{\left(\frac{\phi_b}{\varphi_0}\right)}+\alpha \cos{\left(\frac{\phi_a+\phi_b-\Phi}{\varphi_0}\right)}\right]\,,
\end{aligned}
\label{e. H flux qubit}
\end{equation}
where $\varphi_0 =\Phi_0/2\pi$ is the flux-phase relation discussed in appendix \ref{a. Josephson junction}. The periodic, nonlinear potential $V(\phi_a,\phi_b)$ that governs this Hamiltonian has two competing terms: the potential of the the big junctions, $ \cos{\left(\phi_a\right)}+ \cos{\left(\phi_b\right)}$, produces a minimum at $\phi_a=\phi_b=0$, whereas the potential of the small junction, $ \cos(\phi_a+\phi_b-\Phi)$, produces a minimum that depends on $\Phi=(n\Phi_0-\Phi_\text{ext})$. The qualitative behaviour of the circuit can be explained analyzing the potential across the direction $\phi_a=\phi_b$ (which we will later call the direction $t_1$) for different values of $\Phi_\text{ext}$ as shown in fig. \ref{f. potential vs phi}. Note that $\phi_a=\phi_b$ means that the electric flux differences across the junctions point in the same direction, and result in states of the circuit with a certain induced persistent current flowing in the clockwise / counterclockwise direction, or, in other words, states with a certain induced magnetic flux pointing up or down.

When $\Phi_\text{ext}=0$, fig. \ref{f. potential vs phi}~(a), the potential has a symmetric minimum at $\phi_a=\phi_b=0$, which means that the expected values of the current/flux states of the circuit are zero. As we increase the external flux the potential landscape changes and the ground and first excited states start to depart in their current/flux states. 

When the external flux is around half of a flux quantum—the so-called frustration point—the circuit reaches a regime where the currents of the ground and excited states are as different as possible. If the external flux is slightly below the frustration point, fig \ref{f. potential vs phi}~(b), the ground state opposes $\Phi_\text{ext}$ generating a diamagnetic current in the loop which produces a magnetic flux such that the total magnetic flux threading the loop is exactly zero. The first excited state favors $\Phi_\text{ext}$, generating a slightly larger paramagnetic current in the loop which produces a magnetic flux such that the total magnetic flux threading the loop is exactly one flux quantum, $\Phi_0$. 

When the external flux is slightly above the frustration point, fig \ref{f. potential vs phi}~(d), the circuit enters a regime where it behaves in the opposite way as described above: the ground state has a paramagnetic current that favours $\Phi_\text{ext}$ and the total magnetic flux threading the loop is exactly one flux quantum, whereas the first excited state has a slightly larger diamagnetic current that opposes $\Phi_\text{ext}$ and the total magnetic flux threading the loop is zero.

Exactly at the frustration point, fig \ref{f. potential vs phi}~(c), the maximum of the potential of the small junction sits on top of the minimum of the big junctions, and as a result they produce a degenerate energy landscape with two minima of equal depth located at $\varphi=\pm\varphi^*=\pm\varphi_0\arccos\left(1/2\alpha\right)$. This means that the circuit has two degenerate current ground states: clockwise/counterclockwise current states that favor/oppose the external magnetic flux to ensure that the total magnetic flux threading the loop is one/zero flux quantum. In this situation the clockwise and counterclockwise current states are no longer adequate to describe the ground and excited states of the circuit. 
This is why we say that the circuit is magnetically frustrated, and we call this the qubit point. 

When the circuit is frustrated, or close to the frustration point, quantum tunneling couples the two current states by forming symmetric and anti-symmetric superpositions which will be our qubit states. We will analytically study the qubit states in the following section by defining the current states as those that would result from approximating the minima of the potential with a harmonic potential, however, it is necessary to give a full view of the 2D potential of the 3JJQ and its dependence with $\alpha$ before we proceed any further.

\begin{figure}[H]
    \centering
    \includegraphics[width = 1\linewidth]{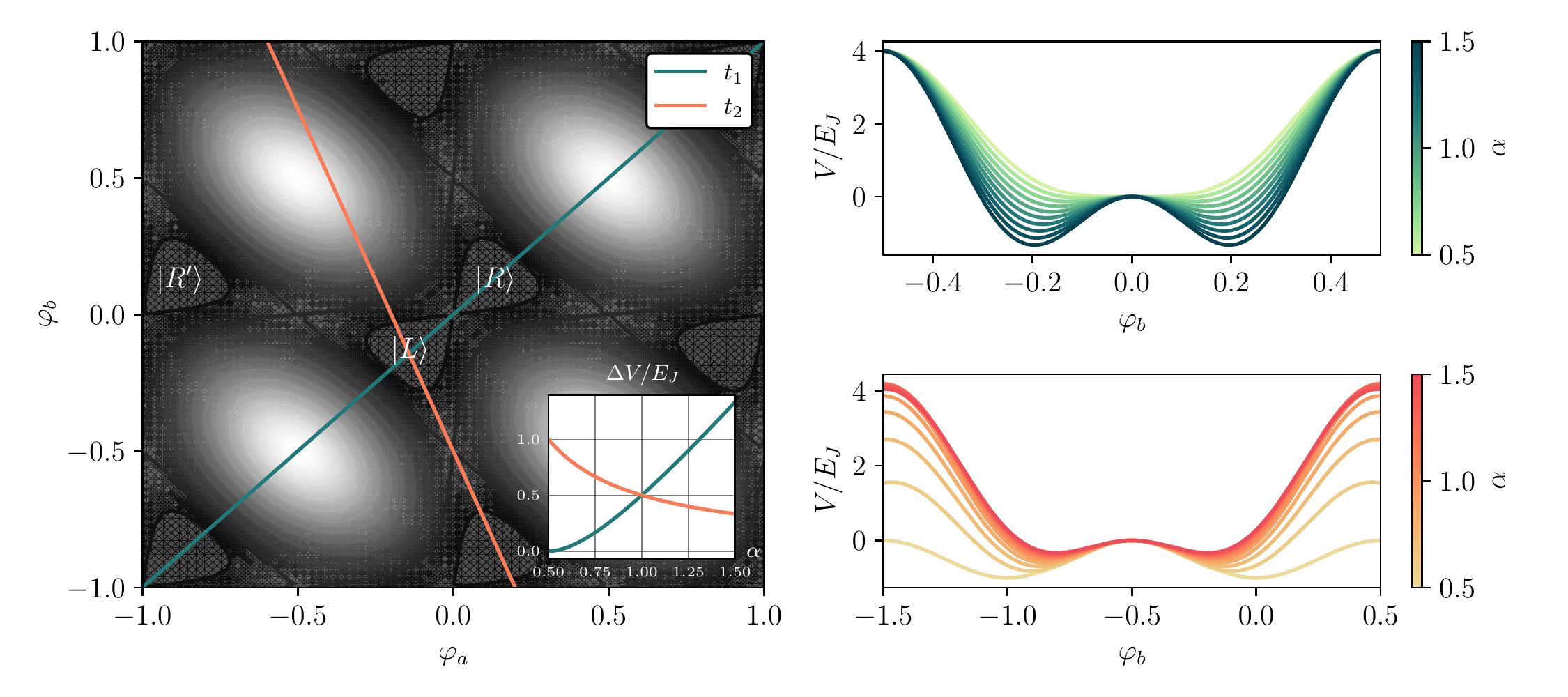}
    \caption{The left figure shows the periodic nonlinear potential landscape of the frustrated 3JJQ, $\Phi_{ext} = \frac{1}{2}\Phi_0$, as expressed in equation \eqref{e. H flux qubit} for $\alpha = 0.8$. The right figures show the 1D sections of the potential landscape across the tunneling directions $t_1$ and $t_2$ for different values of $\alpha$. The inset of the left figure shows the barrier height across the tunneling directions for different values of $\alpha$.}
    \label{f. potencial}
\end{figure}

\begin{figure}[H]
    \centering
    \includegraphics[width = 1\linewidth]{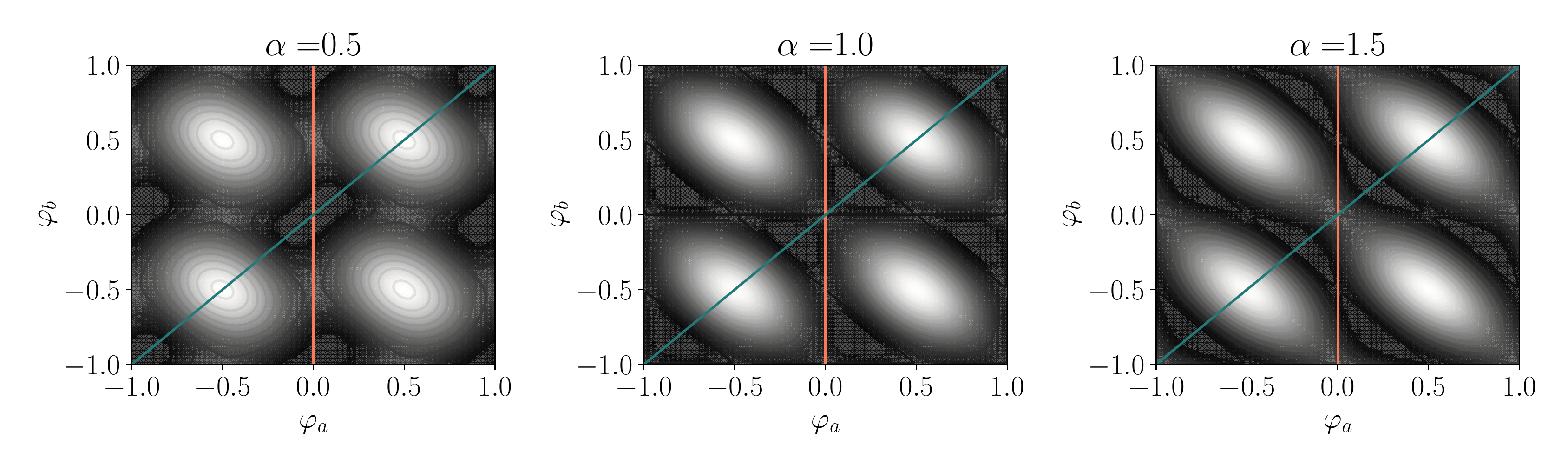}
    \caption{ Periodic nonlinear potential landscape of the frustrated 3JJQ, $\Phi_{ext} = \frac{1}{2}\Phi_0$, for different values of $\alpha$. Note that for $\alpha=0.5$ the landscape only has one minimum located at $\varphi_a=\varphi_b=0$. For $\alpha=1$ the lowest equipotential contour line includes both the maximum barrier heights through $t_1$ and $t_2$, thus, both are the same height, as showed in the inset of fig. \ref{f. potencial}. For $\alpha=1.5$ it is clear that there is a large intra cell barrier, whereas the minima of different unit cells are easily connected thorugh a kind of canyon that facilitates hopping between those states. }
    \label{f. potenciales}
\end{figure}

Figs. \ref{f. potencial} and \ref{f. potenciales} show the potential landscape of the 3JJQ at the frustration point. When  $\alpha<0.5$ the unit cell only has one minima, meaning that it does not have two degenerate ground states of opposite current even at the frustration point. This is the regime of the C-shunted flux qubits \cite{yan2016flux} and it will not be thoroughly discussed in this document. When $0.5<\alpha<1$ the unit cell of the potential is a highly isolated valley with two minima corresponding with the persistent current states. As we mentioned, these two minima are connected through tunneling, mainly in the direction $t_1$, and result in a qubit whose ground and excited energy eigenstates are a symmetric and anti-symmetric superposition of the current states \cite{orlando}. As the parameter $\alpha$ increases from 0.5 the minima drift apart and get deeper. The drift of the minima means that the expected values of the currents increase, and since the minima are deeper the potential barrier through $t_1$ increases the wave-functions of the qubits become more localised and less interacting. As $\alpha$ approaches and exceeds $\alpha=1$ the potential landscape has a qualitative change that can be see in in fig. \ref{f. potenciales}~(c): the intra-cell barrier through $t_1$ becomes larger than the inter-cell barrier through $t_2$, and hence the potential no longer has two close minima in a isolated valley. In this situation tunneling thorugh $t_2$ becomes important and the system gains sensitivity to charge noise \cite{orlando}, thus, we will also try to avoid this regime.

\subsection{Harmonic approximation}
\label{s. harmonic approx 1D}

To ease further manipulations and compress the 3JJQ Hamiltonian shown in eq. \eqref{e. H flux qubit} it is useful to nondimensionalize variables. To nondimensionalzie we switch from the dimensional operators charge and flux to the dimensionless operators number of Cooper pairs $\hat{n_i}=\hat{q}_i/(-2e)$ and phase of the macroscopic wave function $\hat{\varphi_i} = \hat{\phi_i}/\varphi_0$, see appendix \ref{a. Josephson junction} for more details. As before, these observables must obey the commutation relation $[\hat{\varphi_i},\hat{n_j}]=-i\delta_{ij}$. As a result of the nondimensionalization of the charge variables we get a energy scale accompanying the number operators, $E_C=e^2/2C$, the capacitive energy of one of the big Josephson junctions. One of the parameters that will govern the Hamiltonian is the balance between the ``kinetick'' capacitive energy, $E_C$, and ``potential'' inductive energy, $E_J$, stored in the Josephson junctions, thus, we can nondimensionalize the Hamiltonian dividing it by the inductive energy and study the problem as a function of the ratio of the two energies, $r=E_J/E_C$. Additionally, it is useful to change variables such that the kinetic energy term of eq. \eqref{e. H flux qubit} becomes diagonal. Changing the variables to $\varphi_{+} = (\varphi_a+\varphi_b)/2$, $\varphi_{-} = (\varphi_a-\varphi_b)/2$ and $\hat{n}_\pm =i \partial/\partial\varphi_\pm$, allows us to write the Hamiltonian of the three-junction flux qubit as:

\begin{equation}
\frac{\hat{H}}{E_J}=\frac{1}{2}\frac{\hat{n}^2_{+}}{m_{+}}+\frac{1}{2}\frac{\hat{n}^2_{-}}{m_{-}}-\big[2\cos \left( \varphi_{+} \right) \cos \left( \varphi_{-} \right) +\alpha  \cos \left( 2\varphi_{+} - 2\pi f\right) \big]\,,
\label{e. H 3JJQ +-}
\end{equation}
This Hamiltonian has the shape of a Hamiltonian of two charges with different masses in a nonlinear potential, where $f=\Phi/\Phi_0$ is the dimensionless externally induced flux, \textit{i.e.} the externally induced phase, and $m_{+} = (2\alpha+1)r/4$, $m_{-}=r/4$ are the ``dimensionless masses'' of the hypothetical charges. Nevertheless, even though flux and charge satisfy a canonical commutation relation they are not completely equivalent to position and momentum. The first difference is that charge is a discrete operators proportional to the number of Cooper pairs, in contrast to the continuous momentum operator. The second difference is that the flux is related to the superconductor’s phase—a periodic operator remarkably different from the position operator, as explained in appendix \ref{a. Josephson junction}.

\begin{figure}[H]
    \centering
    \includegraphics[width=0.55\textwidth]{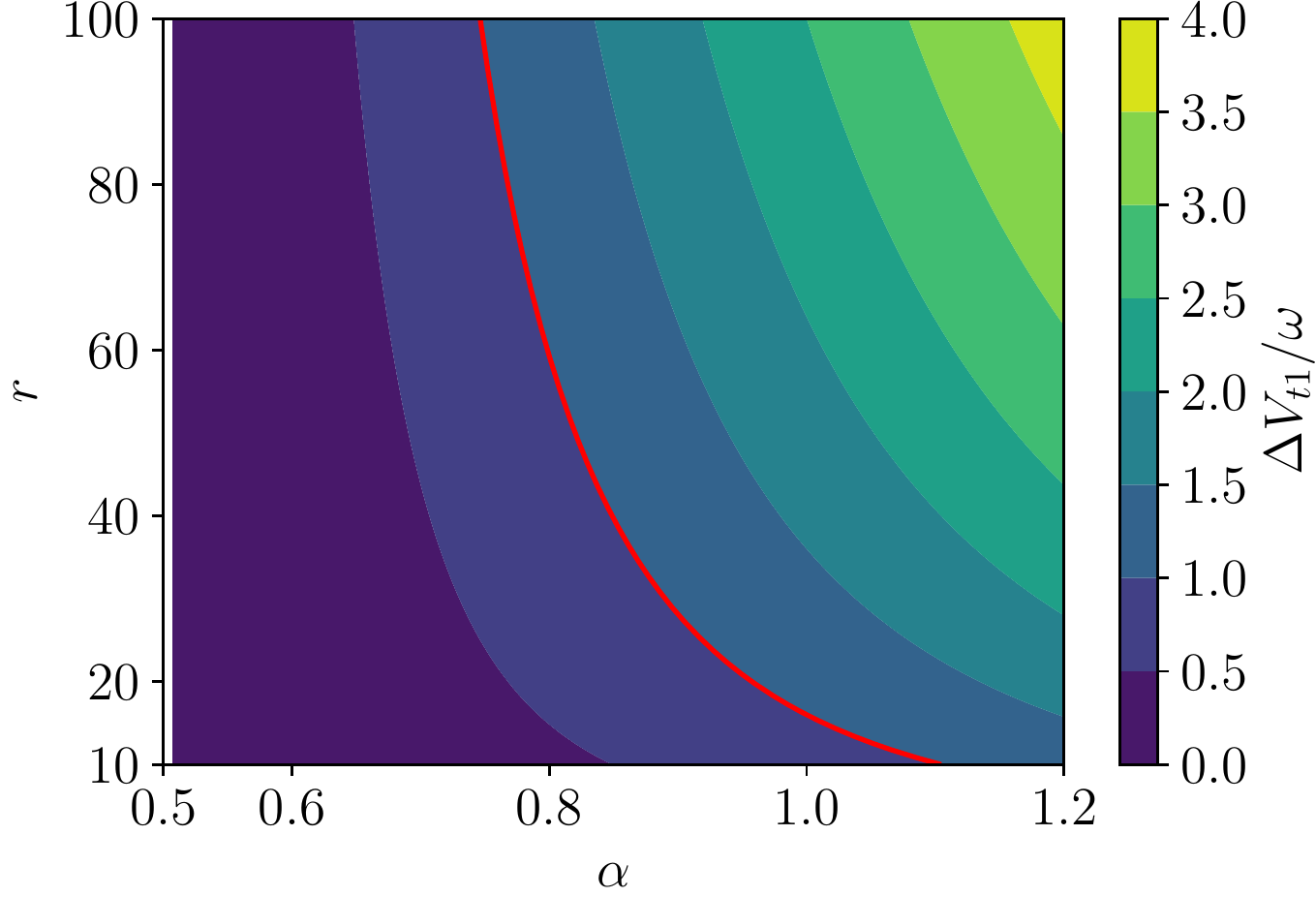}
    \caption{Barrier potential through the tunneling direction compared with the frequency of the harmonic oscillators, $\Delta V_{t_1}/\omega$, as a function of the qubit's parameters $\alpha$ and $r$. The red contour line shows the points where $\Delta V_{t_1}=\omega$.}
    \label{f. barrier height vs frequency oscillator }
\end{figure}

As we have shown, when $\alpha < 1$ the least energetic path to travel between potential minima corresponds to the intra-cell trajectory $\varphi_a =\varphi_b$ which we have called $t_1$. As a consequence, when $\alpha<1$ we can assume that the main phenomena regarding our qubit can be described studying the problem inside the unit cell and along the trajectory $t_1$\footnote{We will later study the validity of this assumption.}, which allow us to simplify the bidimensional Hamiltonian \eqref{e. H flux qubit} for $\varphi_+=\varphi$ and $\varphi_-=0$, obtaining the unidimensional Hamiltonian:
\begin{equation}
\frac{\hat{H}_{t_1}(f=0.5)}{E_J}=\frac{1}{2}\frac{\hat{n}^2_{+}}{m_{+}}-\big[ 2\cos(\varphi) -\alpha \cos\left(2\varphi\right) \big]
\label{e. H t1}
\end{equation}
To simplify the notation we will eliminate the $+$ subindices such that $\hat{n}_{+}\to\hat{n}$ and $m_{+}\to m$. This expression is still not amenable to being analytically studied, thus, we can proceed by approximating the nonlinear potential around its minima with a harmonic potential derived from its second order Taylor series. This results a Hamiltonian for each minima of the potential. The Hamiltonians approximating \eqref{e. H t1} in the minima of the unit cell are:
\begin{align*}
\frac{\hat{H}^h_{t_1,L} }{E_J}& =\frac{1}{2}\frac{\hat{n}^2}{m}+\frac{1}{2}\frac{4 \alpha^2-1}{\alpha} (\hat{\varphi}+\varphi^*)^2 \\
\frac{\hat{H}^h_{t_1,R} }{E_J} & =\frac{1}{2}\frac{\hat{n}^2}{m}+\frac{1}{2}\frac{4 \alpha^2-1}{\alpha} (\hat{\varphi}-\varphi^*)^2
\end{align*}
The harmonic potential in this expression can be written in the usual form, $m\omega^2(\varphi\pm\varphi^*)^2/2$, which allows us to identify the frequency of the oscillator as $\omega=\sqrt{4(2\alpha-1)/(\alpha r)}$. Note that our approximation breaks down for $\alpha<0.5$, which makes sense because for those values of $\alpha$ the energy of the small junction is not large enough as to create a degenerate landscape with two minima. Additionally, we can note that our approximation will be better when the wells are deep enough as to allow us to identify two localized current states, that is, when the frequency of the oscillator is smaller than the barrier potential through the tunneling direction, $\Delta V_{t_1}/\omega\gg1$. We have plotted this ratio in fig. \ref{f. barrier height vs frequency oscillator }.

 As we can see the harmonic approximation will be reasonable for large values of $\alpha$ and $r$. This is problematic, specially for the case of $\alpha$, since we want to stay in the regime $\alpha<1$, thus, we will have to take the results of the harmonic approximation with a grain of salt and use them only to understand the qualitative behaviour of the qubit.

\noindent The frequency of the oscillator establishes the energy scale of the current states, thus, it is useful to specifically write it with dimensions:
\begin{equation}
    \omega E_J = E_J\sqrt{\frac{4(2\alpha-1)}{\alpha r}}=\sqrt{\frac{4(2\alpha-1)}{\alpha }E_JE_C} = \sqrt{\frac{(2\alpha-1)}{2\alpha }} \sqrt{8E_JE_C} 
\end{equation}
Since $\sqrt{8E_JE_C}$ is the frequency of a transmon qubit we can conclude that a 3JJQ made with the same components will always have a smaller energy scale than a transmon. 

With an understanding of the frequency of the oscillator we can turn to the eigenstates of the left and right harmonic approximate Hamiltonians, which can be written as a function of the Hermite polynomials $H_n$:
\begin{equation}
\braket{\hat{\varphi}|\psi_n^h}=\psi_{n}^h(\varphi)=\frac{1}{\pi^{1/4}\sqrt{2^{n} n!\sigma}} e^{\scaleto{-\frac{(\varphi \pm \varphi^*)^{2}}{2\sigma^2 }}{22pt}}  H_{n}\left( \frac{(\varphi \pm \varphi^*)}{\sigma}\right), \quad \sigma=\frac{1}{\sqrt{mw}}\,.
\label{e. general harmonic wavefunctions}
\end{equation}
The most relevant parameters of the wavefunctions are the position of the minima, $\varphi^*=\arccos\left(1/2\alpha\right)$, and its dispersion $\sigma$. The dispersion is the square root of the inverse of $m\omega$, thus, it is useful to do a brief detour to write $m\omega$ explicitly and study its scaling with the qubit parameters:
\begin{equation}
    m\omega = \sqrt{\frac{(2\alpha-1)(2\alpha+1)^2 }{4\alpha}r}\,,\quad\to \quad 
    \left\{\begin{matrix} 
        \begin{aligned}
            &\frac{d}{dr}(m\omega)>0,     \quad  \forall r\,,   \\ 
            &\frac{d}{d\alpha}(m\omega)>0,\quad  \forall\alpha>1/2\,.  
        \end{aligned}
    \end{matrix}\right.
    \label{e. efecto alpha y r}
\end{equation}
From this equation we can conclude that as we increase either $\alpha$ or $r$, $\sigma$ will decrease and the wavefunctions will be more and more localised.

Following from \eqref{e. general harmonic wavefunctions} we can explicitly write the ground and first excited eigenstates of the left and right harmonic approximate Hamiltonians as:
\begin{equation}
\begin{aligned}
\braket{\hat{\varphi}|g_L} &=g_L(\varphi)= \left(\frac{m\omega}{\pi}\right)^{1/4} e^{\scaleto{-\frac{m \omega (\varphi +\varphi^*)^{2}}{2 }}{22pt}} = g(\varphi +\varphi^*) \\
\braket{\hat{\varphi}|g_R} &=g_R(\varphi)=\left(\frac{m\omega}{\pi}\right)^{1/4} e^{\scaleto{-\frac{m \omega (\varphi -\varphi^*)^{2}}{2 }}{22pt}} = g(\varphi -\varphi^*)   
\\ \braket{\hat{\varphi}|e_L} &=e_L(\varphi)=  \left(\frac{m\omega}{4\pi}\right)^{1/4} e^{\scaleto{-\frac{m \omega (\varphi +\varphi^*)^{2}}{2 }}{22pt}}\sqrt{2m\omega}(\varphi +\varphi^*)  = e(\varphi +\varphi^*) \\
\braket{\hat{\varphi}|e_R} &= e_R(\varphi)= \left(\frac{m\omega}{4\pi}\right)^{1/4} e^{\scaleto{-\frac{m \omega (\varphi -\varphi^*)^{2}}{2 }}{22pt}}\sqrt{2m\omega}(\varphi -\varphi^*)  = e(\varphi -\varphi^*) 
\end{aligned}
\end{equation}
How can we use these current states to approximate the eigenstates of the Hamiltonian \eqref{e. H t1} of the 3JJQ? On one hand, if the qubit is not in the frustration point then the minima won't have the same depth—figs. \ref{f. potential vs phi}~(b) and (c)—and the left and right harmonic eigenstates would be good candidates for the ground and first excited eigenstates. On the other hand, if the qubit is in the frustration point the answer will depend on the qubit's parameters. 

\begin{figure}[H]
  \centering
  \hfill%
  \subcaptionbox{}{\includegraphics[width=.437\textwidth]{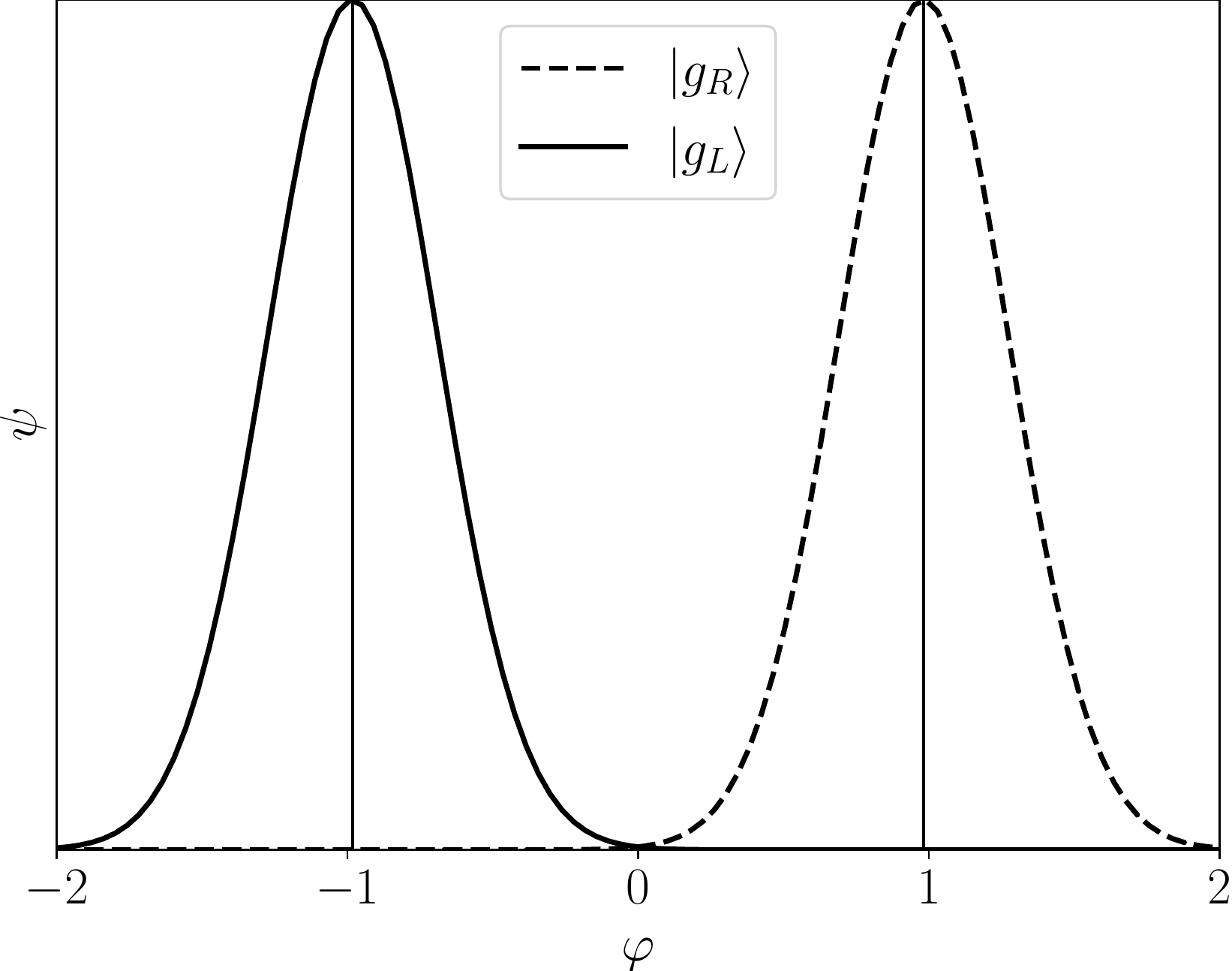}}\hfill%
  \subcaptionbox{}{\includegraphics[width=.475\linewidth]{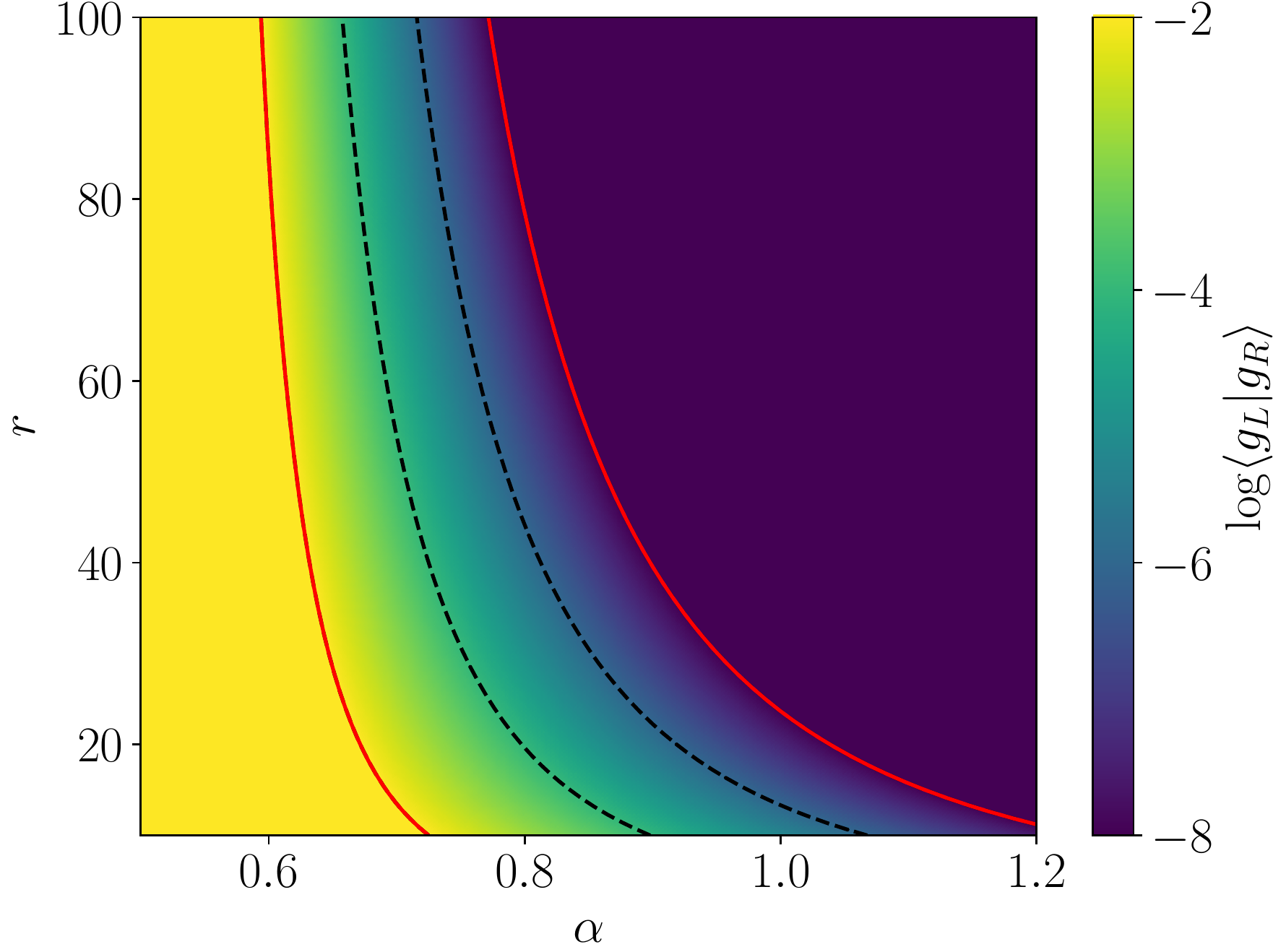}}\hfill%
\medskip
  \caption{\textbf{(a)} Left and right harmonic ground wavefunctions for $\alpha=0.7$ and $r=0.5$, vertical lines are shown for $\varphi=\varphi^*$. \textbf{(b)} Overlap between the left and right wavefunctions: $\braket{g_L|g_R}=\exp(-m\omega\varphi^{*2})$.}
  \label{f. Harmonic wavefunctions}
\end{figure}

\begin{figure}[H]
  \centering
  \hfill%
  \subcaptionbox{}{\includegraphics[width=.45\textwidth]{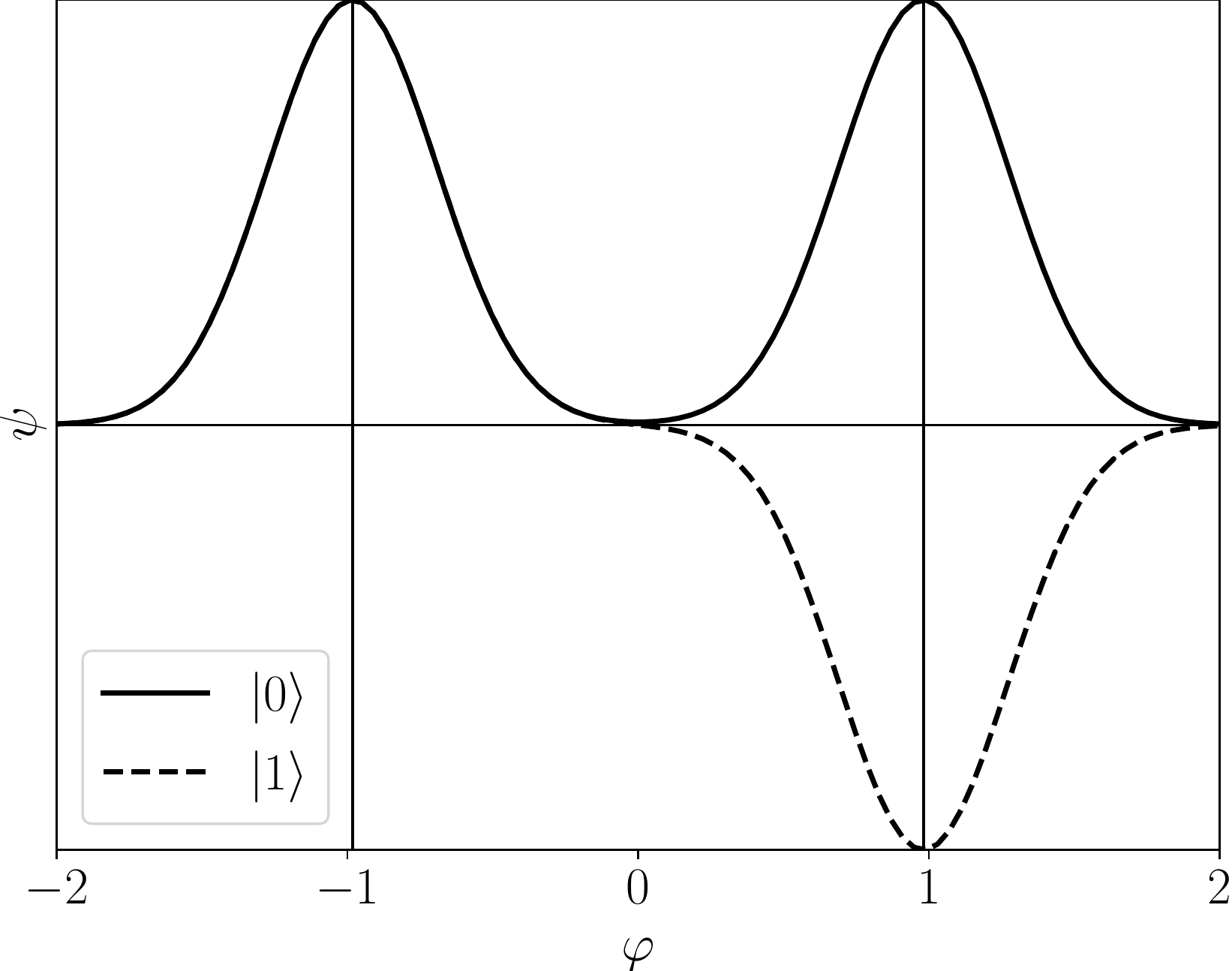}}\hfill%
  \subcaptionbox{}{\includegraphics[width=.45\linewidth]{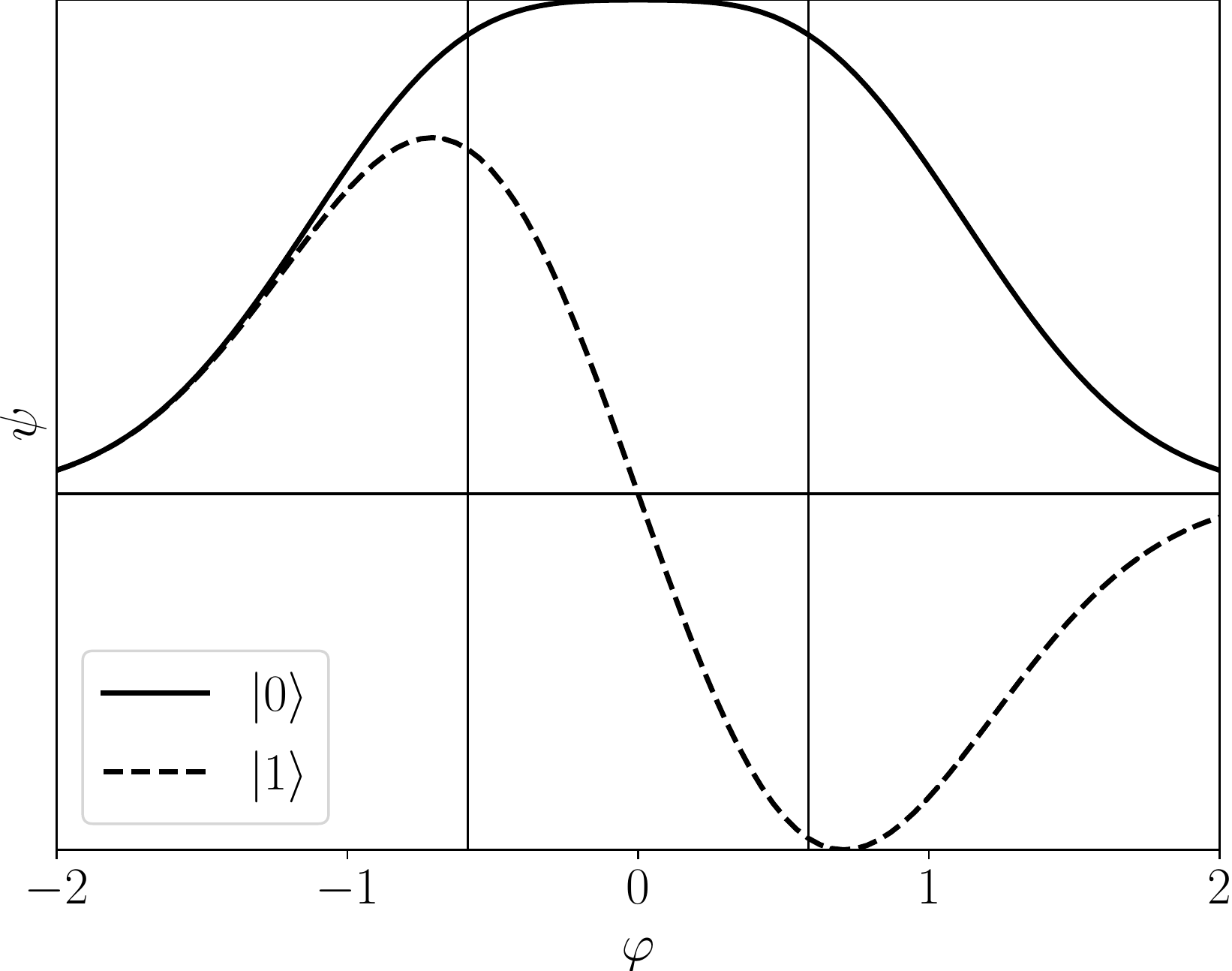}}\hfill%
\medskip
  \caption{Ground and excited qubit approximate wavefunctions for \textbf{(a)} very localised state with $\alpha=0.9$ and $r=80$, and \textbf{(b)} very interacting state $\alpha=0.6$ and $r=20$. Vertical lines are shown at the potential minima $\varphi=\varphi^*$.}
  \label{f. Qubit harmonic wavefunctions}
\end{figure}

If the potential barrier separating the two minima in the $t_1$ direction is large compared to the energy of the harmonic eigenstates, or if the effective mass of the qubit in the $t_1$ direction is large, then the left and right harmonic wavefunctions will have a negligible overlap in the $t_1$ direction, fig. \ref{f. Qubit harmonic wavefunctions}(a), and would result in ground and excited wavefunctions with very similar properties. In this case we would have to consider the full 2D potential to understand the problem. As we have just seen in equation \eqref{e. efecto alpha y r} increasing either of our design parameters, $\alpha$ and $r$, has this effect: increasing $r$ increases the effective mass of the qubit, increasing $\alpha$ increases the effective mass of the qubit, the potential barrier height and the distance between the minima.

When the harmonic eigenstates are not too localised in the $t_1$ direction they will have a significant overlap and interact via tunneling, fig. \ref{f. Qubit harmonic wavefunctions}(b), hybridizing and inviting us to approximate the 3JJQ eigenstates as a linear combination of the harmonic eigenstates. A rule-of-thumb region where this overlap is sufficiently big as to not require a 2D analysis, but not so big as to make the harmonic approximation invalid due to $\Delta V_{t_1}/\omega\ll1$, is show in fig \ref{f. Harmonic wavefunctions} (b) as a region between the solid red contour lines.

Since the 3JJQ potential is symmetric it is only reasonable to demand that the wavefunctions must obey parity symmetry, \textit{i.e.} to be even or odd in the phase space.  The only problem left is to decide whether the ground state of the 3JJQ is a symmetric or antisymmetric superposition of the left and right harmonic wavefunctions. Since the ground eigenstate must have no nodes we can conclude that the ground state will be the symmetric superposition, leaving the antisymmetric case for the excited state. Experimental \cite{orlando2} and numerical \cite{orlando} data confirm this approach, hence, we can write the approximate qubit eigenstates as:
\begin{equation}
\begin{aligned}
    \ket{0} &= \frac{1}{\sqrt{2}}\big(\ket{g_L}+\ket{g_R} \big) \quad \to \quad \braket{\hat{\varphi}|0} = \psi_0(\varphi)= \frac{1}{\sqrt{2}}\big(g_L(\varphi )+g_R(\varphi )  \big)\,, \\
    \ket{1} &= \frac{1}{\sqrt{2}}\big( \ket{g_L}-\ket{g_R} \big) \quad \to \quad \braket{\hat{\varphi}|1} = \psi_1(\varphi)=  \frac{1}{\sqrt{2}}\big(g_L(\varphi )-g_R(\varphi )  \big)\,. 
    \label{e. Qubit harmonic wavefunctions}
\end{aligned}
\end{equation}

Now that we have the approximate qubit eigenstates we can use them to compute a two-level effective Hamiltonian as described in section \ref{s. Heff}. Defining the Hamiltonian \eqref{e. H t1} as the unperturbed Hamiltonian, $\hat{H}_{t_1}=H_0$, we can obtain the low-energy Hamiltonian, \textit{i.e.} the qubit Hamiltonian, of $\hat{H}_{t_1}$ in the harmonic approximation, $H_q^h$, simply projecting it into it's first two eigenstates as we did in eq. \eqref{e. qubit hamiltonian} to obtain :
\begin{equation*}
H_\text{eff}^h=\frac{\Delta_{01}^h}{2}\sigma^z
\end{equation*}
Since the Hamiltonian is proportional to a $\sigma^z$ operator we can see that indeed the two eigenstates of the system are separated by a energy gap, but since the Hamiltonian does not have any off diagonal term it means that the system has zero probability amplitude to flip between eigenstates. 

To gain a deeper understanding of this result it is interesting to write the effective Hamiltonian in the basis of left and right persistent currents, which is a basis that can be more intuitively understood if one imagines a qubit whose state is obtained measuring the direction of the current. One can obtain the left and right states as a function of the eigenstates from eq. \ref{e. Qubit harmonic wavefunctions} and conclude that in this basis the effective Hamiltonian can be written as $H_\text{q,LR}^h=\frac{\Delta_{01}^h}{2}\sigma^x$. It can be shown that the transformation that takes us from the eigenstate basis to the current basis and vice versa is simply a switch between the $\sigma^z$ and $\sigma^x$ operators, thus, in this case there is no $\sigma^z$ operator, which means that the left and right current states have the same energy as one would naturally expect. The presence of the $\sigma^x$ operator means that now the system has a $\Delta_{01}^h$ probability amplitude to switch from one current state to the other via tunneling. This is precisely the intuition we wanted to obtain: the energy gap of the qubit eigenstates is exactly the tunneling amplitude to go travel between the left and right current states.

\begin{figure}[H]
    \centering
    \includegraphics[width=\linewidth]{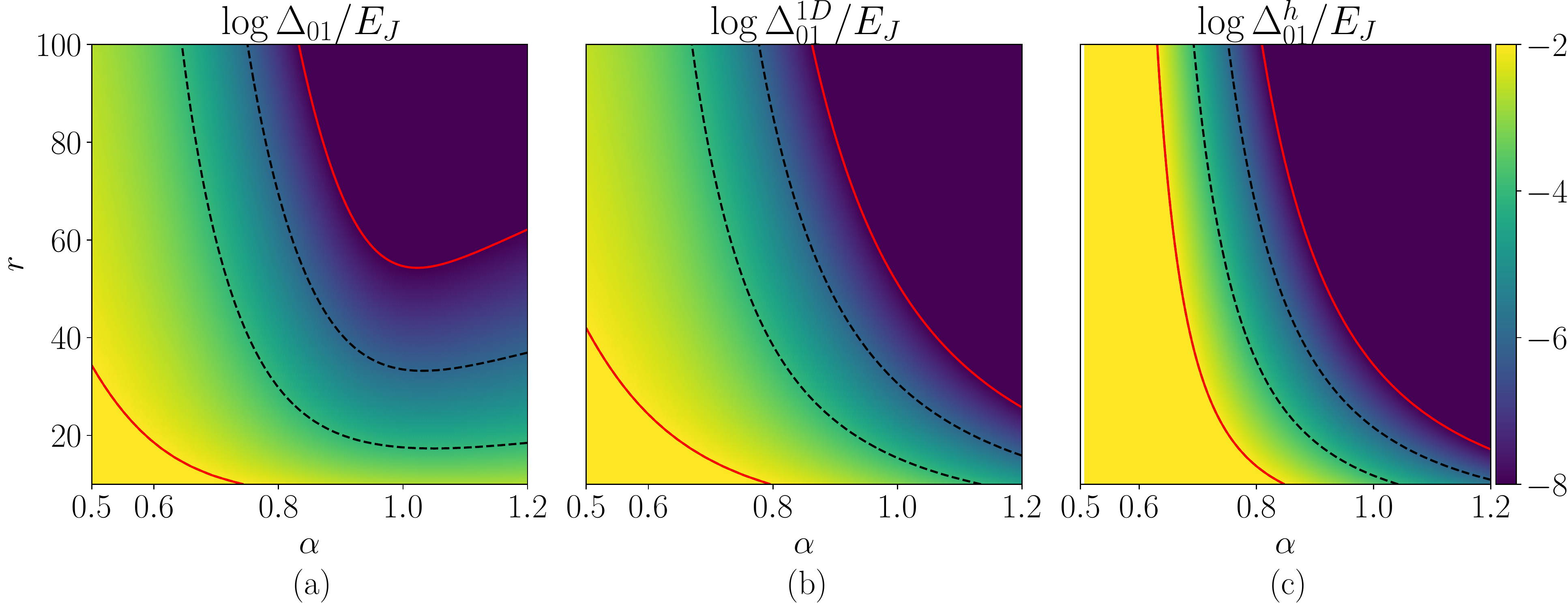}
    \caption{Gap of the 3JJQ, $\Delta_{01}=E_1-E_0$, as a function of $\alpha$ and $r$. The results have been obtained with three approaches: \textbf{(a)}, exact numeric gap of the 2D full Hamiltonian \eqref{e. H 3JJQ +-}; \textbf{(b)}, exact numeric gap of the 1D approximate Hamiltonian \eqref{e. H t1}; \textbf{(c)}, approximate analytic gap of the the harmonic approximation \eqref{e. Gap harmonic}. The contour lines correspond to the ticks of the color bar. The red contour lines correspond to the limits -2 and -8, the color map has been truncated above and below these values.}
    \label{f. gap}
\end{figure}

The energy gap is a relevant parameter which we can calculate and compare with the exact numerical gap to obtain a measure of the goodness of the harmonic approximation.  To be fair in this comparison one has to remember that the harmonic approximation truly contains two approximations: first, neglecting $\varphi_{-}$ (and hence tunneling in the $t_2$ direction), which we will call the 1D approx.; second, approximating the potential with a harmonic well, the harmonic approx. For this reason we will consider three qubit gaps: the gap of 2D full Hamiltonian \eqref{e. H 3JJQ +-}, $\Delta_{01}$; the gap of the 1D approximate Hamiltonian \eqref{e. H t1}, $\Delta_{01}^{1D}$ ; the gap of the harmonic approximation \eqref{e. Gap harmonic}, $\Delta_{01}^h$. The gap of the qubit in the harmonic approximation can be calculated as:
\begin{equation}
\begin{aligned}
    &\Delta_{01}^h=E_1^h-E_0^h=\braket{1|\hat{H}_{t_1}|1} - \braket{0|\hat{H}_{t_1}|0} \\ &= \frac{1}{2} \big(\bra{g_L}-\bra{g_R}\big)\hat{H}_{t_1}\big(\ket{g_L}-\ket{g_R}\big)-\frac{1}{2} \big(\bra{g_L}+\bra{g_R}\big)\hat{H}_{t_1}\big(\ket{g_L}+\ket{g_R}\big) \\ &= -\braket{g_L|\hat{H}_{t_1}|g_R} - \braket{g_R|\hat{H}_{t_1}|g_L} = -2\braket{g_L|\hat{H}_{t_1}|g_R} \\
    &= -2E_J\left[\braket{g_L|\frac{1}{2}\frac{\hat{n}^2}{m}|g_R}+\braket{g_L|\alpha\cos(2\varphi)|g_R}+\braket{g_L|-2\cos(\varphi)|g_R} \right] \\
    &= -2E_J\left[\int\frac{-1}{2m}g_L(\varphi )\frac{d^2g_R(\varphi )}{d\varphi^2} d\varphi + \int\alpha\cos(2\varphi)g_L(\varphi)g_R(\varphi ) d\varphi+\int-2\cos(\varphi)g_L(\varphi ) g_R(\varphi ) d\varphi \right] \\ &=-2E_J\left[ \left( \frac{\omega}{4}-\frac{m\omega^2\varphi^{*2}}{2} \right) + \left( \alpha e^{-1/m\omega}\right) + \left(-2e^{-1/4m\omega} \right) \right]e^{-m\omega\varphi^{*2}}\sim E_Je^{-m\omega\varphi^{*2}} =E_J\braket{g_L|g_R}\,.
    \label{e. Gap harmonic}
\end{aligned}
\end{equation}
This result has been plotted in fig. \ref{f. gap} as a function of $\alpha$ and $r$. The reasonable way to compare this results is first to judge the goodness of the 1D approx., comparing figs. \ref{f. gap} (a) and (b), and then judge the goodness of the harmonic potential approx., comparing figs. \ref{f. gap} (b) and (c).

For $\alpha<1$ all of the gaps are qualitatively similar. On one limit high values of $\alpha$ and $r$ produce a potential landscape with deep minima and high effective masses in the $t_1$ direction. As a result the harmonic wavefunctions have exponentially small interactions, the ground and excited qubit eigenstates become exponentially indistinguishable, and the qubit gap becomes exponentially small, fig.  \ref{f. Qubit harmonic wavefunctions} (a). On the other limit the result is the opposite: a flat landscape with low effective masses produces highly interacting harmonic wavefunctions, very different ground and excited qubit eigenstates, and large qubit gaps, fig. \ref{f. Qubit harmonic wavefunctions} (b).  It is important to note that, as stated in the last line of eq. \eqref{e. Gap harmonic}, the qubit gap of the harmonic approximation is very well described by the harmonic wavefunction overlap, fig. \ref{f. Harmonic wavefunctions}. Indeed, this overlap also captures rather well the behaviour of the 1D approx. gap, fig. \ref{f. gap} (b) except for very low values of $\alpha$. For $\alpha>1$ the 1D approx. and the harmonic approx. break down and give qualitative wrong results for the dependence with $\alpha$. This is because when $\alpha\sim1$ the barrier height through $t_1$ starts to be comparable to that through $t_2$, as we can see in fig. \ref{f. potencial}, and hence the tunneling does not get exponentially smaller but rather switches directions. Since this feature is not included in the 1D approx. it predicts a decrease in the interactions and in the gap due to the increasing localization of the wavefunctions with increasing $\alpha$.

\begin{figure}[H]
  \centering
  \subcaptionbox{$\Delta_{01}(\alpha,r=50)$}{\includegraphics[width=.5\textwidth]{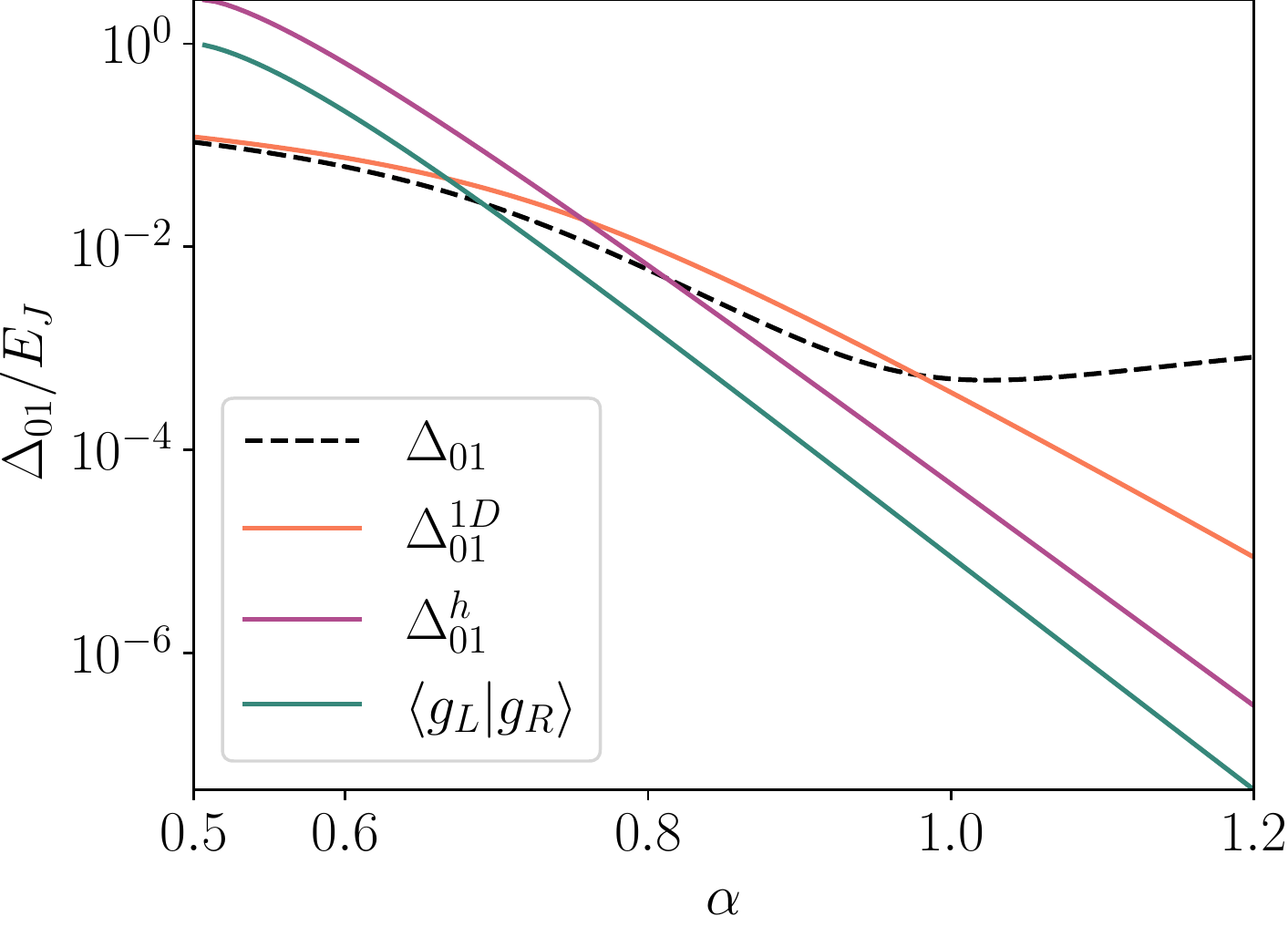}}\hfill%
  \subcaptionbox{$\Delta_{01}(\alpha=0.7, r)$}{\includegraphics[width=.48\textwidth]{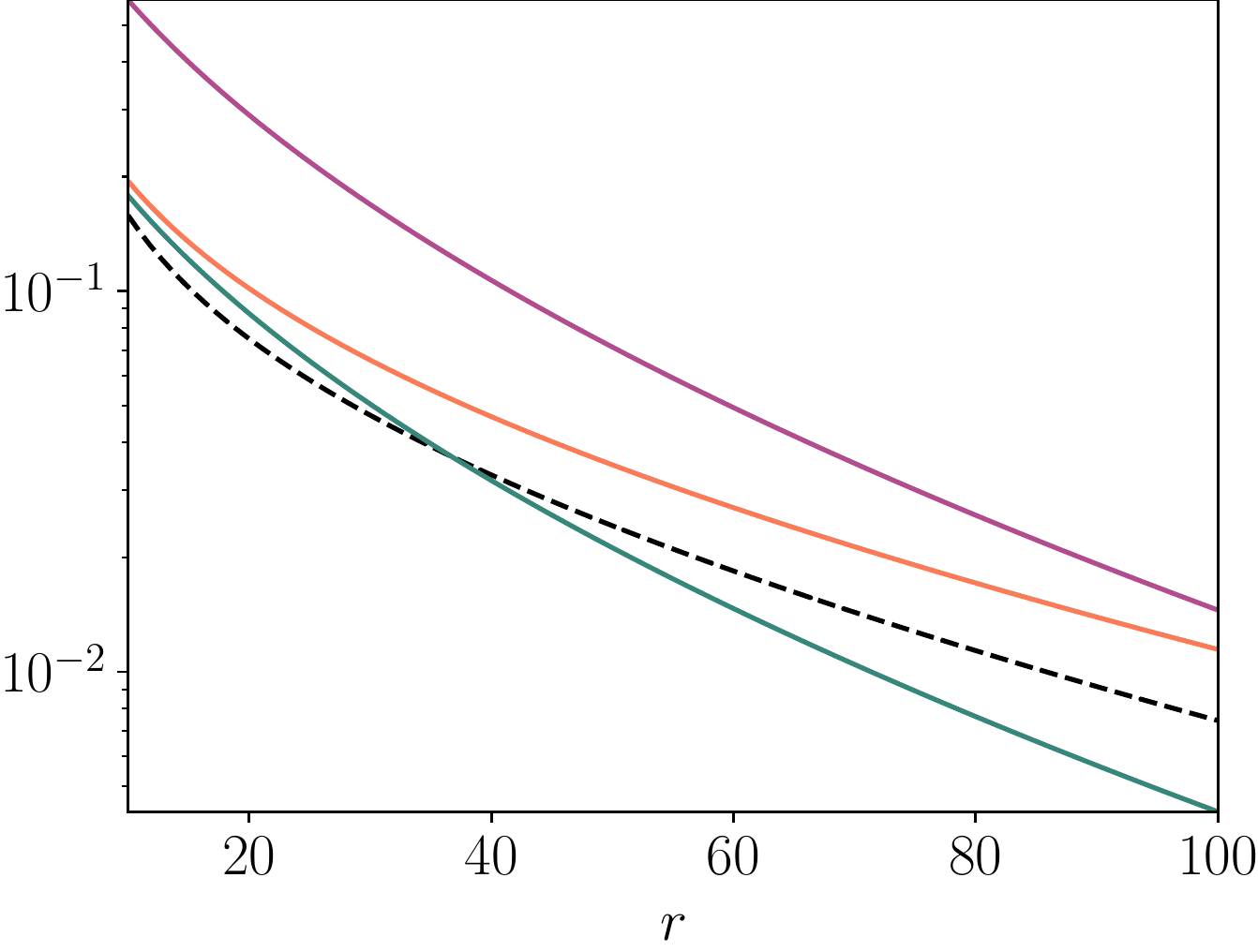}}
\medskip
  \caption{Gap of the 3JJQ versus $\alpha$ and $r$.}
  \label{f. 1D gap vs alpha and r}
\end{figure}

To do a more quantitative analysis in fig. \ref{f. 1D gap vs alpha and r} we have plotted the qubit gap for a constant value of $\alpha$ and for a constant value of $r$. We start by comparing the 1D approx. with the exact results, orange and black-dashed lines respectively. We can see that the 1D approx. captures very well the general behavior of the gap but makes two errors. First, for a given $\alpha$ and $r$ the 1D approx. constantly overestimates the gap. This is because the overlap of two Gaussian wavefunctions in 1D will always be a superior limit to the overlap of those Gaussian wavefunctions in 2D, and hence the 1D approx. predicts higher interactions and higher energy splitting. Second, as we have just seen for $\alpha>1$ the 1D approx. breaks down. 

We now compare the 1D approx. with the harmonic approx. The harmonic approx. roughly captures the quantitative behaviour of the 1D approx. In the dependence with $\alpha$ the harmonic approx. makes a good prediction of the gap in the region shown in fig. \ref{f. Harmonic wavefunctions}. In the dependence with $r$ the harmonic approx. predicts a too-large exponential decay with $r$ but captures the general shape very well. We can see that indeed $\braket{g_L|g_R}$ captures very well the dependence of $\Delta_{01}^h$ with $\alpha$ and $r$. 

\subsection{Capacitive loading}
\label{s. capactive loading}
The model that we have developed in the previous section captures very well the scaling of the gap of the 3JJQ with $\alpha$ and $r$. The results are particularly good in the region $0.6<\alpha<0.9$, which is great because that is the regime where 3JJQs operate. As we have see, for values of $\alpha$ above 1 the model fails because the scaling of the gap changes dramatically, however, this should not be a great concern because for those values the qubit gains sensitivity to charge noise and the region must be avoided.

Nonetheless, there is situation that will render our model unusable: the capacitive couplings between qubits. In this scheme two 3JJQs are connected with a capacitor. The problem of the extra capacitor is that it modifies the capacitance matrix of the qubit and therefore largely modifies the kinetic energy of the Hamiltonian, which is obtained inverting the capacitance matrix as explained in appendix \ref{a. circuit hamiltonian}. This phenomenon is known as `capacitive loading' or `renormalization' of the charge variables, and one of its effects is that the kinetic term is no longer diagonal with the variables $\hat{n}_{\pm}$, which was rather useful because those where commuting with the phases $\hat{\varphi}_{\pm}$ which allowed to write the harmonic potential approximation easily in diagonal form. In this section we will study the effect of the extra capacitor.

If we connect a capacitor of capacitance $\gamma C$ to the node 2 of the 3JJ Flux qubit show in fig. \ref{f. circuito 3JJ} the Hamiltonian of the circuit at the degeneration point becomes
\begin{equation}
    \frac{\hat{H}}{E_J} = \frac{1}{rd} \left[ (\gamma+2)n^2_{+}+(4\alpha+2+\gamma)n^2_{-} + 2\gamma n_{+}n_{-}  \right] -\big[2\cos \left( \varphi_{+} \right) \cos \left( \varphi_{-} \right) -\alpha  \cos \left( 2\varphi_{+}\right) \big]\,,
    \label{e. renormalized hamiltonian}
\end{equation}
where $d=|\hat{C}|=2\alpha+1+\gamma(\alpha+1)$ is the determinant of the \textit{renormalized} capacitance matrix $\hat{C}$. This problem is rather complicated because the kinetic term is no longer diagonal, however, if we stick to our 1D approximation and assume that $\varphi_{-}=0$ the only effect of the extra capacitance is to modify the effective mass of the qubit in the $t_1$ direction. This approx. is only reasonable as long as $\gamma\ll1$, since otherwise the kinetic energy will be very different and it will not be reasonable to neglect the $\varphi_{-}$ direction. In the 1D approx. the resulting Hamiltonian is:
\begin{equation}
    \frac{\hat{H}_{t_1}}{E_J} =\frac{1}{2}\frac{\hat{n}^2_{+}}{\widetilde{m}_{+}}-\big[ 2\cos(\varphi_{+}) -\alpha \cos\left(2\varphi_{+}\right) \big]\,,  \quad \widetilde{m}_{+} = \frac{r \left(2 \alpha +1+ \gamma \left(\alpha + 1\right) \right)}{2 \gamma + 4}\,.
\end{equation}
Note that due to this effect the oscillator frequency in the harmonic approximation will also be renormalized, $\tilde{\omega}=\sqrt{(4\alpha^2-1)/2\tilde{m}}$. The most important effect of the additional capacitance on the qubit is that its gap will be modified.

\begin{figure}[H]
    \centering
    \includegraphics[width=\textwidth]{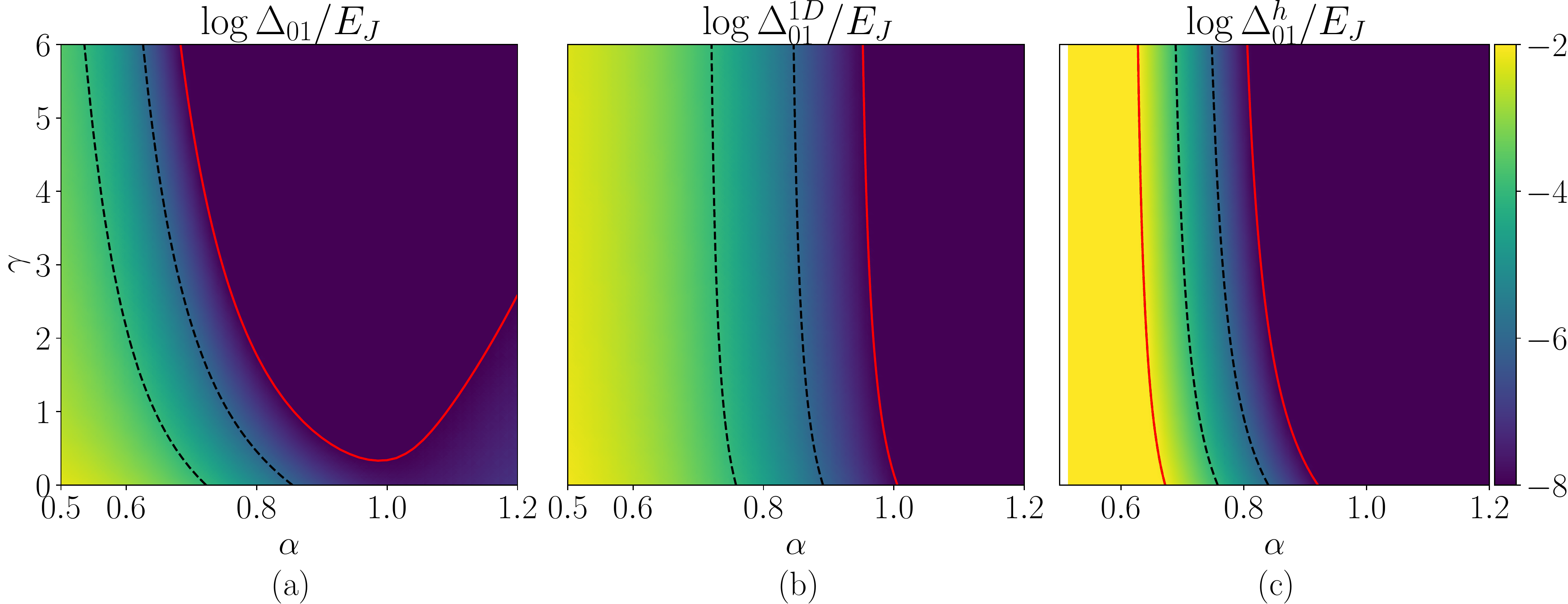}
    \caption{$\Delta_{01}(\alpha,\gamma)$ for $r=50$. The calculation methods and contour lines are described in fig. \ref{f. gap}.}
    \label{f. gap renormalization 2D}
\end{figure}

\begin{figure}[H]
    \centering
    \includegraphics[width=0.5\linewidth]{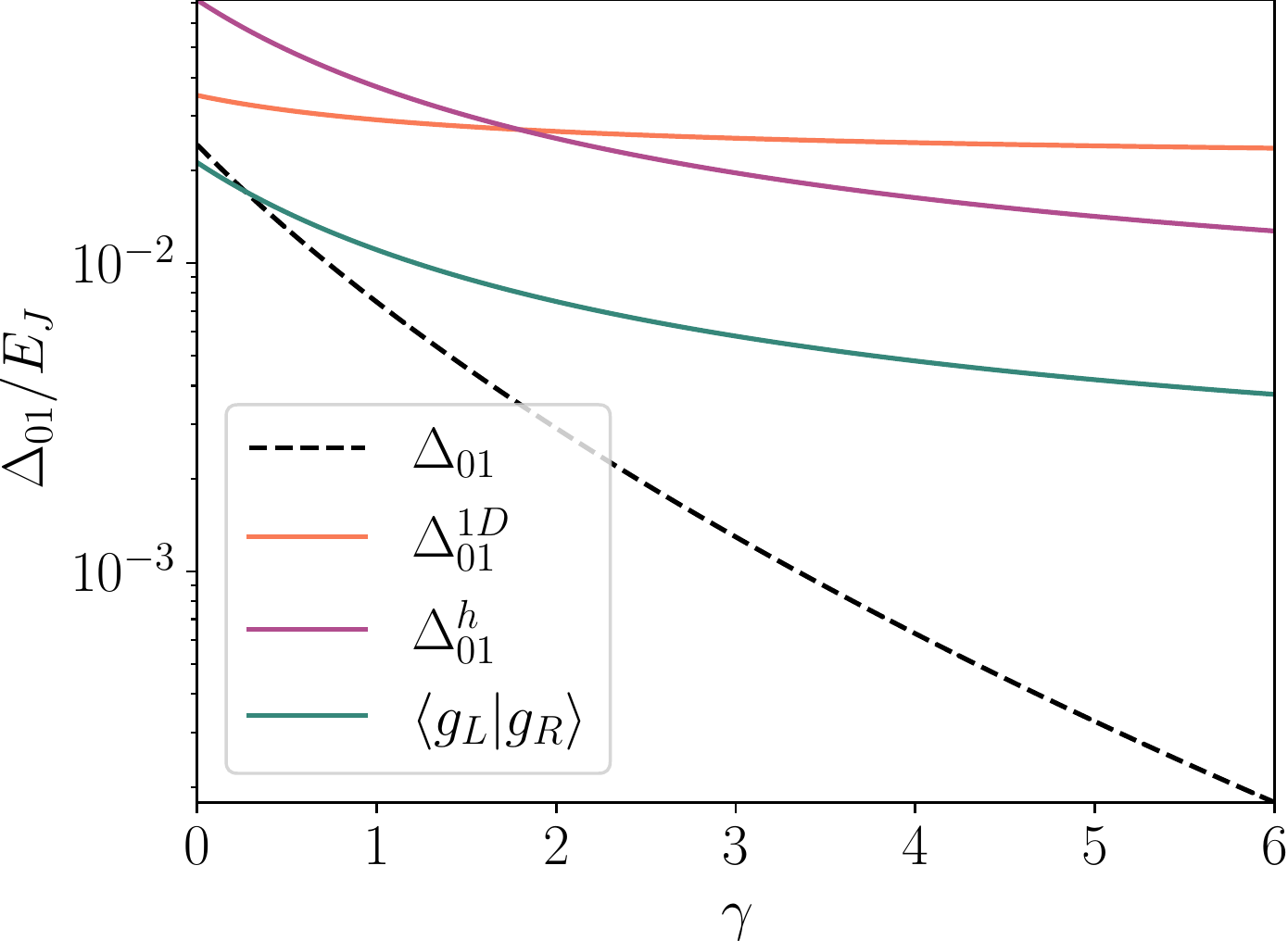}
    \caption{$\Delta_{01}(\gamma)$ for $\alpha=0.7$, $r=50$. }
    \label{f. gap vs gamma 1D}
\end{figure}

Fig. \ref{f. gap renormalization 2D} and \ref{f. gap vs gamma 1D} show that the effect of the renormalization of $\gamma$ is qualitatively similar to that of $r$. Increasing $\gamma$ increases the effective mass of the qubits, making the left and right harmonic states more localised and hence ground and excited qubit states with a smaller energy gap. Comparing figures \ref{f. gap renormalization 2D}(a) and (b) we can see that again the bidimensionality of the potential is essential to understand the effect of $\gamma$ on the qubit's gap for $\alpha>1$. In a simile manner to fig. \ref{f. 1D gap vs alpha and r}, the harmonic approximation looks like the 1D approx. compressed in the $\alpha$ axis. 
Fig. \ref{f. gap vs gamma 1D} gives us a more quantitative view of the scaling of the gap with $\gamma$. As we can see the models only give reasonable results for very low values of $\gamma$, since it does not capture well the exponential decay of the gap. As we can see, we now have two problems derived from the 1D approximation: we miss the behaviour change for $\alpha>1$ and we miss the effect of the renormalization for large values of $\gamma$. To solve this problem we have transformed the full Hamiltonian \eqref{e. renormalized hamiltonian} to a diagonal form which allows to obtain a 2D Harmonic approximation, as explained in appendix \ref{a. 2D Harmonic approximation}. 

Figs. \ref{f. gap vs alpha and r 2D harm} and \ref{f. gap vs alpha and gamma 2D harm} show the overlaps obtained with the 2D Harmonic approximation. In this case we have to distinguish two types of overlap, depending on which minima we consider: the intra-cell overlap along $t_1$, which we have been studying in the previous sections; and the inter-cell overlap along $t_2$ (see fig. \ref{f. potencial}). Comparing this with figures \ref{f. gap vs gamma 1D} and \ref{f. gap} we can see that indeed the 2D Harmonic approximation captures the change of behaviour for $\alpha>1$. We can intuitively understand these figures by looking at the wavefunctions obtained with the 2D Harmonic approximation, fig. \ref{f. 2D Harmonic wavefunctions}.

\begin{figure}[H]
    \centering
    \includegraphics[width=\textwidth]{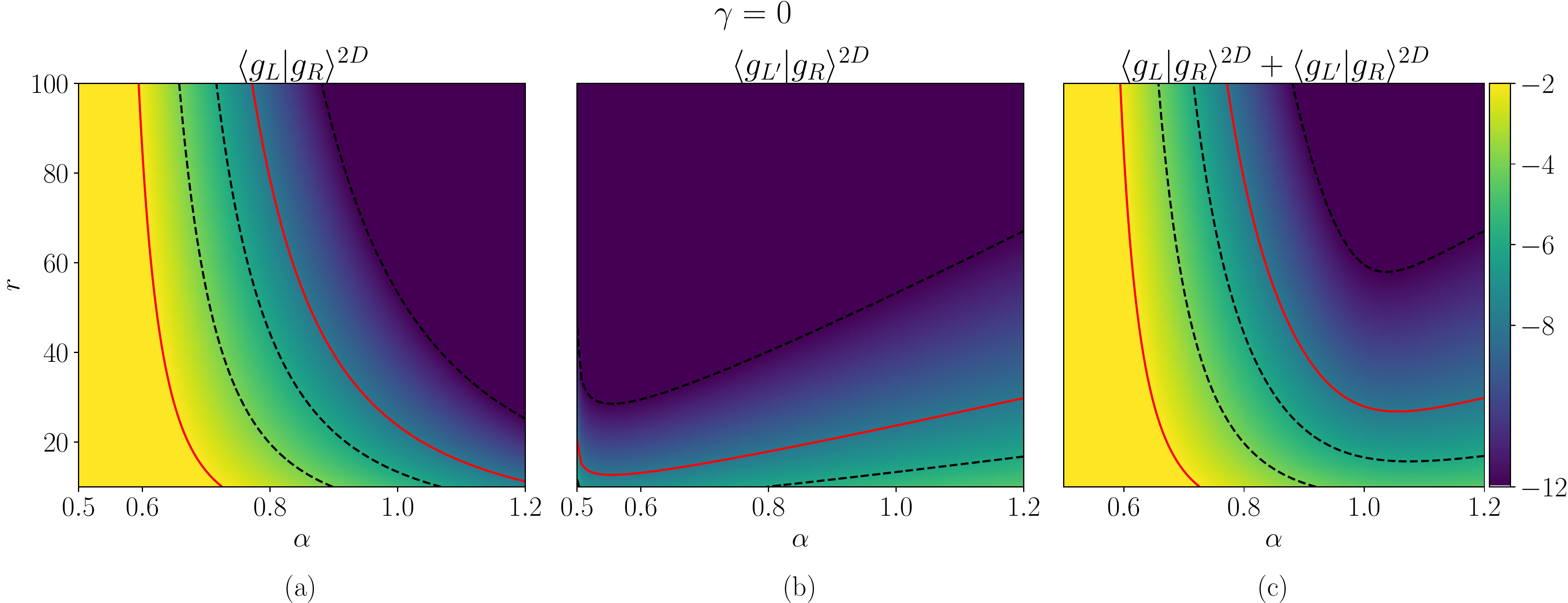}
    \caption{Overlaps obtained with the 2D Harmonic approximation discussed in appendix \ref{a. 2D Harmonic approximation} for different values of $\alpha$ and $r$, $\gamma=0$. \textbf{(a)} Intra-cell overlap, \textbf{(b)} inter-cell overlap, \textbf{(c)} sum of both overlaps. The contour lines are described in fig. \ref{f. gap}. }
    \label{f. gap vs alpha and r 2D harm}
\end{figure}

\begin{figure}[H]
    \centering
    \includegraphics[width=\textwidth]{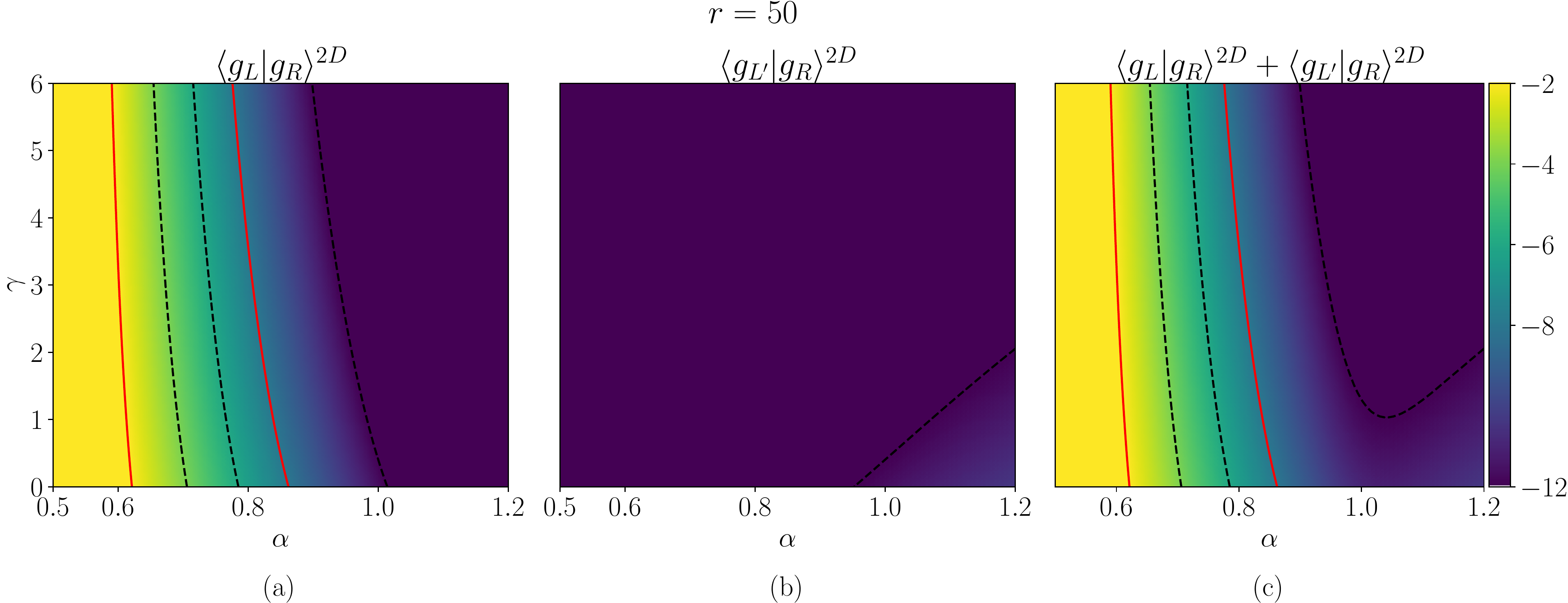}
    \caption{Replica of fig. \ref{f. gap vs alpha and r 2D harm} but for different values of $\alpha$ and $\gamma$, $r=50$.}
    \label{f. gap vs alpha and gamma 2D harm}
\end{figure}

\begin{figure}[H]
    \centering
    \includegraphics[width=\linewidth]{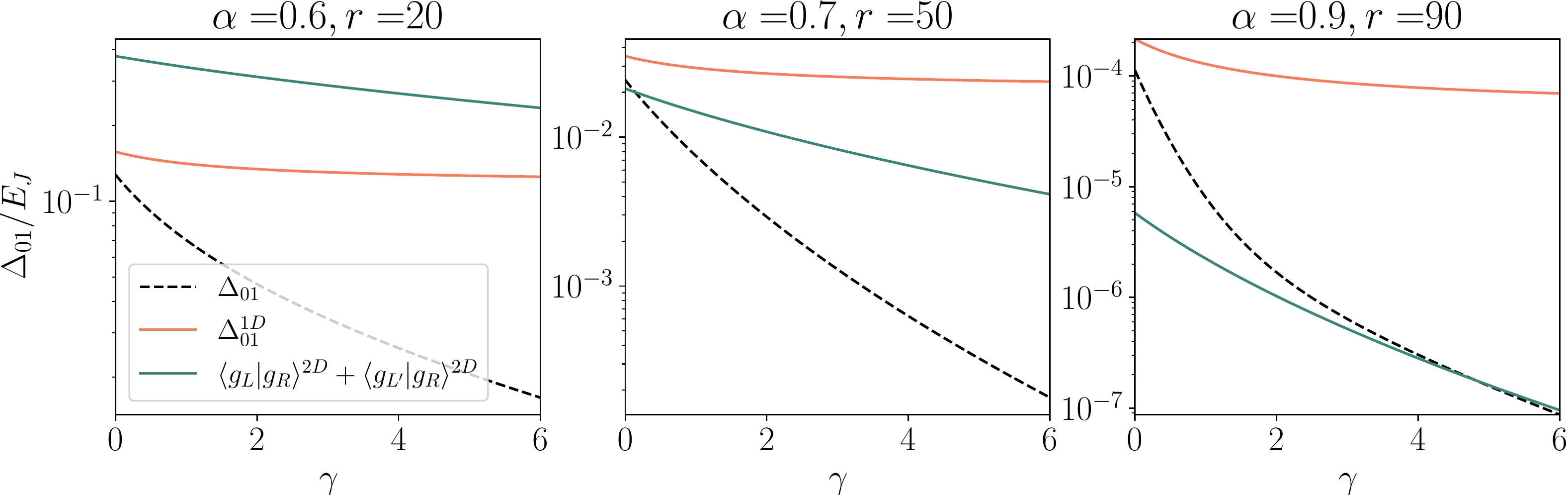}
    \hspace*{3.3cm} (a) \hfill (b) \hfill (c) \hspace*{2.12cm}
    \caption{$\Delta_{01}(\gamma)$ for different values of alpha and gamma.}
    \label{f. gap vs gamma 2D harm}
\end{figure}

Again, to do a more quantitative analysis we have plotted the overlap obtained with the 2D harmonic approximation along with the gap of the 3JJQ of the full Hamiltonian and the 1D approx in fig. \ref{f. gap vs gamma 2D harm}. The results show that we can capture the exponential decay for large values of $\gamma$ only for large values of $\alpha$ and $r$, fig. \ref{f. gap vs gamma 2D harm}~(c), which makes sense because as we showed in fig. \ref{f. barrier height vs frequency oscillator } this is the deep-well limit where the harmonic approximation works better.

\pagebreak

\section{Electromagnetic interaction with flux qubits}
\label{s. Electromagnetic interaction with flux qubits}
Now that we understand the functioning of the 3JJQ we turn to seek the fundamental ways to interact with it. Since a qubit is a two level system we can express any operator acting on the qubit space as a linear combination of Pauli matrices, hence, a certain type of interaction between qubits will be fundamentally different from another one if we can express them on the qubit space and they give rise to different Pauli matrices. These qubit-qubit interactions would be fundamental in the sense that if we were able to find three ways to make two qubits interact that resulted on three distinct Pauli matrices, then we would be able to reconstruct any possible interaction between qubits as a linear combination of those. In an ideal world these interactions would also be highly tunable and strictly orthogonal.

\subsection{Electromagnetic observables of flux qubits}
The first step to seek an answer to our problem is to understand how different operators are expressed on the qubit space. In a way we have already started to answer this question—the effective representation of the Hamiltonian operator on the qubit space is proportional to $\sigma^z$. Since the eigenbase of the qubit (symmetric and antisymmetric superposition of current states) was the result of quantum tunneling we can conclude that in this basis the tunneling operator is associated to $\sigma^z$. Being in the domain of quantum electrodynamics it would be reasonable to think that each degree of freedom (each Pauli matrix) is associated with each of the three pillars of the triad quantum-electricity-magnetism. If quantum tunneling is associated with $\sigma^z$, are magnetic-like and electric-like operators associated with the other degrees of freedom, $\sigma^x$ and $\sigma^y$?

Since the flux qubit can be understood as a magnetic dipole—a pseudo spin—we can start by considering a magnetic-like operator. Noting that are expressing the qubit states in a base of phase states we can indeed start by considering the phase operator. Thanks to the electric flux-phase relation any result regarding the phase operator can be immediately associated to the electric flux operator. Additionally, the electric flux can be associated with a current of charges and hence with a magnetic flux, thus, all these phenomena fall in the same category of magnetic-like operators. The first order contribution of the effective representation of the flux operator on the qubit basis can be expressed as:
\begin{equation}
\hat{\varphi}_\text{eff,1} = P_0\hat{\varphi}P_0 =\left[
\begin{array}{cc}
  \braket{0|\hat{\varphi}|0}  & \braket{0|\hat{\varphi}|1} \\
  \braket{1|\hat{\varphi}|0}  & \braket{1|\hat{\varphi}|1}
\end{array}
\right] =   \braket{1|\hat{\varphi}|0} \sigma^x
\end{equation}
Without invoking any approximation we can predict the shape of this operator. Since the qubit states are a symmetric and antisymmetric superposition of the current / flux states it is clear that the overall expected value of these operators in the ground or excited state will be exactly zero, $\braket{0|\hat{\varphi}|0}=\braket{1|\hat{\varphi}|1}=0$. We can also understand this considering that $\hat{\varphi}$ appears as an antisymmetric function in the integrals of the expected values in the phase base, thus, the only nonzero matrix elements will be those where $\hat{\varphi}$ multiplies another antisymmetric function. Since our wavefunctions are real we know that $\braket{1|\hat{\varphi}|0} = \braket{0|\hat{\varphi}|1}$ and hence we can conclude that  $\hat{\varphi}_\text{eff,1} \propto \sigma^x$.

We can do a similar analysis for electric-like operators. In this case we can study the operator number of cooper pairs, because we have a simple expression for it in the base of phase states, $\hat{n}=i\partial/\partial_\varphi$. Again, we can relate this operator to the charge operator and hence any result regarding $\hat{n}$ can be translated to $\hat{q}$ up to a constant factor. For the first order contribution, since $\hat{n}$ changes the symmetry of the state it is acting upon we can foresee that $\braket{0|\hat{n}|0}=\braket{1|\hat{n}|1}=0$. Expanding the qubit states in the $|L>$ and $|R>$ basis we can convince ourselves that $\braket{0|\hat{n}|1}=-\braket{1|\hat{n}|0} = -ai$, thus, we can conclude that up to first order $\hat{n}\propto\sigma^y$. In this case we can use finite differences to write the number operator in the base of phase states as an imaginary antisymmetric matrix with zeros in the diagonal and decaying matrix elements as we move away from the diagonal, thus, we can expect contributions from higher order perturbation theory that might result in other Pauli matrices.  

In the following section we will study two particular cases of magnetic-like and electric-like interactions where we expect to find these operators.

\subsection{Dipolar interaction with a magnetic field}
\begin{figure}[H]
    \centering
    \includegraphics[width=0.5\linewidth]{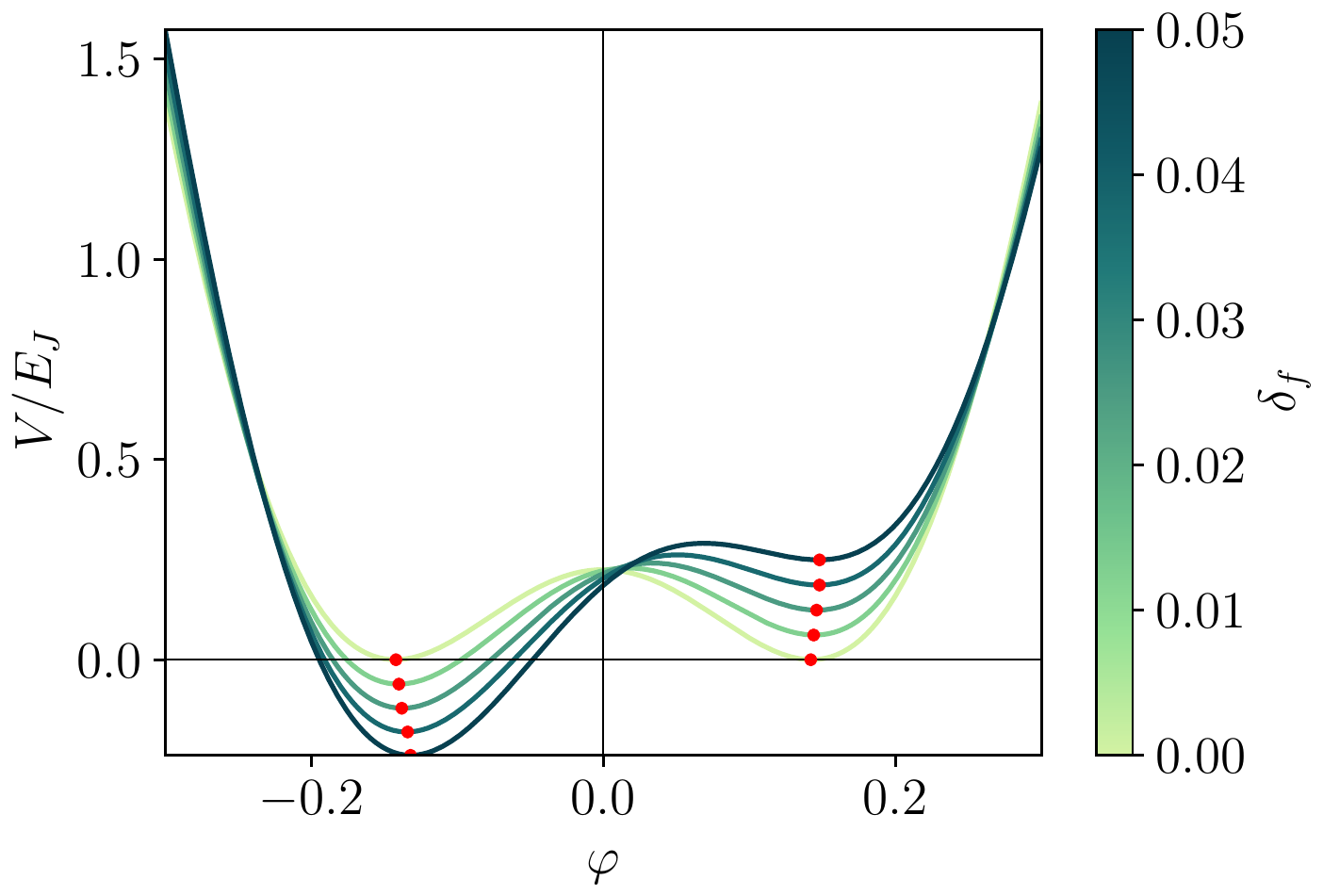}
    \caption{3JJQ potential in the $t_1$ direction for different values of $\delta_f$, eq. \eqref{e. perturbation flux}. The red dots show the position of the minima of the potential}
    \label{f. perturbation in flux potential}
\end{figure}

Lets consider the effect of a perturbation in the external flux threading the loop of the 3JJQ. This perturbation deviates the qubit from the frustration point and as a result the 3JJQ potential looses its symmetry with respect to the line $\varphi_{-}=0$, fig. \ref{f. perturbation in flux potential}.

We can qualitatively predict the effect of such perturbation in the effective Hamiltonian of the qubit. In the current basis it is easy to see the effect of a perturbation in the external flux: the currents required to trap an integer number of flux quanta inside the loop are no longer opposite currents of the same intensity, since one of them will be more energetically favourable and hence less intense. We can check this by looking at fig. \ref{f. perturbation in flux potential}: as we increase $\delta_f$ the minimum corresponding to negative $\varphi$ (counter clockwise current) drifts towards smaller-magnitude values of $\varphi$ (less intense currents) and consequently lowers its energy. The minimum corresponding to the opposite current behaves in the opposite way, shifting towards larger intensities and energies.  This means that in the current base this perturbation will appear as a $\sigma_z$ operator, creating a energy gap between the current states. 

In section \ref{s. harmonic approx 1D} we showed that the transformation that takes us from the current base to the eigenstate base switches the $\sigma_x$ and $\sigma_z$ operators, thus, as expected the perturbation in flux will appear in the effective Hamiltonian written in the eigenstate base as a $\sigma_x$ operator. This means that shift away from the frustration point controls the probability amplitude of the qubit to travel between the ground and first excited state, and since it is reasonably easy to create external magnetic fluxes with good precision one can understand why this is the control parameter of flux qubits.  

Our strategy to study this perturbation quantitatively is to write the resulting Hamiltonian in perturbation form as in  eq. \eqref{e. perturbed hamiltonian}, and then apply first order perturbation theory to calculate the expected values of the perturbation operator using the approximate harmonic wavefunctions. A perturbation in the external phase that deviates the qubit from the frustration point can be written as $f = 0.5 + \delta_f$, and results in a perturbed version of the Hamiltonian \eqref{e. H t1}:
\begin{equation}
\frac{\hat{H}_{t_1}(f = 0.5 + \delta_f)}{E_J}=\frac{1}{2}\frac{\hat{n}^2_{+}}{m_{+}}-\big[ 2\cos(\varphi) -\alpha \cos\left(2\varphi-2\pi\delta_f\right) \big]\,.
\label{e. perturbation flux}
\end{equation}
Extracting the perturbation from the cosine and expanding up to first order in $\delta_f$
\begin{equation}
\cos\left(2\varphi-2\pi\delta_f\right)  = \cos(2\varphi)\cos(2\pi\delta_f) + \sin(2\varphi)\sin(2\pi\delta_f)  \approx \cos(2\varphi)+ 2\pi\delta_f \sin(2\varphi)+ O(\delta_f^2) 
\label{e. taylor expansion perturbation flux}
\end{equation}
allows us to write
\begin{equation}
\frac{\hat{H}_{t_1}(f = 0.5 + \delta_f)}{E_J}=\hat{H}_{t_1} + 2\pi\alpha\sin(2\varphi)\delta_f  =\hat{H}_{t_1}+V_{f}\delta_f \,.
\end{equation}
With this expression we can obtain the two level effective Hamiltonian via first order perturbation theory as explained in section \ref{s. Heff perturbation}, eq. \eqref{e. first order correction}:
\begin{equation}
\frac{H_{\text{eff},1}^h(\delta_f)}{E_J} =\frac{1}{E_J}\left( P_0\hat{H}_{t_1}P_0 + \delta_f P_0V_f P_0 \right)= \frac{\Delta_{01}^h}{2}\sigma_z + \delta_f\sum_{i,j}\braket{i|V_{f}|j}\ketbra{i}{j} \,.
\end{equation}
Due to the symmetries of the functions we can see that $\braket{0|V_{f}|0}=\braket{1|V_{f}|1}=0$, $\braket{0|V_{f}|1}=\braket{1|V_{f}|0}$. Using the harmonic approximating to calculate the expected value of this operator gives:
\begin{equation}
\epsilon=\braket{0|V_{f}|1} = \int 2\pi\alpha\sin(2\varphi) \psi_0(\varphi)\psi_1(\varphi) d\varphi = \pi\frac{ \sqrt{4 \alpha^{2} - 1}}{\alpha} e^{- \frac{1}{m \omega}}\,,
\end{equation}
thus, the effective Hamiltonian of a perturbation in flux according to first order perturbation theory and up to first order in $\delta_f$ is:
\begin{equation}
\frac{ H_{\text{eff},1}^h(\delta_f)}{E_J} = \frac{\Delta_{01}^h}{2}\sigma_z + \epsilon\delta_f\sigma_x\,.
\end{equation}

\begin{figure}[H]
    \centering
    \includegraphics[width=\linewidth]{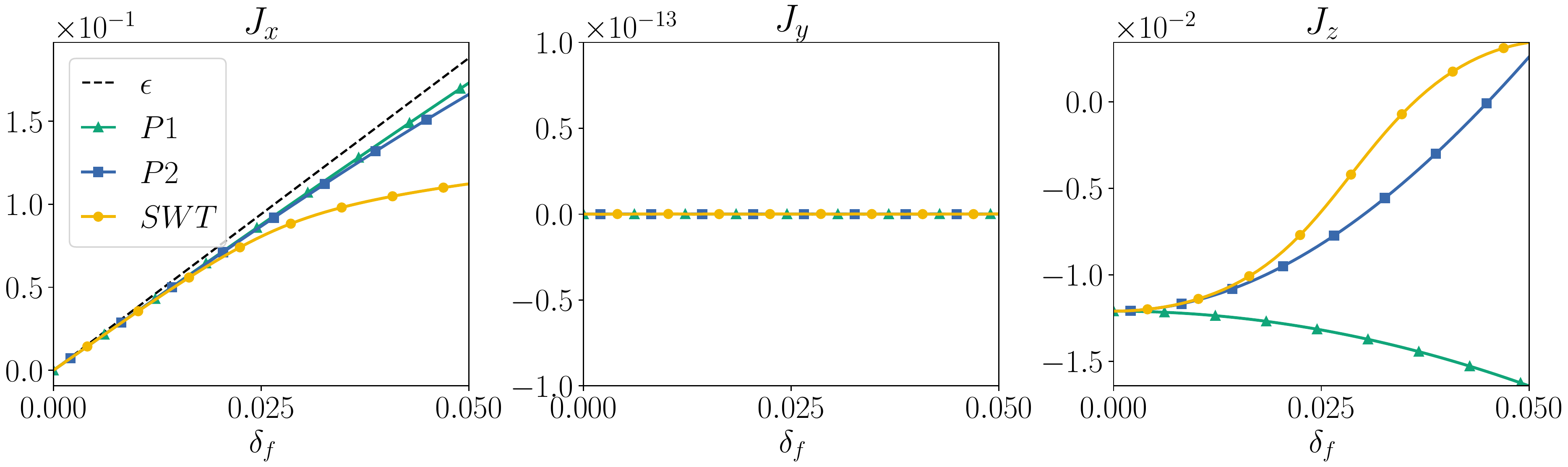}
    \caption{Effective Hamiltonian of a 3JJQ, $\alpha=0.7$ and $r=50$,, with a perturbation in flux, eq. \eqref{e. perturbation flux}, calculated according to first(second) order perturbation theory, $P1$($P2$), and the Schrieffer-Wolff transformation, $SWT$. The $J_i$ values correspond to the constants that accompany the $\sigma_i$ operators in the effective Hamiltonian. }
    \label{f. perturbation in flux }
\end{figure}

Fig. \ref{f. perturbation in flux } compares this result with those obtained by means of a numerical first and second order perturbation theory, along with the exact results of the SWT. As we can see the model predicts correctly coefficient of the $\sigma^x$ for low values of $\delta_f$, giving a result very similar to that of first order perturbation theory. However, we can see that we are missing some information: all the numerical models predict a modification to the $\sigma_z$ operator—the gap of the qubit—with a scaling of order $\delta_f^2$ or higher. Our application of the first order perturbation theory with the harmonic approximation did not yield this term because we neglected $\delta_f^2$ terms and higher from the Taylor expansion \eqref{e. taylor expansion perturbation flux}. Nonetheless we can understand qualitatively this term by looking at fig. \ref{f. perturbation in flux potential}: the perturbation $\delta_f$ not only shifts up and down the minima, but also makes the ground minima deeper and the excited minima flatter. This results in a modification of the left and right wavefunctions and hence a modification in their overlap and consequently in the gap of the qubit.

\subsection{Dipolar interaction with an electric field}
\label{s. Perturbation in charge}
To electrically perturb the 3JJQ we have to connect one of it's notes to a voltage source thorugh a mediating capacitor. As we saw in section \ref{s. capactive loading} simply including a capacitor renormalizes the effective mass of the 3JJQ and consequently changes its gap.  Having isolated the effect of the additional capacitance, we are ready to study the effect of adding a perturbing voltage $\delta_{V}$ thorugh the new capacitance. According to reference \cite{consani2020effective} we have to include the perturbation 
$$
\Delta H(\tilde{\delta}_V) = \gamma C \delta_{V} \sum_{i \neq 2}\left(\tilde{\mathbf{C}}^{-1}\right)_{a i} q_{i} + \frac{1}{2}\left(\tilde{\mathbf{C}}^{-1}\right)_{a a}\left(\gamma C \delta_{V}\right)^{2}
$$
to the Hamiltonian, where $a$ is the node where we are connecting the voltage source. Assuming that we connect the capacitance to the node 2 of the 3JJQ, ignoring the constant term and substituting the values from our inverse capacitance matrix gives
\begin{equation}
\Delta H_{t_1}(\tilde{\delta}_V)=\gamma C \delta_{V} \sum_{i \neq 2}\left(\tilde{\mathbf{C}}^{-1}\right)_{2 i} q_{i} = \frac{\alpha\gamma}{d}\delta_{V}\hat{q}_1\,.
\end{equation}
where$d=|\hat{C}|=2\alpha+1+\gamma(\alpha+1)$ is the determinant of the \textit{renormalized} capacitance matrix $\hat{C}$. 

\noindent Since we are working with dimensionless variables it is useful to write the voltage and charge operator of this equation in terms of a dimensionless perturbation in voltage, $\tilde{\delta}_V=\frac{C\delta_{V}}{e}$ —which is essentially a perturbation in the number of cooper pairs—and the number of cooper pairs operator:
\begin{equation}
     \delta_{V}\hat{q}_1 =-4\frac{C\delta_{V}}{e}\frac{e^2}{2C}\hat{n}_1 =-4\frac{C\delta_{V}}{e}E_C\hat{n}_1 =-4 E_C\tilde{\delta}_V\hat{n}_1\,.
\end{equation}
To proceed with our analysis we have to write the charge number operator on the node 1 in terms of our $\pm$ variables. Since $\hat{n}_a$ = $\hat{n}_1-\hat{n}_0$ and $\hat{n}_b$ = $\hat{n}_0-\hat{n}_2$, setting the node 0 to ground and applying the 1D approx. allows us to write $\hat{n}_1\approx\hat{n}_{+}$, $\hat{n}_2\approx-\hat{n}_{+}$ .The perturbed Hamiltonian can be written as
\begin{equation}
\frac{\hat{H}_{t_1}(\tilde{\delta}_V)}{E_J}  = \frac{1}{E_J}\big(\hat{H}_{t_1}+\Delta H(\tilde{\delta}_V)\big) = \frac{\hat{H}_{t_1}}{E_J} -4\frac{\alpha \gamma }{rd}\hat{n}_{+}\tilde{\delta}_V = \frac{\hat{H}_{t_1}}{E_J} + V_{V}\tilde{\delta}_V \,.
\end{equation}
First order perturbation theory gives an approximate effective Hamiltonian of the voltage-perturbed circuit:
\begin{equation}
\frac{H_{\text{eff},1}(\tilde{\delta}_V)}{E_J} = P_0\frac{\hat{H}_{t_1}}{E_J}P_0 + \tilde{\delta}_V P_0V_VP_0\,.
\label{e. perturbation voltage}
\end{equation}
One important note must be made here. When calculating the effective Hamiltonian we must choose which base to use as the unperturbed base $P_0$. For the perturbation in flux the choice was obvious, however, in this case we must decide whether or not to include the additional capacitance $\gamma$ and its renormalization in the unperturbed base. In a experimental setup the capacitance $\gamma$ is something hard to control, it's either there or not, thus, we have decided to take the renormalized Hamiltonian as the unperturbed Hamiltonian.

Using the harmonic approximation we can estimate the value of the matrix elements of \eqref{e. perturbation voltage}, however, since the perturbation is proportional to a number operator we can predict that the first order correction will be a $\sigma_y$. The harmonic approximation yields the following matrix elements:
\begin{equation}
\eta =\frac{\braket{0|V_{V}|1}}{i} = -4\frac{\alpha \gamma}{rd}\int  \psi_0(\varphi)\frac{d\psi_1(\varphi)}{d\varphi} d\varphi = 4\frac{\alpha \gamma }{rd} \widetilde{m}_{+} \widetilde{\omega}_{+} \varphi_{0} e^{- \widetilde{m}_{+} \widetilde{\omega}_{+} \varphi^{*2}}\,,
\label{e. eta}
\end{equation}
which result in a first order effective Hamiltonian
\begin{equation}
\frac{H_{\text{eff},1}^h(\tilde{\delta}_V)}{E_J} = \frac{\Delta_{01}^h}{2}\sigma_z  + \eta\tilde{\delta}_V\sigma_y \,.
\end{equation}

\begin{figure}[H]
    \centering
    \includegraphics[width=\textwidth]{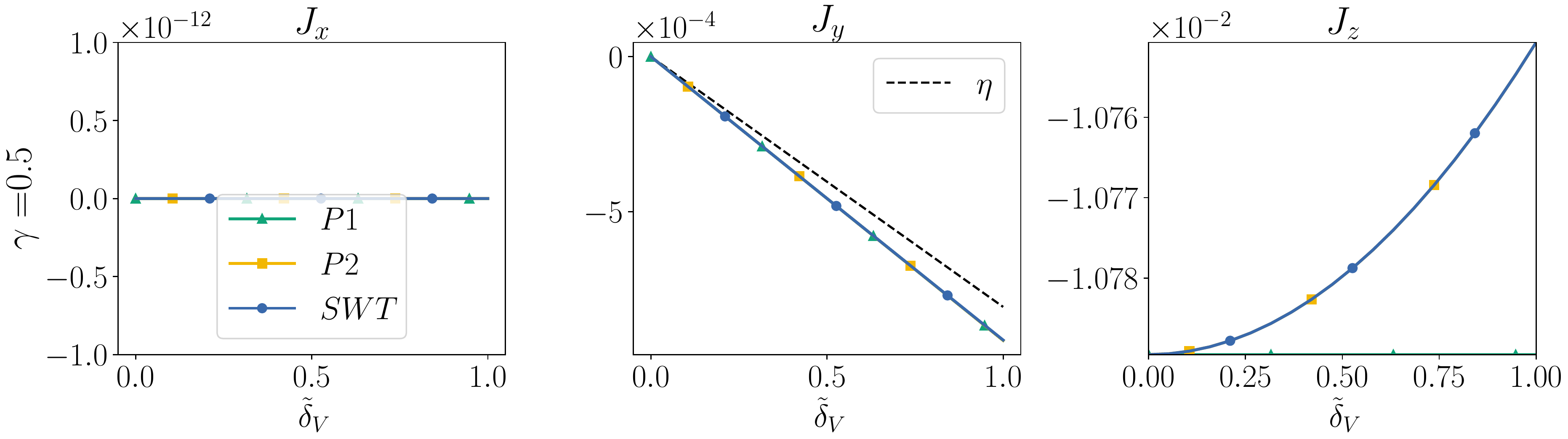}
    \includegraphics[width=\textwidth]{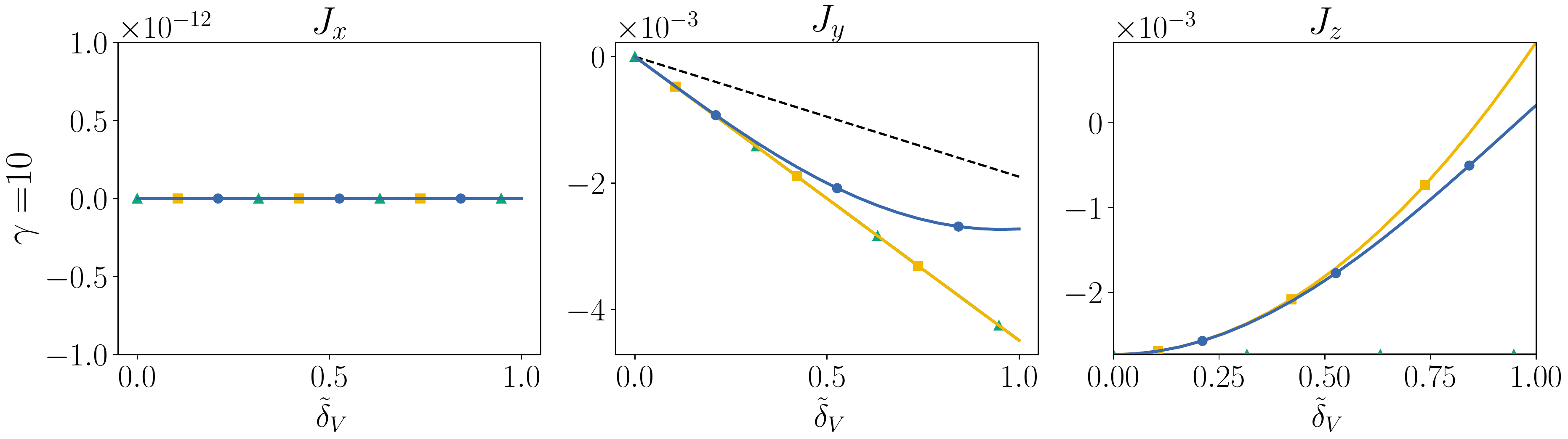}
    \caption{Effective Hamiltonian of a 3JJQ, $\alpha=0.7$ and $r=50$, with a perturbation in voltage $\tilde{\delta}_V$ connected to the node 2 through a capacitor of capacitance $\gamma C$.}
    \label{f. perturbation in voltage}
\end{figure}

Fig. \ref{f. perturbation in voltage} compares this result with those obtained numerically for two different values of $\gamma$. For a small value $\gamma=0.5$, top row fig. \ref{f. perturbation in voltage}, the model predicts correctly the coefficient of $\sigma^y$. First order perturbation theory gives the same results as second order and SWT for this coefficient, however, it fails to capture the apparition of a $\sigma_z$ operator. The second order correction neatly captures this therm, thus, we can conclude that it originates from interactions among the high energy states of the circuit. For a large value of $\gamma=10$, bottom row fig. \ref{f. perturbation in voltage}, the results are qualitatively similar, however, in this case the prediction of the harmonic approximation worsens and the $SWT$ departs from first and second order perturbation theory for large values of $\tilde{\delta}_V$. 

\pagebreak
\section{Analysis and design of couplings between flux qubits}
\label{s. Couplings}
\subsection{Design of couplings between flux qubits}

\begin{figure}[H]
    \centering
    \includegraphics[width=0.7\textwidth]{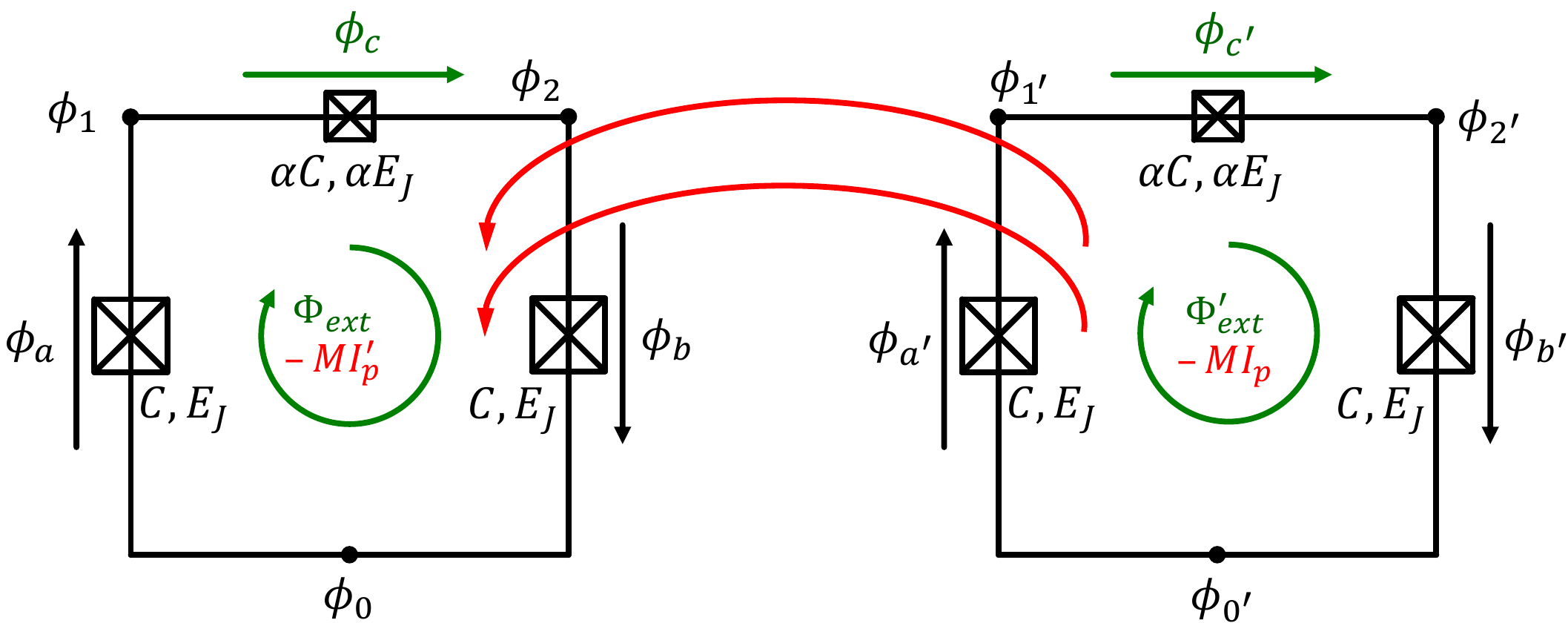}
    \caption{Coupling between two identical 3JJQ thorugh their mutual inductance.}
    \label{f. acoplo inductivo}
\end{figure}

Since flux qubits are superconducting circuits with well-defined flux/phase/current states the natural way to make two flux qubits interact is via their magnetic fluxes, and the simplest way to achieve this coupling is to place the qubits close together such that their magnetic fluxes thread each others loop \cite{orlando, ind1}. As a result of this configuration the external flux threading each qubit will have an additional contribution proportional to the magnetic flux generated by the persistent current of the other qubit, as shown in fig. \ref{f. acoplo inductivo}. For instance, the external flux of the first qubit will have a contribution $MI^{'}_p$, where $M$ is the qubit's mutual inductance coefficient and $I^{'}_p$ is the persistent current of the other qubit. 

Thanks to Kirchoff's current law we know that the persistent current circulating in each qubit will be the current through any of the Josephson junction. Since we are going to adapt the 1D approximation it is useful to work with the current circulating thorough the small junction. We make this choice because we know that the flux through the small junction is related to our 1D variable, $\varphi_c=2\varphi - 2\pi f$, and hence the persistent current of each qubit will be $I_p=\alpha I_c\sin(\varphi_c)=\alpha I_c\sin(2\varphi - 2\pi f)$, where $I_c=\varphi_0E_J$. At the frustration point the persistent current simplifies to $I_p=-\alpha I_c\sin(2\varphi)$. Following from the previous section, the Hamiltonian of a 3JJQ in the frustration point with a perturbation in flux can be written as
\begin{equation}
    \hat{H}_{t_1}\left(\Phi_\text{ext} = \frac{1}{2}\Phi_0 + \delta_\Phi\right)=\hat{H}_{t_1}\left(\Phi_\text{ext} =\frac{1}{2}\Phi_0\right) +  \alpha E_J\sin(2\varphi)\frac{\delta_\Phi}{\varphi_0}\,,
\end{equation}
thus, if the perturbation is $\delta_\Phi=-MI^{'}_p$ and identifying in the previous equation $\alpha E_J\sin(2\varphi)=\varphi_0 \alpha I_c\sin(2\varphi)= -\varphi_0I_p $ we can finally write:
\begin{equation}
    \hat{H}_{t_1}\left(\Phi_\text{ext} = \frac{1}{2}\Phi_0 -MI^{'}_p\right)=\hat{H}_{t_1}^0 + \hat{H}_{int}\,,\quad \hat{H}_{int}=  MI_pI_p^{'}=M\alpha^2I_c^2\sin(2\varphi)\sin(2\varphi^{'})\,.
\end{equation}
Since the two qubits are identical we could apply the same analysis to the second qubit and write the the Hamiltonian of the coupled system is
\begin{equation}
    \hat{H}^M_{t_1} = \hat{H}_{t_1}^0 + \hat{H}_{t_1}^{'0} + 2MI_pI_p^{'}\,.
\end{equation}
Noting that $\sin(2\varphi)$ is a antisymmetric real function we know from the previous section that first order perturbation theory will yield an effective $\sigma_x$ operator, as expected from a magnetic interaction, with a magnitude
$$
\braket{0|\sin(2\varphi)|1}=\sin(2\varphi^{*})e^{-\frac{1}{4m\omega}} \sim 1\,,
$$
thus, we can conclude that the first order effective Hamiltonian in the local qubit base can be written as
\begin{equation}
    \frac{\hat{H}^M_{t_1}}{E_J} = \frac{\Delta_{01}}{E_J}\sigma^z_1 + \frac{\Delta_{01}}{E_J}\sigma^z_2 + 2\frac{MI_c^2}{E_J} \alpha^2\sigma^x_1\sigma^x_2\,.
\end{equation}
One problem of this scheme is that it only favours an anti-ferromagnetic alignment\footnote{Note that $MI_pI_p^{'}$ will be negative if the currents have opposite sign.} of the qubits, \textit{i.e.} it is not tunable. To overcome this problem multiple works \cite{ind2,ind3,ind4,ind5,ind6,ind7} have shown that one can place a mediating circuit between the qubits— for instance a rf-SQUID or another flux qubit—and control the magnetic flux threading the mediating circuit to obtain a tunable $\sigma^x\sigma^x$ interaction in sign and magnitude. This interaction can also be tuned in time, which is why it has been used by D-Wave to perform quantum annealing \cite{DWAVEannealing}.

\begin{figure}[H]
    \centering
    \includegraphics[width=0.8\textwidth]{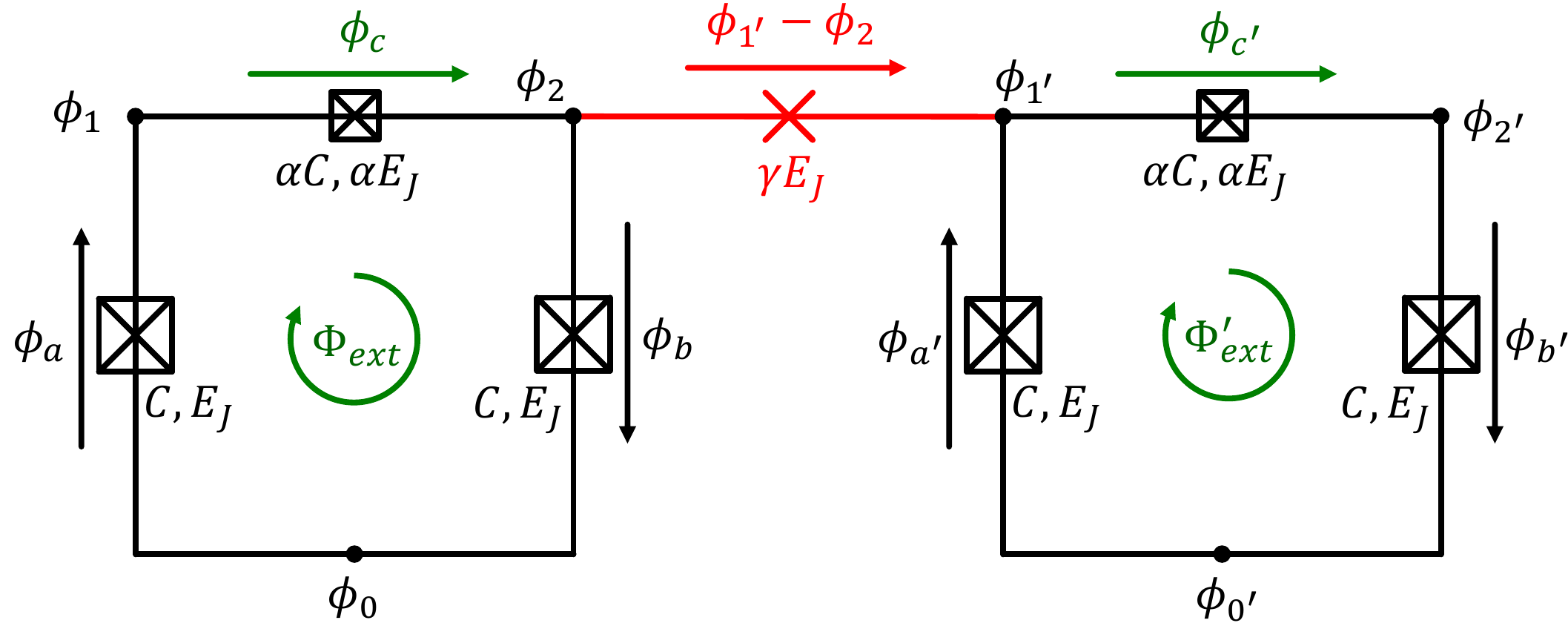}
    \caption{Coupling between two identical 3JJQ thorugh a Josephson Junction of energy $\gamma E_J$.}
    \label{f. acoplo JJ}
\end{figure}

Another problem of this scheme is that it relies on the proximity and shape of the qubits and the mediating circuit, imposing severe geometric constrains in the design and not allowing for interactions between distant qubits. Additionally, it is hard to apply magnetic field to a localized region, and hence it is difficult to achieve individual addressability of all qubits, because the magnetic field may affect not only the target qubits but also other qubits as well. Here we propose and analyze another coupling scheme to obtain a tunable inductive $\sigma^x\sigma^x$ interaction which overcomes these limitations. The proposed scheme consists on connecting two identical 3JJQs with a Josephson junction\footnote{To fully model a Josephson junction one has to consider its capacitance, nonetheless, to better understand the inductive coupling we will assume an ideal junction and study the capacitive coupling in the next scheme.}, as shown in fig. \ref{f. acoplo JJ}.

\noindent To obtain Hamiltonian of the coupled system one has to add to the Hamiltonians of the 3JJQs the energy of the junction, which depends on the flux difference across it, $\phi_{1'}-\phi_2$. This flux difference can be written in terms of our $\pm$ variables, $\phi_{1'}=\phi^{'}_{+}+\phi^{'}_{-}$ and $\phi_{2}=\phi_{-}-\phi_{+}$. Neglecting $\phi^{'}_{-}$ and $\phi_{-}$ according to the 1D approx. and forgetting the $+$ subindex gives
\begin{equation}
    \hat{H}^{JJ}_{t_1} = \hat{H}_{t_1}^0 + \hat{H}_{t_1}^{'0} + \gamma E_J \cos\left(\frac{\phi'+\phi}{\varphi_0}\right)\,,
\end{equation}
where the interaction term can be expanded as
\begin{equation}
    \hat{H}^{JJ}_{int}=\gamma E_J \cos\left(\frac{\phi'+\phi}{\varphi_0}\right)=\gamma E_J \left[\cos(\varphi)\cos(\varphi') +\sin(\varphi)\sin(\varphi')\right]\,.
\end{equation}
The first order effective Hamiltonian can be calculated in two parts. The $\sin(\varphi)$ operators will give a $\sigma^x$ as expected, with a magnitude of:
$$
\braket{0|\sin(\varphi)|1}=\sin(\varphi^{*})e^{-\frac{1}{4m\omega}} \sim 1\,.
$$
The $\cos(\varphi)$ operators are symmetric and hence they will yield a  $\sigma^I$ and a $\sigma^z$ operator. The constant energy shift can be ignored and the magnitude of the $\sigma^z$ operator can be calculated as
$$
 \frac{1}{2}\big(\braket{1|\cos(\varphi)|1} - \braket{0|\cos(\varphi)|0}\big) = e^{-m\omega\varphi^{*}-\frac{1}{4m\omega}}= \braket{g_L|g_R}e^{-\frac{1}{4m\omega}} \sim \frac{\Delta_{01}}{E_J}\,,
$$
thus, we can expect this term to be quite smaller than $J_x^{JJ}$. The Hamiltonian of the coupled system hence can be estimated as
\begin{equation}
    \frac{\hat{H}^{JJ}_{t_1}}{E_J} \approx \frac{\Delta_{01}}{E_J}\sigma^z_1 + \frac{\Delta_{01}}{E_J}\sigma^z_2 + \gamma \sigma_1^x\sigma_2^x + \gamma\left[\frac{\Delta_{01}}{E_J}\right]^2\sigma_1^z\sigma_2^z \sim \Delta_{01}\sigma^z_1 + \Delta_{01}\sigma^z_2 + \gamma \sigma_1^x\sigma_2^x 
\end{equation}

\noindent As it is right now this coupling is not tunable, however, we can easily solve this by substituting the Josephson junction for a dc-SQUID: two Josephson junctions in parallel, which behave as a Josephson junction whose inductive energy can be controlled with an external flux threading their loop
$$
E_{J}(\Phi_\text{ext})_\text{dc-SQID}=2 I_{c} \varphi_{0} \cos \left(\Phi_\text{ext} / 2 \varphi_{0}\right)=E_{J}(0) \cos \left(\Phi_\text{ext} / 2 \varphi_{0}\right)\,.
$$
As we can see, this coupling will produce a tunable and very similar effective interaction to that of two qubits magnetically coupled, with the advantage of not imposing any constrain on the geometry or position of the qubits, enabling magnetic interactions between distant qubits in the chip, and enhancing qubit addressability, since one can put the coupling junctions well spaced in a remote section of the chip and ensure a minimal unwanted cross-talk between qubits.

\begin{figure}[H]
    \centering
    \includegraphics[width=0.8\textwidth]{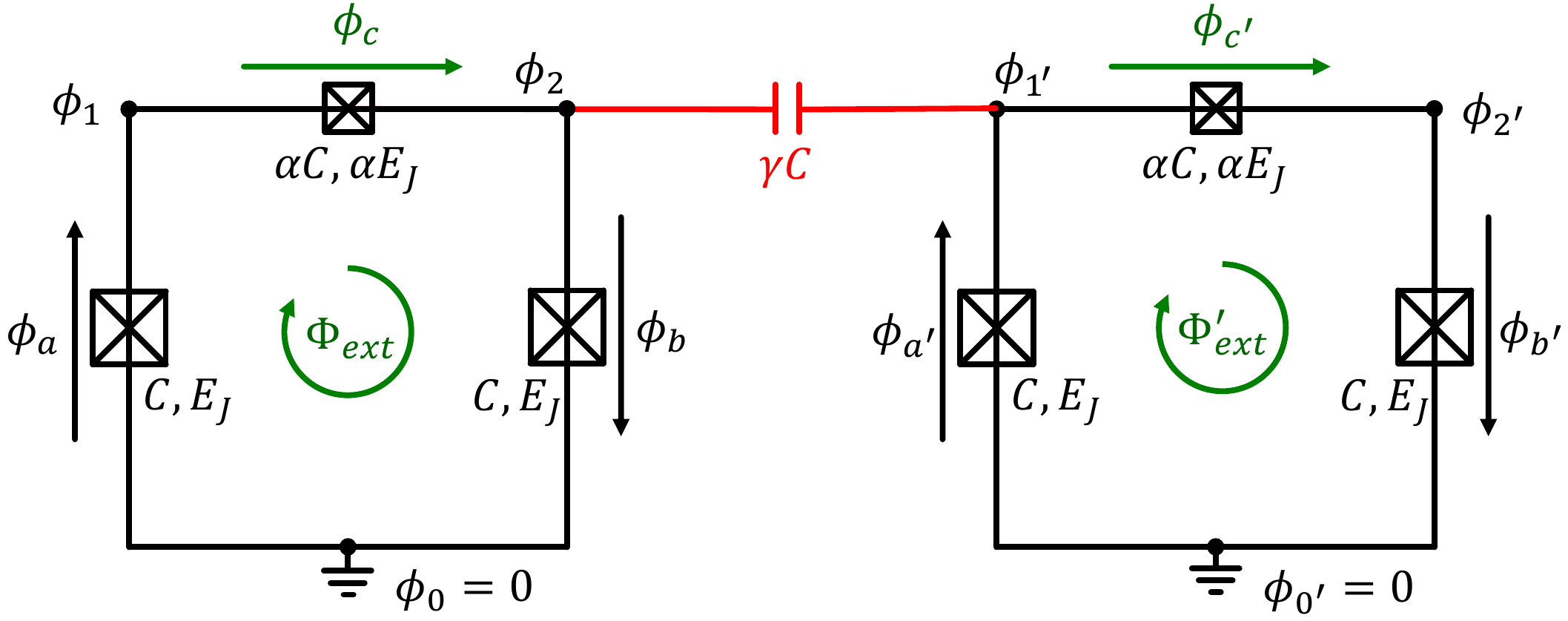}
    \caption{Coupling between two identical 3JJQ thorugh a capacitor of capacitance of $\gamma C$.}
    \label{f. acoplo C}
\end{figure}

At last we consider the capacitive coupling between two identical 3JJQs, depicted in the fig.~\ref{f. acoplo C}. Experiments with capacitively coupled flux qubits\ \cite{satoh2015,ozfidan2020demonstration} have demonstrated an effective interactions along more than one direction\ \cite{consani2020effective},  but the coupling strength seems to be limited and there is no analytical framework to understand the range of available interactions. In this section we will try to analytically predict the shape and scaling of this interactions with the harmonic approximation, and in the following section we will numerically study its strength.

We know that one of the effects of the capacitance will be to renormalize the gaps of the qubits, however, since we have already studied this phenomena we can work with the renormalized singe quibt Hamiltonians and concentrate on the effective interactions produced by capacitive coupling.

\noindent Calling $\tilde{H}_{t_1}^0$ and $\tilde{H}_{t_1}^{'0}$ to the renormalized Hamiltonians of the single qubits we can write the Hamiltonian of the coupled system as:
\begin{equation}
    \hat{H}^{C}_{t_1} = \tilde{H}_{t_1}^0 + \tilde{H}_{t_1}^{'0} + \mathbf{Q}^T\mathbf{C}_C^{-1}\mathbf{Q}^{'}\,,
\end{equation}
where $\mathbf{Q}^T=[\hat{q}_1, \hat{q}_1]$ is the matrix containing the charge variables of the first qubit, $\mathbf{Q}^{'}$ of the second, and $\mathbf{C}_C^{-1}$ is the section of the inverse renormalized capacitance matrix, $\tilde{\mathbf{C}}$, that refers to the product of charges of different qubits (\textit{i.e.} the off-diagonal block of the matrix) as explained in appendix \ref{a. circuit hamiltonian}. For the scheme depicted in fig.~\ref{f. acoplo C} the capacitive interaction can be written as:
\begin{equation}
\hat{H}_{int}^C = \frac{\alpha^2\gamma+\alpha\gamma}{d}\left(\hat{q}_1\hat{q}_1' + \hat{q}_2\hat{q}_2'\right) + \frac{2\alpha^2\gamma+2\alpha\gamma+\gamma}{d}\hat{q}_1\hat{q}_2'\,.
\end{equation}
where $d=|\tilde{\mathbf{C}}|=C(4\alpha^2\gamma+4\alpha^2+6\alpha\gamma+4\alpha+2\gamma+1)$. Introducing the $\pm$ variables and applying the 1D approx simplifies this expression to:
\begin{equation}
\hat{H}_{int}^C \approx -2\frac{2\alpha^2+2\alpha+1}{4\alpha^2\gamma+4\alpha^2+6\alpha\gamma+4\alpha+2\gamma+1}\gamma \frac{\hat{q}\hat{q}'}{C} = A(\alpha,\gamma)\frac{\hat{q}\hat{q}'}{C} \,.
\end{equation}
The coefficient $A(\alpha,\gamma)$ is $O\big(\gamma/(\gamma+1)\big)$. Substituting $ \hat{q}\hat{q}'/C = \hat{n}\hat{n}'4e^2/C = \hat{n}\hat{n}'E_C$  gives
\begin{equation}
\frac{\hat{H}_{int}^C}{E_J} \approx  \frac{A(\alpha,\gamma)}{r}\hat{n}\hat{n}' \,.
\end{equation}
Since we know how to calculate the first order contribution of the number operator in the harmonic approximation we can conclude that the first order effective Hamiltonian of the capacitive coupling will be
\begin{equation}
\frac{\hat{H}_{int}^C}{E_J} \approx  \frac{A(\alpha,\gamma)}{r}\eta^2\sigma^y_1\sigma^y_2  \,,
\end{equation}
where we have defined $\eta$ in eq. \eqref{e. eta}. As we saw in fig. \ref{f. perturbation in voltage}, we can also predict that second order perturbative corrections will yield a $\sigma^z_1\sigma^z_2$ which appears from interactions between they high energy subspace of the qubit.

We have showed the origin and scaling of the low order perturbation terms of the effective Hamiltonian of two not-widely studied coupling mechanisms between 3JJQs, the Josephson junction coupling and the capacitive coupling. However, our analysis is limited to small interaction intensities. In the next section we will apply the numerically-exact SWT to verify our predictions and analyze the coupling schemes far from the perturbative regime. For the numerical analysis we will not limit ourselves to the circuits shown in figs. \ref{f. acoplo JJ} and \ref{f. acoplo C}, but rather we will consider all possible coupling topologies that connect one node of one qubit to another node of the other quibt, considering also different ground configurations. Additionally, since in the numerical analysis we do not seek to find compact analytic expressions we have decided to add another parameter to the 3JJQs, a shunting capacitor of capacitance $\beta C$ placed in parallel to the small junction of the qubit. The effect of this capacitance is to change the effective capacitance of the small junction, which will go from $\alpha C \to (\alpha+\beta)C$. Note, however, that the shunting capacitance will not modify at all the inductive potential of the 3JJQ.

The general Hamiltonian for these kind of circuits is $H=H_1+ H_2 +H_{int}$. Here, $H_{1(2)}$ is the single qubit Hamiltonian of the first (second) qubit, and $H_{int}$ describes the interaction between the qubits, including contributions that act both on the first and second qubit. This has the typical form of a perturbed Hamiltonian, where the sum of the \textit{renormalized} (or not) qubit Hamiltonians acts as the unperturbed Hamiltonian and the interaction term gives the perturbation. If we don't consider single qubit external magnetic fields that deviate the qubits from the frustration point, the effective Hamiltonian of the coupled system can be written in two-local qubit basis as:
\begin{equation}
    H_{\text{eff}}=\frac{\Delta_1}{2}\sigma_1^z+\frac{\Delta_2}{2}\sigma_2^z+\sum_{i,j=x,y,x} J_{ij}\sigma_1^i\sigma_2^j
    \label{e. Final Hamiltonian}
\end{equation}

\subsection{Analysis of 3JJQs coupled thorugh capacitors}
\label{s. numeric C coupling}
\noindent In this section, we study coupling schemes between two identical 3JJQs mediated by capacitors, as shown in the example Fig. \ref{f. acoplo C}. We have considered in total twelve possible configurations of the system (nine when taking into account the fact that both qubits are identical): two different ground configurations, $\varphi_0=\varphi_{0'}=0$ and $\varphi_1=\varphi_{2'}$, and all possible couplings of one or two capacitors connecting the remaining nodes. We will focus on the configuration shown in Fig. \ref{f. acoplo C}, but the following comments and conclusions can be applied qualitatively to any of the studied configurations.

\begin{figure}[H]
    \centering
    \includegraphics[width=\linewidth]{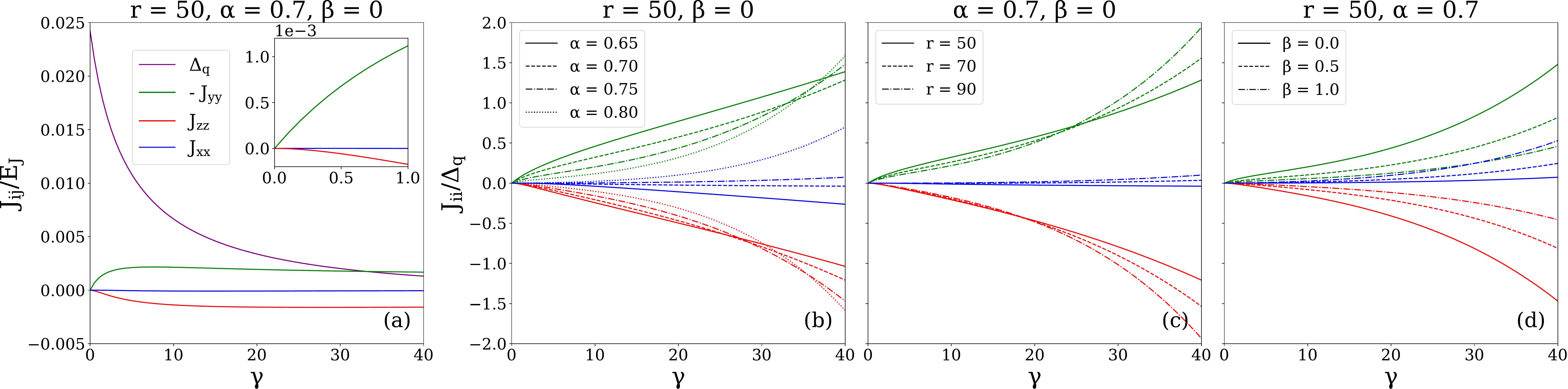}
    \caption{Coupling strengths for the reference circuit : two 3JJQs with ground in $\phi_0$($\phi_0'$) coupled through a capacitor connecting nodes $\phi_2-\phi_1'$. (a) Effective Hamiltonian parameters ($J_{ij}$) as a function of $\gamma$ for $\alpha=0.70$, $r=50$ and $\beta=0$. (b-d) Ratios between the coupling strengths ($J_{ii}$) and the qubit gap ($\Delta_q$): (b) for fixed $r$ and $\beta$,  (c) for fixed $\alpha$ and $\beta$, (d) for fixed $\alpha$ and $r$. Note that we represent $-J_{yy}$ for the sake of clarity.}
    \label{f. capacitive coupling}
\end{figure}

We have shown thorugh perturbation theory that we can expect two types of interactions. First-order corrections predict a $\sigma^y\sigma^y$ interaction that scales linearly with $\gamma$ and can be explained as the result of the direct interaction between charges on both qubits. Second-order corrections predict a $\sigma^z\sigma^z$ that scales quadratically $\gamma$, and that can be explained as interactions between qubit states mediated by states of the high energy subspace $\mathcal{Q}$ (exited states). It can be shown\cite{hita2021ultrastrong} that third-order corrections predict a $\sigma^x\sigma^x$ operator that scales cubically with $\gamma$. Perturbation theory agrees qualitatively and semi-quantitatively with the numerically exact results shown in the inset of fig.~\ref{f. capacitive coupling}(a) for small interaction strengths $\gamma$. However, at moderate interactions these predictions fail. The $J_{yy}$ and $J_{zz}$ couplings then reach a maximum and slowly start to decay, becoming equal in magnitude.

Figs.~\ref{f. capacitive coupling}(b-d) display the growth of the relative interaction strength $J/\Delta$ for several design parameters, illustrating the crossover from weak $J/\Delta\ll 1$ to strong coupling regime $J/\Delta\approx 1$. We have studied $J/\Delta$ because if one wants to fully appreciate the real strength of the coupling, the relevant magnitude is not the interaction itself but the interaction in units of the qubit's gap. For small $\gamma$, the behavior of the coupling is dominated by the perturbative tendencies in $J_{ii}$. For larger couplings, the growth of $J/\Delta$ is dominated by the exponential decrease of the gap with the renormalized qubit capacitance, which grows with $\gamma,\beta$ and $\alpha$, as we showed in section \ref{s. capactive loading}. This competition explains the non-monotonical behaviour found in $J_{zz}/\Delta,J_{yy}/\Delta$ with respect to $\alpha,r$ [cf. Figs.~\ref{f. capacitive coupling}(b-c)], as $J$ decreases while $1/\Delta$ increases with those parameters. Finally, for the limited range of $\gamma$ where the gap is not negligible, $J_{zz}/\Delta$ and $J_{yy}/\Delta$ always decrease with the shunting $\beta$.

Note that, at the same time that the intra-cell tunneling is renormalized, the inter-cell tunneling may get activated. This phenomenon is strongly conditioned by the renormalization of the capacitances along different directions and is thus dependent on the qubit's parameters and the circuit topology, as we show in appendix \ref{a. 2D Harmonic approximation}. A consequence of this activation is the fast growth of the $J_{xx}\sigma^x_1\sigma^x_2$ interaction. This is, in our opinion, a regime to be avoided. First, because the $J_{xx}$ can be obtained by other (inductive) means. And second, because the activation of the inter-cell tunneling is accompanied by a greater sensitivity to electrostatic field fluctuations \cite{orlando}.

Having an understanding of the interactions shown in Fig. \ref{f. capacitive coupling}(a), we turn to Figs. \ref{f. capacitive coupling}(b), (c) and (d). If one wants to fully appreciate the real strength of the coupling, the relevant magnitude is not the interaction itself but the interaction in units of the qubit's gap, $J_{ii}/\Delta_q$. Looking at Fig. \ref{f. capacitive coupling}(a) we can check the interactions remain more or less constant after a certain $\gamma$, thus, after this point the defining parameter will be the qubit's gap. As noted, increasing $\gamma$ produces an exponential decay in the qubit's gap due to the renormalization of the mass. This phenomenon is rather helpful, up to a reasonable value of the gap, since it enormously increases the ratios $J_{ii}/\Delta_q$. When considering the ratios $J_{ii}/\Delta_q$ the effect of the design parameters is not obvious, since increasing any of them decreases the interaction but also decreases the gap. For this reason, we have studied the dependency of these ratios as a function of the qubits parameters. 

Different coupling topologies produce qualitatively similar plots, although the relative coupling strength $J_{ii}/\Delta=1$ may be reached for lower or higher values of the capacitance $\gamma$, and the relative sign of the interactions might change. We have also studied different grounding schemes. Topologically, there are two distinct combinations: we can place the grounds in the upper corners---e.g. $\phi_1=0$, $\phi_{2}=0$ or similar for the other qubit---or we can place them at the bottom $\phi_0=\phi_{0'}=0$. Choosing between $\phi_1=0$ or $\phi_2=0$ is equivalent to flipping the flux passing through the qubit, and changes the sign of the $\sigma^y$ and $\sigma^x$ operators. If we choose topologically equivalent grounds for both qubits, we obtain coupling strengths with similar magnitude as the ones seen before. However, there are somewhat pathological choices---e.g. $\phi_1=\phi_{2'}=0$ connecting nodes 0 and $1'$---where the qubits experience different renormalizations, their gaps differ as interaction grows and the resulting interactions loose the symmetry. 

It must be remarked that for all choices of connecting nodes and ground nodes we always obtain \emph{both} $J_{yy}\sigma^y_1\sigma^y_2$ and $J_{zz}\sigma^z_1\sigma^z_2$ interactions simultaneously, with very similar magnitude. This means that we can engineer effective qubit-qubit interactions of the approximate form $J(\sigma_1^z\sigma_2^z\pm\sigma_1^y\sigma_2^y)$, with  $J\approx \Delta$, where the sign depends on the topology.

\begin{figure}[H]
    \centering
    \includegraphics[width=\linewidth]{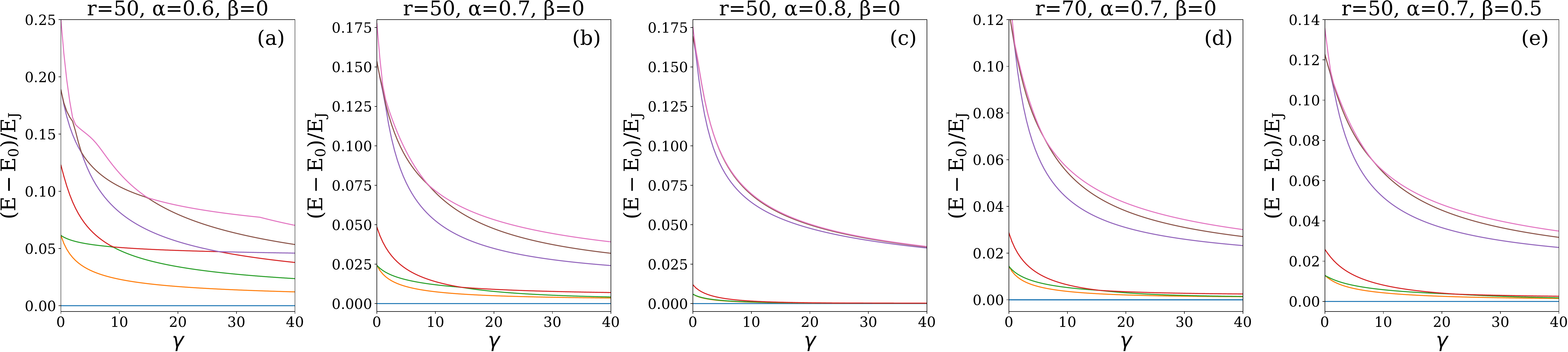}
    \caption{Low energy subspace as a function of $\gamma$ for the reference circuit, two 3JJQs with ground in $\phi_0$($\phi_0'$) coupled through a capacitor connecting nodes $\phi_2-\phi_1'$, and different qubit parameters: (a) $r=50$, $\alpha=0.6$, $\beta=0$, (b) $r=50$, $\alpha=0.7$, $\beta=0$, (c) $r=50$, $\alpha=0.8$, $\beta=0$, (d) $r=70$, $\alpha=0.7$, $\beta=0$, (e) $r=50$, $\alpha=0.7$, $\beta=0.5$.}
    \label{f. capacitive coupling multiple parameters}
\end{figure}

Finally, in fig.~\ref{f. capacitive coupling multiple parameters}, we show the energy spectra obtained from the reference capacitive coupling configuration (Fig.~\ref{f. acoplo C}) for different values of the qubit parameters. As shown in Fig.~\ref{f. capacitive coupling multiple parameters}(a) for small values of the parameter $\alpha$ the qubit's anharmonicity is not sufficiently large and hence there is no distinction between the qubit subspace (four lower energy levels) and the exited subspace. Increasing $\alpha$ results in an increase of the qubit's anharmonicity avoiding this problem but significantly reducing the magnitude of the interaction and qubit's gap, this is shown in Figs.~\ref{f. capacitive coupling multiple parameters} (b) and (c). The effect in the spectrum of increasing $\beta$ and $r$ is really similar to that of $\alpha$: the subspace of the coupled qubits gains definition while the interactions and the qubit's gap decrease (Fig.~\ref{f. capacitive coupling multiple parameters}(d) and (e)).  The results commented here motivate the election of an value of $\alpha$ between $0.6$ and $0.9$ for our study and are in concordance with the results presented.

\subsection{Analysis of 3JJQs coupled thorugh Josephson junctions}
\label{s. numeric JJ coupling}
We have performed a similar study for two identical 3JJ flux qubits, coupled inductively by a single junction. We will now discuss the topology shown in Fig.~\ref{f. acoplo JJ}, but with grounds $\phi_1=\phi_{2'}=0$ (notice again that we neglect the junction's capacitance). This interactions is so strong that around $\gamma\approx 0.1$ it produces a full hybridization of the low and high energy subspaces, a point at which we cannot identify the qubit subspace in the coupled system and hence we cannot extract an effective Hamiltonian with the SWT.

\begin{figure}[H]
    \centering
    \includegraphics[width=0.5\linewidth]{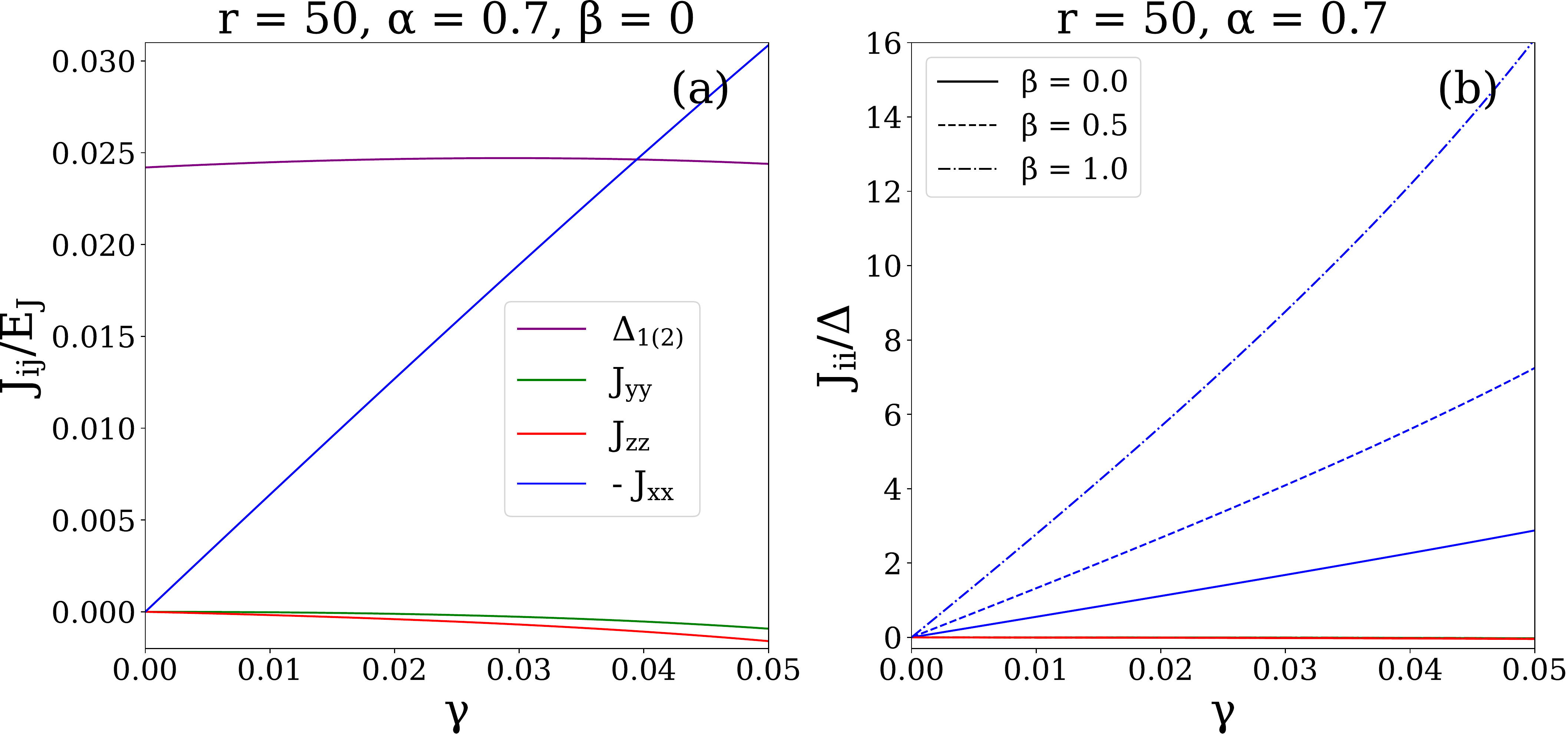}
    \caption{Coupling strengths for two 3JJQs with ground in $\phi_1$($\phi_2'$) coupled through a Josephson junction connecting nodes $\phi_2-\phi_1'$. (a) Effective Hamiltonian parameters as a function of $\gamma$ for $\alpha=0.70$, $r=50$ and $\beta=0$. (b) Ratios between the coupling strengths ($J_{ii}$) and the qubit gap ($\Delta$) for fixed $\alpha$ and $r$. Note that we represent $-J_{xx}$ for the sake of clarity.}
    \label{f. inductive coupling}
\end{figure}

Before this regime, $0<\gamma<0.05$, as illustrated by Fig.~\ref{f. inductive coupling}(a), interactions are dominated by the coupling $J_{xx}\sigma_1^x\sigma_2^x$ between the effective dipolar magnetic moments of both qubits. In addition to this, we find some residual $J_{zz}\sigma^z_1\sigma^z_2$ and $J_{yy}\sigma^y_1\sigma^y_2$ contributions, that are up to three orders of magnitude weaker and can be neglected, as we expected from our previous analysis.

The dependency of the coupling strengths on the 3JJQ parameters offers a simple picture, where the dominant inductive term $J_{xx}\sigma^x_2\sigma^x_2$ grows with $\alpha$ and $r$ (data not shown). This tendency is accompanied by a reduction of the qubit gap for increasing $\alpha$ and $r$. Finally, as it can be extracted from Fig.~\ref{f. inductive coupling}(b), adding a shunting capacitor to the 3JJQs reduces the qubits gap while strengthening the $\sigma^x_2\sigma^x_2$ inductive coupling. This allows for arbitrarily large ratios between the coupling strength and the gap of the qubit leading to stronger couplings but also favoring the crossing between levels inside and outside the qubit subspace for increasingly small values of $\gamma$.

Similar to the capacitive circuit, changing the circuit topology does not affect the qualitative behavior of the interaction with the coupling strength $\gamma$. At most, the choice of coupling points and ground nodes can speed up or slow down the growth of interactions, or change the sign of the corresponding qubit operator---equivalent to changing the flux that threads the loop.

\begin{figure}[H]
    \centering
    \includegraphics[width=\linewidth]{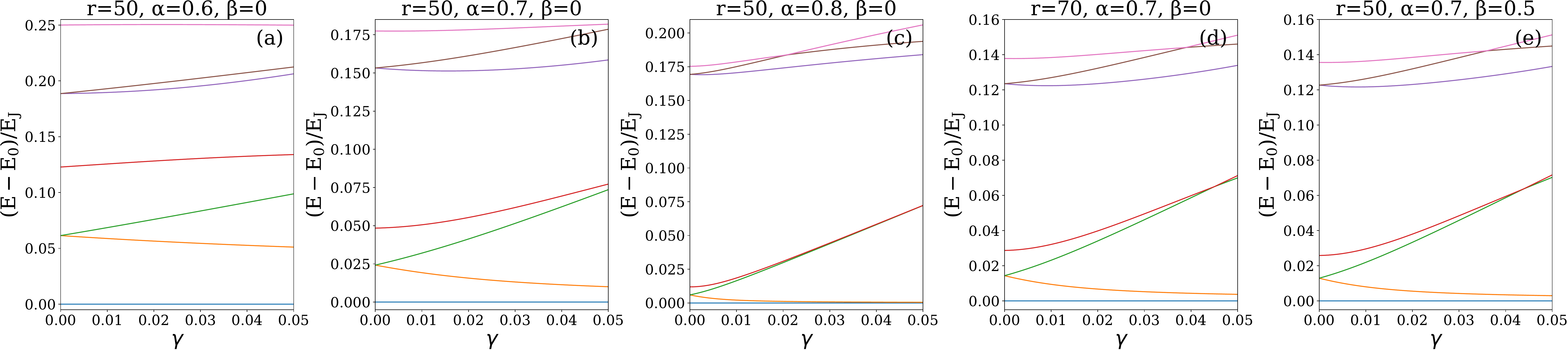}
    \caption{Effective Hamiltonian parameters ($J_{ij}$) as a function of $\gamma$ for different configurations of a circuit compound by two 3JJQs with $\alpha=0.7$ and $r=50$ coupled through a Josephson junction. (a-e) Ground in $\phi_1$($\phi_2'$): (a) coupled connecting $\phi_2-\phi_1'$, (b) coupled connecting coupled connecting $\phi_0-\phi_0'$, (c) coupled connecting $\phi_0-\phi_1'$, (d) coupled connecting $\phi_2-\phi_1'$ and $\phi_0-\phi_0'$, (e) coupled connecting $\phi_2-\phi_0'$ and $\phi_0-\phi_1'$.(f-i) Ground in $\phi_0$($\phi_0'$): (f) coupled connecting $\phi_1-\phi_1'$, (g) coupled connecting $\phi_1-\phi_2'$, (h) coupled connecting $\phi_1-\phi_1'$ and $\phi_2-\phi_2'$, (i) coupled connecting $\phi_1-\phi_2'$ and $\phi_2-\phi_1'$.}
    \label{f. inductive coupling multiple configurations}
\end{figure}

To motivate the range of parameters that we have studied in Fig.~\ref{f. inductive coupling} we show the lower energy spectra of the reference Josephson junction coupling circuit for different values of the qubit parameters. For the range of $\gamma$ considered none of the values of the qubit parameters selected result in an extreme reduction of the coupling and qubit gaps or compromise the application of the SWT (the two subspaces remain separated), nevertheless, there are two important considerations that we have to make when interpreting these graphs. On one hand, it is shown in Fig.~\ref{f. inductive coupling multiple configurations}(a) that, even though no levels of the qubits subspace cross with levels in the high energy subspace, the two subspaces are not clearly differentiated in practice. The reduction of the qubit's anharmonicity for small values of $\alpha$ makes the distance between the two subspaces comparable to the qubits gaps.

\pagebreak

\section{Conclusions}

The aim of this thesis was to introduce and analyze coupling mechanisms between superconducting flux qubits. In the introduction we showed that to comprehend the usefulness and the fundamental differences between these couplings one has to write them in a local qubit basis, that is, as the tensor product of two \textit{effective} operators, each one defined in its respective qubit subspace. In section~\ref{s. Heff} we introduced the  Schrieffer-Wolff transformation (SWT) as a means to obtain these effective operators. The SWT translates operators from a well defined Hilbert subspace to a different subspace of the same dimension, which in our case is the non-interacting qubit subspace. We showed that the SWT can be performed perturbatively—opening an analytic framework to study the interactions—or in an exact numerical manner, and we introduced a novel numerical scheme \cite{hita2021} to perform this transformation more efficiently than current methods \cite{consani2020effective}. 

With a general method to obtain effective operators we switched to a more specific focus and began the analysis of flux qubits. In section~\ref{s. 3JJQ} we reviewed the three Josephson junctions flux qubit (3JJQ) and showed how one can approximate its nonlinear potential with harmonic wells to obtain analytic expressions for its qubit wavefunctions and energy gap, paying special attention to understand where and why this approximation breaks down. We then used these tools in section \ref{s. Electromagnetic interaction with flux qubits} to find the effective representation of the electromagnetic observables of the 3JJQ, and to understand how the 3JJQ interacts as an electric/magnetic dipole with external fields. 

The goal of these sections was to develop an intuition and a general toolbox to study couplings between flux qubits. In section \ref{s. Couplings} we first analyzed the most widely used coupling between flux qubits: coupling through mutual inductance. We showed that this coupling imposes strict conditions on the design of the qubits, does not allow to couple distant qubits and also has qubit-addressability issues when scaled to large systems. As a solution to these obstacles we have proposed to inductively couple flux qubits through a tunable Josephson junction, \textit{i.e.} a dc-SQUID. This coupling preserves the inductive interaction of the mutual inductance coupling without imposing restrictions on the design of the qubits and allowing for a better qubit-addressability.

Both the mutual inductance coupling and the Josephson coupling produce a dominant $\sigma^x\sigma^x$ magnetic interaction, thus, to obtain a two-local non-stoquastic Hamiltonian we had to seek fundamentally different interactions. For this reason we proposed the capacitive coupling of 3JJQs. The capacitive coupling of flux qubits had been proposed for four Josephson junctions flux qubits\cite{satoh2015} and experimentally tested for rf-SQUIDS\cite{ozfidan2020demonstration}. We have explained the origin and scaling of these interactions in the perturbative regime \cite{hita2021ultrastrong}. In sections \ref{s. numeric C coupling} and \ref{s. numeric JJ coupling} we have verified the predictions of our analysis in the perturbative and we have showed that one can achieve different interactions with $J/\Delta\sim1$.

This study suggests the possibility of coupling two flux qubits with a capacitively-shunted dc-SQUID, producing arbitrary interactions of the form $J^\text{cap}(\sigma_1^y\sigma_2^y\pm\sigma_1^z\sigma_2^z)+(J_{xx}^\text{cap}+J_{xx}^\text{JJ})\sigma_1^x\sigma_2^x$, where $J^\text{cap}$ are fixed by design, and $J_{xx}^\text{JJ}$ can be tuned using the magnetic flux that is passes through the SQUID.

Our results confirm the idea that flux qubits may be used to simulate strong non-stoquastic spin Hamiltonians. However, we have found that not all the interactions are independent as found by the simultaneous appearance of $\sigma^y_1\sigma^y_2$ and $\sigma^z_1\sigma^z_2$ terms of equal magnitude in the capacitive coupling. This may have consequences for the interpretation works that argue the classical simulability of superconducting quantum circuits~\cite{stoqciani2021stoquasticity,halverson2020efficient}.

Finally, this work leaves open questions, such as the application of said interactions in the context of quantum computation, where the tunability of the capacitive couplings may become relevant. We expect to analyze this question in future works, combining the capacitive coupling with mediating circuits, such as microwave resonators~\cite{hita2021ultrastrong} or other qubits~\cite{arute2019quantum}.

\bibliographystyle{unsrt}
\bibliography{Bibliography}

\pagebreak

\appendix
\section{Superconducting circuits and Josephson junctions}
\label{a. Josephson junction}
In this section we will briefly cover the necessary concepts to understand superconducting circuits. We encourage the interested reader to check references \cite{vool2017introduction, krantz2019quantum} for a deeper immersion in the subject. When the characteristic length of a circuit is much smaller than the wavelength of its operating frequency one can use a lumped-element model to describe its dynamics. This is the case of the superconducting circuits described in this document, which typically have a size of $\mu m$ and operate in the range of microwaves with wavelengths in the order of $cm$. The lumped-element model of the circuit consists of a series of nodes connected by branches that composed by one or more two-pole elements in parallel. The elements of a branch $b$ at a time $t$ are characterized by two classical variables: the voltage $v_b(t)$ across the elements and the current $i_b(t)$ flowing thorough it. These variables are defined by the electric and magnetic fields in the elements
\begin{equation}
    \begin{aligned}
        v_{b}(t)&=\int_{\text {beginning of } b}^{\text {end of } b} \vec{E}(\vec{r}, t) \cdot \overrightarrow{d \ell}\,, \\
        i_{b}(t)&=\frac{1}{\mu_{0}} \oint_{\text {around } b} \vec{B}(\vec{r}, t) \cdot \overrightarrow{d s} \,,
    \end{aligned}
\end{equation}
where the loop integral of $i_{b}(t)$ is done along a closed curve that encircles the element. Since the power absorbed by an element is the product of its voltage and intensity we can define the energy stored in a element as:
\begin{equation}
    E_b(t) = \int_{-\infty}^{t} v_{b}\left(t^{\prime}\right) i_{b}\left(t^{\prime}\right) d t^{\prime}\,.
\end{equation}
The Hamiltonian description of the circuit requires the introduction of branch fluxes and branch charges, which are defined by
\begin{equation}
    \begin{aligned}
\Phi_{b}(t) &=\int_{-\infty}^{t} v_{b}\left(t^{\prime}\right) d t^{\prime}\,, \\
q_{b}(t) &=\int_{-\infty}^{t} i_{b}\left(t^{\prime}\right) d t^{\prime}\,,
\end{aligned}
\end{equation}
The lower bound of the integrals implies a time in the past where the circuit was at rest, \textit{i.e.} with zero voltages and currents. The branch fluxes and variables allow us to distinguish two types of circuit elements. Capacitive elements are those for which the voltage is only a function of the charge and not directly of the time or any other variable:
\begin{equation}
    v_b(t)= f\left(q_b(t)\right)\,. 
\end{equation}
Linear capacitors are those for which $v^C_b(t)= q_b/C$, where $C$ is the capacitance of the element and is independent of the charge in the capacitor:
\begin{equation}
   v^C_b(t)= \frac{q_b}{C} = \frac{1}{C} \int_{-\infty}^{t} i^C_{b}\left(t^{\prime}\right) d t^{\prime}\,,\quad \to\quad  i^C_{b}(t) = C \frac{d v^C_b(t)}{dt} = C\ddot\phi\,,
\end{equation}
and hence the energy stored in them can be expressed as:
\begin{equation}
    E_b^C(t) = \frac{1}{2C}\left(q_b(t)-q_\text{offset}\right)^2\,.
\end{equation}
Inductive elements are those for which the current is only a function of the flux:
\begin{equation}
    i_b(t) = g\left( \phi_b(t) \right)\,.
\end{equation}
Linear inductors are those for which $i^L_b(t) = \phi_b/L $, where $L$ is the inductance of the element and is independent of the flux across the inductor:
\begin{equation}
    i^L_b(t) = \frac{\phi_b}{L} = \frac{1}{L}\int_{-\infty}^{t} v^L_{b}\left(t^{\prime}\right) d t^{\prime} \quad \to \quad v^L_{b}(t) = L \frac{di_b^L(t)}{dt}\,.
\end{equation}
and hence the energy stored in them can be expressed as:
\begin{equation}
    E_b^L(t) = \frac{1}{2L}\left(\phi_b(t)-\phi_\text{offset}\right)^2\,.
\end{equation}
So far this discussion has been classical with no regard for quantum effects. Nevertheless, if we build the circuits with a superconducting material, such as aluminum, and cool them in a dilution refrigerator to a temperature $k_BT\ll\hbar\omega_0$, the thermal fluctuations become negligible in comparison with the quantum fluctuations associated with the resonant frequency of the circuit $\omega_0$. These circuits are usually cooled to a temperature around $20mK$, ensuring that quantum fluctuations dominate for circuits operating at microwave frequencies of the order of $1-20$ GHz. In the Hamiltonian description of the circuit this condition means that we can promote the classical variables to quantum operators
$$
\begin{aligned}
\phi & \rightarrow \widehat{\phi} \\
q & \rightarrow \widehat{q} \\
H & \rightarrow \widehat{H}
\end{aligned}
$$
where the flux and charge operators of the nodes must satisfy the commutation relation
$$
\left[\phi_i,q_j\right] = i\hbar\delta_{ij}\,.
$$
The only problem of our discussion so far is that with the linear capacitors and inductors that we have introduced the type of circuits that one can build are, for instance, LC circuits. These circuits behave as quantum harmonic oscillators, where the average value of the flux (position) and charge (momentum) operators follow the classical equations of motion. Quantum mechanics is only revealed when one considers higher moments such as $\braket{\phi^2}$ or $\braket{q^2}$, which have the problem of being considerably harder to measure than the averages. Another problem of LC circuits is that they present a harmonic spectrum, and hence are useless if one wants to build a qubit with a clearly differentiable two level energy subspace.

The key to build superconducting circuits with directly observable macroscopic quantum effects and anharmonic spectrums is to use non-linear components. The most commonly used non-linear and non-dissipative electrical component is the Josephson junction. It consists of a ``sandwich'' of two superconductors separated by an insulator, usually two aluminum electrodes separated by a 1nm-thin oxide (alumina) layer. This component is modeled as a non-linear ideal inductor in parallel with a capacitor, which accounts for the capacitor formed by the parallel plates of the superconducting electrodes. The non-linear Josephson inductance is characterized by the current-flux relation:
\begin{equation}
    i_b^J(t) = I_0\sin\left(\varphi_b(t)\right)
\end{equation}
where $\varphi_b(t)$ is the phase difference between the macroscopic wavefunction at each side of the inductance. Here the macroscopic wavefunction is that of the Bose-Einsten condensate formed by the Cooper pairs in the superconductor. One can show \cite{tinkham2004introduction,orlando1991foundations} that this phase is related to the electric flux
\begin{equation}
    \partial_{t} \varphi_b(\mathbf{x}, t)=\frac{2 \pi}{\Phi_{0}} \partial_{t} \phi_b(\mathrm{x}, t)\,,
\end{equation}
where $\Phi_0$ is the magnetic flux quantum
\begin{equation}
    \Phi_0=\frac{h}{2e}\,.
\end{equation}
Note that the presence of $2e$ instead of $e$ in the equation is because the charge unit in a superconductor is the charge of a Cooper pair, fact that was discovered in the first experimental measure of the magnetic flux quantum \cite{deaver1961experimental}. The flux-phase relation allows us to write the current across a Josephson junction as
\begin{equation}
    i_b^J(t) = I_c\sin\left(\frac{2\pi}{\Phi_0}\phi_b(t)\right)
\end{equation}
where $I_c$ is the Josephson junction's critical current, usually in the range of $I_c\sim400-600$ nA, current above which the model presented here breaks down. Notice how we can obtain from this equation the clearly non-linear inductance of the Josephson junction
\begin{equation}
    L_J = \frac{v_b^J(t)}{di_b^J(t)/dt} = \frac{\Phi_0}{2\pi}\frac{1}{I_c\cos\left(2\pi\phi_b(t)/\Phi_0\right)}
\end{equation}
and the energy stored in a Josephson junction
\begin{equation}
    E_b^J(t) = E_J\left[1-\cos\left(\frac{\phi_b(t)-\phi_\text{offset}}{\Phi_0/2\pi}\right)\right]
\end{equation}
where $E_J$ is the Josephson energy and is related to its critical current by $E_J=I_c\Phi_0/2\pi$. Note that this energy has a constant term that we will ignore because it only produces an undetectable phase shift in the wavefunctions.

Another condition of the flux-phase relation is called the fluxoid quantization, and states that the total magnetic flux thorugh a superconducting loop must be an integer number of the magnetic flux quantum. The total magnetic flux through a loop is usually decomposed as the flux through the branch elements of the loop plus any external flux, and hence the fluxoid quantization can be written as:
\begin{equation}
    \oint_C\nabla\phi\cdot dl +\Phi_\text{ext} = n\Phi_0 
\end{equation}

\section{Hamiltonian of a superconducting circuit}
\label{a. circuit hamiltonian}
Following the procedure presented in \cite{circuitquantization}  for circuit quantization, we find that the general Lagrangian for the circuits contemplated in this document reads:
\setcounter{equation}{0}
\begin{equation}
    \label{eq:Lagrangian}
\mathcal{L}=\mathcal{L}_0 +\mathcal{L}_{int}
\end{equation}
where $\mathcal{L}_0=\mathcal{L}_1(\boldsymbol{\phi},\dot{\boldsymbol{\phi}})+\mathcal{L}_2(\boldsymbol{\phi}',\dot{\boldsymbol{\phi}'})$ is the sum of the Lagrangians of the single flux qubits \cite{orlando}, fig. \ref{f. circuito 3JJ}, which ca be derived from the energy stored in the components that we derived in the previous chapter:
\begin{align}
    \mathcal{L}_q(\boldsymbol{\phi},\dot{\boldsymbol{\phi}})
    =&\frac{1}{2}\dot{\boldsymbol{\phi}}\textbf{C}_q\dot{\boldsymbol{\phi}} + E_J\cos{\left(\frac{\phi_1-\phi_0}{\varphi_0}\right)} + \\
    & + E_J\cos{\left(\frac{\phi_2-\phi_0}{\varphi_0}\right)} + \alpha E_J\cos{\left(\frac{\phi_2-\phi_1-\Phi}{\varphi_0}\right)}\notag
\end{align}
with $E_J$ the characteristic Josephson energy of the junctions,  $\Phi$ the externally induced flux, $\phi_i$ the flux variables in the nodes of the circuit, and $\textbf{C}_q$ the capacitance matrix of the flux qubit which depends on the characteristic capacitance of the junctions $C$ and the qubits parameter $\alpha$.
$\mathcal{L}_{int}$ gives the interaction between qubits and depends on the specific coupling. For a Josephson junction connecting two nodes, $i$ and $j'$, of different flux qubits we obtain a contribution of the form:
\begin{equation}
\mathcal{L}_{int}^{JJ}= \gamma_{ij'}^{JJ} E_J \cos{\left(\frac{\phi'_j-\phi_i}{\varphi_0}\right)}
\end{equation}
While a capacitor coupling gives a contribution:
\begin{equation}
\mathcal{L}_{int}^{cap}= \gamma_{ij'}^{cap} \frac{C}{2} (\dot{\phi'_j}-\dot{\phi_i})^2
\label{eq:capintL}
\end{equation}
In both cases, $\gamma_{ij'}$ represents the proportionality constant, between coupling Josephson energy (capacitance)  and the flux qubits reference Josephson energy, $E_J$ (capacitance, $C$).
Using canonical variables, $Q_i=\frac{\partial \mathcal{L}}{\partial\dot{\phi}_i}$, and the Legendre transformation, $H(\mathbf{Q}, \boldsymbol{\phi})=\mathbf{Q}\dot{\boldsymbol{\phi}}-\mathcal{L}(\boldsymbol{\phi},\dot{\boldsymbol{\phi}})$, we can conclude from the previous Lagrangian~\eqref{eq:Lagrangian} that the general Hamiltonian for two coupled 3JJQ is:
\begin{equation}
    H=H_0+H_\text{int}
    \label{eq:Hamiltonian}
\end{equation}
Here,  $H_0=H_1(\mathbf{Q},\boldsymbol{\phi})+H_2(\mathbf{Q}',\boldsymbol{\phi}')$ is the sum of the single flux qubits Hamiltonians (c-shunted or not) whose inverse capacitance matrix may be modified by the action of the coupling  (\textit{renormalization} or \textit{capacitive loading}~\cite{consani2020effective}), $\widetilde{\mathbf{C}}_q^{-1}$,
\begin{equation}
    \begin{aligned}
        H_q(\mathbf{Q},\boldsymbol{\phi})=\frac{1}{2}\mathbf{Q}\widetilde{\mathbf{C}}_q^{-1}\mathbf{Q}- E_J\cos{\left(\frac{\phi_1-\phi_0}{\varphi_0}\right)} \\
        - E_J\cos{\left(\frac{\phi_2-\phi_0}{\varphi_0}\right)} - \alpha E_J\cos{\left(\frac{\phi_2-\phi_1-\Phi}{\varphi_0}\right)}
    \end{aligned}
\end{equation}
Here ${H}_{int}$ describes the interaction between pairs of different flux qubits. For the inductive coupling mediated by a Josephson junction, we find that
\begin{equation}
    H_\text{int}^{JJ}= - \gamma^{JJ}_{ij'} E_J \cos{\left(\frac{\phi'_j-\phi_i}{\varphi_0}\right)}
\end{equation}
where only the two connected nodes, $i$ and $j$, are implicated and no renormalization for the qubits Hamiltonians has to be considered. However, for the electrostatic interaction mediated by a capacitor we get an interaction term of the form
\begin{equation}
     H_\text{int}^{cap}= \mathbf{Q}\mathbf{C}_c^{-1}\mathbf{Q}'
     \label{eq:capint}
\end{equation}
which gives connections between all node charges of different qubits and depends on the inverse mutual capacitance matrix, $\mathbf{C}_c^{-1}$. The single qubits Hamiltonians are renormalized by rescaling the inverse of the capacitance matrix, $\mathbf{C}_q^{-1}(\alpha)\rightarrow\widetilde{\mathbf{C}}_q^{-1}(\alpha, \gamma^{cap}_{ij})$. To fully understand this procedure it is necessary to define the capacitance matrices and inverse capacitance matrices that have been mentioned during the explanation. The full capacitance matrix for our system (including all nodes in both qubits) is defined as follows
\begin{equation}
    \mathbf{C}=
        \begin{pmatrix}
        \widetilde{\mathbf{C}}_q & -\mathbf{C}_c \\
        -\mathbf{C}^{T}_c & \widetilde{\mathbf{C}}_q'
        \end{pmatrix}.
\end{equation}
For example, the full capacitance matrix for the circuit in Fig.~\ref{f. acoplo C}(a) is
\begin{equation}
    {\mathbf{C}}=C
        \begin{pmatrix}
        1+\alpha+\beta & -(\alpha+\beta)        &0                       &  0 \\
        -(\alpha+\beta) & 1+\alpha+\beta+\gamma & -\gamma               &  0 \\
        0                & -\gamma             & 1+\alpha+\beta+\gamma  &  -(\alpha+\beta) \\
        0                & 0                     & -(\alpha+\beta)       & 1+\alpha+\beta
        \end{pmatrix}.
\end{equation}
Thus, $\widetilde{\mathbf{C}}_q$ is the renormalized (or not) 3JJQs capacitance matrix whose elements have the form
\begin{equation}
    \begin{aligned}
        & (\widetilde{\mathbf{C}}_q)_{ii}=(\mathbf{C}_q)_{ii}+\sum_{j'}\gamma_{ij'}C \\
        & (\widetilde{\mathbf{C}}_q)_{ij}=(\mathbf{C}_q)_{ij} \\
        & (\widetilde{\mathbf{C}}_q')_{i'i'}=(\mathbf{C}_q)_{i'i'}+\sum_{j}\gamma_{ji'}C \\
        & (\widetilde{\mathbf{C}}_q')_{i'j'}=(\mathbf{C}_q)_{i'j'} \\
     \end{aligned}
\end{equation}
Here $\mathbf{C}_q=\mathbf{C}_q'$ is the non-renormalized single 3JJQ capacitance matrix. The elements $(C_q)_{ii}$ are given by the sum of all the capacitances connected to the node $i$ when uncoupled, and the elements $-(C_q)_{ij}$ are given by the sum of all capacitances connecting nodes $i$ and $j$.The $\gamma_{ij'}$ are the coupling parameters in~\eqref{eq:capintL} which are $0$ if there's no capacitive coupling involving the corresponding nodes. And $\mathbf{C}_c$ is the mutual capacitance which accounts for the capacitive couplings between nodes
\begin{equation}
(\mathbf{C}_c)_{ij'}=\gamma_{ij'}C.
\end{equation}
This way, we find the renormalized inverse capacitance matrix of the qubits, $\widetilde{\mathbf{C}}_{q}^{-1}$ and $\widetilde{\mathbf{C}}_{q}^{'-1}$, and the inverse mutual capacitance matrix, $\mathbf{C}_c^{-1}$, by performing the inversion of the full capacitance matrix of the system:
\begin{equation}
    \mathbf{C}^{-1}=
        \begin{pmatrix}
        \widetilde{\mathbf{C}}_q^{-1} & \mathbf{C}_c^{-1} \\
        (\mathbf{C}_c^{-1})^T & \widetilde{\mathbf{C}}_q^{'-1}
        \end{pmatrix}.
\end{equation}

\section{Numerical method to solve a superconducting circuit}
We have written the Hamiltonians of our circuits in terms of the charge/number and flux/phase operators, and we have represented the wavefunctions of the circuits in a base of eigenstates of the flux/phase operators. This can be convenient for instance when one works with the harmonic approximation, however, if we could represent the Hamiltonians of the circuits in the discrete base of eigenstates of the number operator we could write the Hamiltonian as an infinite matrix, which can be truncated to a cutoff, and use any of the available eigensolvers to find the eigenstates and energies of the circuit.

One can show \cite{pegg1989phase} the canonical commutation relations between $\hat{\phi}$ and $\hat{q}$ can be manipulated to find that
\begin{equation}
    e^{i \hat{\phi} / \varphi_{0}} \hat{q}=(\hat{q}-2 e) e^{i \hat{\phi} / \varphi_{0}}\,.
\end{equation}
This means that the exponential of the phase operator is the generator of displacements in the space of charges. Since the charge operator can be easily written in the number basis
\begin{equation}
    \hat{q} = -2e\hat{n}=\sum_n(-2en)\ketbra{n}{n}\,,\quad n\in\mathbb{Z}\,,
\end{equation}
we can conclude that the effect of the exponential of the phase operator is $e^{i \hat{\varphi}}\ket{n}=\ket{n-1}$, \textit{i.e.} the exponential of the phase operator is a ladder operator in the number base:
\begin{equation}
    e^{i \hat{\varphi}}=\sum_{n}|n-1\rangle\langle n|
\end{equation}
The energy stored in a Josephson depends on $\cos(\hat{\varphi})$, which can be decomposed in terms of exponentials and written in the number base as
\begin{equation}
\cos (\hat{\varphi})=\frac{1}{2}\sum_n\ketbra{n+1}{n} + \frac{1}{2}\ketbra{n}{n+1}\,,
\end{equation}
which means that the Josephson junction allows processes in which a Cooper pair tunnels in or out of one of its electrodes. With this we have all of the necessary ingredients to write the Hamiltonian of a 3JJQ.

\section{2D Harmonic approximation of the 3JJQ}
\label{a. 2D Harmonic approximation}
\begin{figure}[H]
    \centering
    \includegraphics[width=\linewidth]{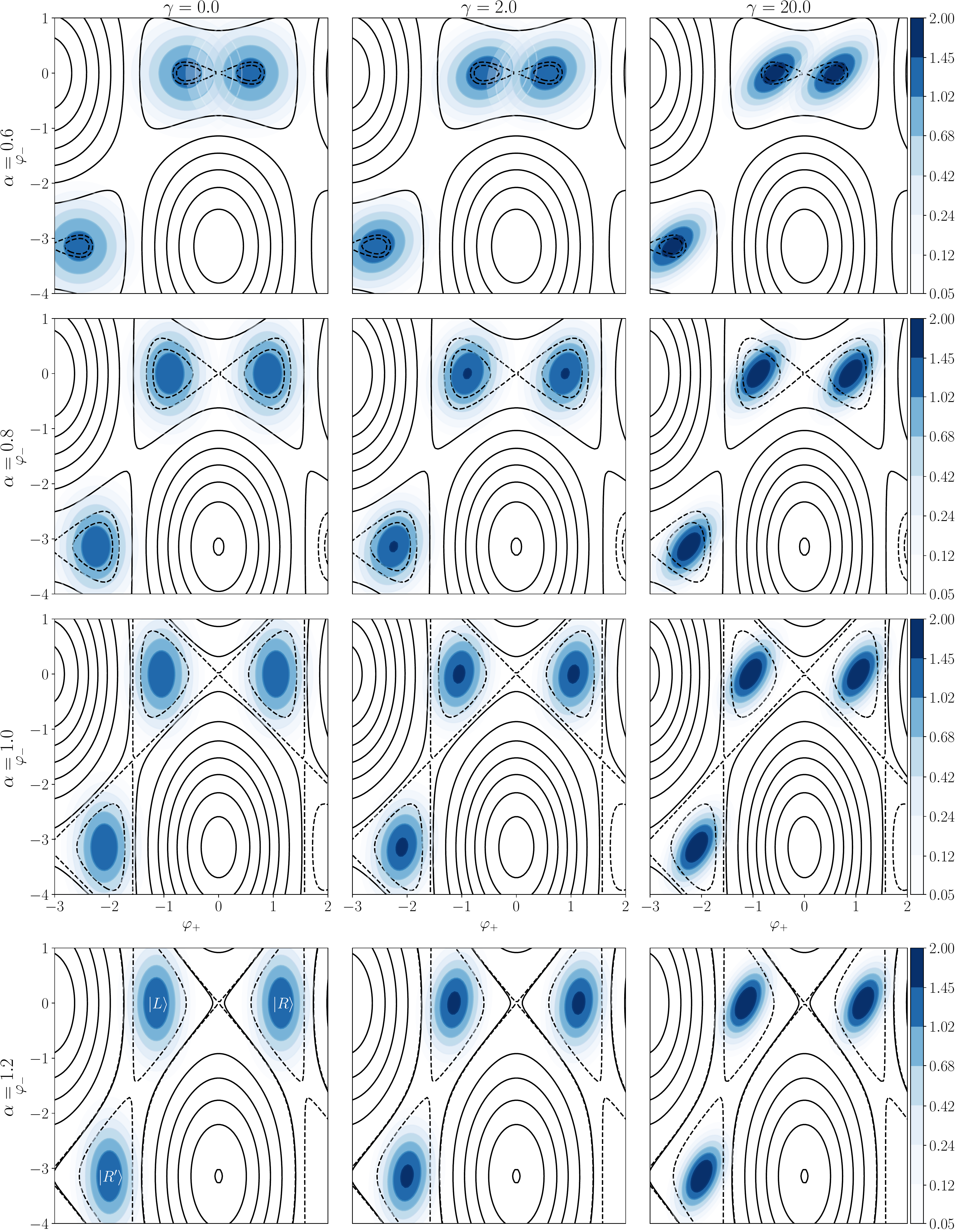}
    \caption{Harmonic wavefunctions, eq. \eqref{e. 2D Harmonic wavefunctions}, for three potential minima as a function of $\alpha$ and $\gamma$. The blue-filled contour plots show the wavefunctions. Note in the colorbar that the contour levels have not been distributed linearly along the height of the wavefunction. on top of a black contour plot of the 3JJQ potential. The solid-black contour lines show the 3JJQ potential and are distributed linearly along the height of the potential. The dashed-black contour lines also show the 3JJQ potential but are distributed linearly along the height of the potential barrier through $t_1$. }
    \label{f. 2D Harmonic wavefunctions}
\end{figure}

If we connect a capacitor of capacitance $\gamma C$ to the node 2 of the 3JJ Flux qubit show in fig. \ref{f. circuito 3JJ}, the Hamiltonian of the circuit at the degeneration point becomes
\begin{equation}
    \hat{H} = \frac{1}{rd} \left[ (\gamma+2)n^2_{+}+(4\alpha+2+\gamma)n^2_{-} + 2\gamma n_{+}n_{-}  \right] -\big[2\cos \left( \varphi_{+} \right) \cos \left( \varphi_{-} \right) -\alpha  \cos \left( 2\varphi_{+}\right) \big]\,,
\end{equation}
where $d=|C|=2\alpha+1+\gamma(\alpha+1)$ is the determinant of the new capacitance matrix. This Hamiltonian can be written in matrix form for any of the potential minima if we introduce the Harmonic approximation, \textit{i.e.} expanding the potential in powers of $\varphi_{+}$ and $\varphi_{-}$ and keeping the quadratic terms. The approximate Hamiltonian for any minimum located at $(\varphi_{+}\!=\!\varphi_{+}^*,\,\varphi_{-}\!=\!\varphi_{-}^*)$ is:
\begin{equation}
    \hat{H}^h =\frac{1}{2} \mathbf{n}^\text{T}\mathbf{ Tn} + \frac{1}{2} \mathbf{\boldsymbol\varphi}^\text{T}\mathbf{ V \boldsymbol\varphi}\,,\vspace{-5mm}
    \end{equation}
where,
\begin{equation}
\mathbf{n}=
    \left[\begin{array}{c}
    n_{+} \\ n_{-} \\ 
    \end{array}\right]
,\quad \boldsymbol\varphi=
    \left[\begin{array}{c}
    \varphi_{+} -\varphi_{+}^* \\ \varphi_{-}-\varphi_{-}^* \\ 
    \end{array}\right]
,\quad \mathbf{T}=\frac{2}{rd}
    \left[\begin{array}{cc}
    \gamma+2 & \gamma \\ \gamma & 4\alpha+2+\gamma \\ 
    \end{array}\right]
,\quad \mathbf{V}=
    \left[\begin{array}{cc}
    \frac{4\alpha^2-1}{\alpha} & 0 \\ 0 & \frac{1}{\alpha} \\ 
    \end{array}\right]\,.
\end{equation}
To solve this Hamiltonian we have to find a set of transformations that allow us to write the kinetic and potential energies in diagonal form. Since the potential energy is already diagonal our first step will be to use a scale transformation to define new phase variables $\boldsymbol{\varphi}_{1}$ that absorb the matrix \textbf{V} and allow us to write the potential energy as $\frac{1}{2}\boldsymbol{\varphi}_{1}^T\boldsymbol{\varphi}_{1}$. Since \textbf{V} is a positive-definite diagonal matrix we can always find its square root and use it as a scale transformation. To ensure that our variables remain canonically commuting we have to apply the inverse scale transformation to the number variables. Applying these transformations:
\begin{equation*}
\boldsymbol{\varphi}_{1} = \mathbf{V}^{\frac{1}{2}}\boldsymbol\varphi\,, \quad \mathbf{n}_{1}=\mathbf{V}^{-\frac{1}{2}}\mathbf{n}\,, \quad \mathbf{T}_{1}=\mathbf{V}^{\frac{1}{2}}\mathbf{T}\mathbf{V}^{\frac{1}{2}}\,,\quad\to\quad \hat{H}^h =\frac{1}{2} \mathbf{n}_{1}^\text{T}\mathbf{T}_{1}\mathbf{n}_{1} + \frac{1}{2} \boldsymbol{\varphi}_{1}^\text{T} \boldsymbol{\varphi}_{1}\,.
\end{equation*}
The second and final step is to apply a rotation to the number variables which diagonalizes the kinetic energy matrix. Since a rotation is a unitary transformations the potential energy will remain diagonal. We can define this rotation as:
\begin{equation}
    \mathbf{R}=    
    \left[\begin{array}{cc}
    \cos(\theta) & -\sin(\theta) \\ \sin(\theta) & \cos(\theta) \\ 
    \end{array}\right]\,, \quad \tan(2\theta) =\frac{\mathbf{T}_{1}(1,2)+\mathbf{T}_{1}(2,1) }{\mathbf{T}_{1}(2,2)-\mathbf{T}_{1}(1,1)}\,.
    \label{e. rotacion}
\end{equation}
Applying this rotation we can finally write the Hamiltonian in diagonal form
\begin{equation}
\widetilde{\boldsymbol{\varphi}} = \mathbf{R}\boldsymbol{\varphi}_{1}\,,\quad \widetilde{\mathbf{n}} = \mathbf{Rn}_{1}\,, \quad \widetilde{\mathbf{T}}=\mathbf{R}\mathbf{T}_1\mathbf{R}^\text{T}\,, \quad\to\quad \hat{H}^h =\frac{1}{2}\widetilde{\mathbf{n}}^{\text{T}}\widetilde{\mathbf{T}} \widetilde{\mathbf{n}} + \frac{1}{2} \widetilde{\boldsymbol{\varphi}}^{\text{T}} \widetilde{\boldsymbol{\varphi}}\,,
\end{equation}
and its ground state in the transformed phase variables:
\begin{equation*}
\braket{\widehat{\widetilde{\boldsymbol{\varphi}}}|g} =g(\boldsymbol{\widetilde{\varphi}})= \left(\pi^{-2}|\widetilde{\mathbf{T}}|^{-\frac{1}{2}}\right)^{\frac{1}{4}}\exp{\left(-\frac{1}{2} \widetilde{\boldsymbol{\varphi}}^\text{T}\widetilde{\mathbf{T}}^{-\frac{1}{2}} \widetilde{\boldsymbol{\varphi}}\right)}\,.
\end{equation*}
We can  undo the transformations to rewrite this eigenstate as a function of the original $\boldsymbol{\varphi}$ variables:
\begin{equation}
\braket{\widehat{\boldsymbol{\varphi}}|g} =g(\boldsymbol{\varphi})= \left(\pi^{-2}|\mathbf{A}|\right)^{\frac{1}{4}}\exp{\left(-\frac{1}{2} \boldsymbol{\varphi}^\text{T}\mathbf{A}\boldsymbol{\varphi}\right)}\,,\quad \mathbf{A} = \mathbf{V}^{\frac{1}{2}}\mathbf{R}^\text{T}\mathbf{T}_2^{-\frac{1}{2}}\mathbf{R}\mathbf{V}^{\frac{1}{2}}\,,\quad \mathbf{|A|}=\sqrt{\frac{|V|}{|T|}}\,.
    \label{e. 2D Harmonic wavefunctions}
\end{equation}
Fig. \ref{f. 2D Harmonic wavefunctions} shows these wavefunctions for different combinations of $\alpha$ and $\gamma$. Finally, we can calculate the overlap between two current states at different wells:

\begin{equation}
\begin{aligned}
    \braket{g_L|g_R} &= 
    \frac{\sqrt{|\mathbf{A}|}}{\pi}
    \int d^2\varphi
    \exp\left({-\frac{1}{2}\big[ (\boldsymbol{\varphi}-\boldsymbol{\varphi}_L)^\text{T}\mathbf{A}(\boldsymbol{\varphi}-\boldsymbol{\varphi}_L) + (\boldsymbol{\varphi}-\boldsymbol{\varphi}_R)^\text{T}\mathbf{A}(\boldsymbol{\varphi}-\boldsymbol{\varphi}_R) \big]}\right) \\
    &= 
    \frac{\sqrt{|\mathbf{A}|}}{\pi} 
    \exp\left({-\frac{1}{2}\big[ \boldsymbol{\varphi}_L^\text{T}\mathbf{A}\boldsymbol{\varphi}_L +  \boldsymbol{\varphi}_R^\text{T}\mathbf{A}\boldsymbol{\varphi}_R \big]}\right)
    \int d^2\varphi 
    \exp\left({-\frac{1}{2} \boldsymbol{\varphi}^\text{T}2\mathbf{A}\boldsymbol{\varphi} + (\boldsymbol{\varphi}_L+\boldsymbol{\varphi}_R)^\text{T}\mathbf{A}\boldsymbol{\varphi}}\right) \\
    &= 
    \frac{\sqrt{|\mathbf{A}|}}{\pi} 
    \exp\left({-\frac{1}{2}\big[ \boldsymbol{\varphi}_L^\text{T}\mathbf{A}\boldsymbol{\varphi}_L +  \boldsymbol{\varphi}_R^\text{T}\mathbf{A}\boldsymbol{\varphi}_R \big]}\right)
    \frac{2\pi}{\sqrt{|2\mathbf{A}|}}
    \exp\left({\frac{1}{4} (\boldsymbol{\varphi}_L+\boldsymbol{\varphi}_R)^\text{T}\mathbf{A}(\boldsymbol{\varphi}_L+\boldsymbol{\varphi}_R) }\right) 
    \\
    &= \exp\left({-\frac{1}{4} (\boldsymbol{\varphi}_L-\boldsymbol{\varphi}_R)^\text{T}\mathbf{A}(\boldsymbol{\varphi}_L-\boldsymbol{\varphi}_R) }\right) 
\end{aligned}
\end{equation}

\end{document}